\documentclass[11pt,a4paper]{article}

\usepackage[utf8]{inputenc}
\usepackage{graphicx}
\usepackage{float}
\usepackage{afterpage}
\usepackage{epstopdf}
\usepackage{epsfig,cite}
\usepackage{amssymb}
\usepackage{amsmath}
\usepackage{dsfont}
\usepackage{multirow}
\usepackage{url}
\usepackage[colorlinks=false]{hyperref}

\usepackage{url}
\usepackage{hyperref}

\newcommand{\be}{\begin{equation}}
\newcommand{\ee}{\end{equation}}
\newcommand{\bea}{\begin{eqnarray}}
\newcommand{\eea}{\end{eqnarray}}
\newcommand{\bi}{\begin{itemize}}
\newcommand{\ei}{\end{itemize}}
\newcommand{\ben}{\begin{enumerate}}
\newcommand{\een}{\end{enumerate}}
\newcommand{\la}{\left\langle}
\newcommand{\ra}{\right\rangle}
\newcommand{\lc}{\left[}
\newcommand{\rc}{\right]}
\newcommand{\lp}{\left(}
\newcommand{\rp}{\right)}

\def\frac#1#2{{{#1}\over {#2}}}
\def\gsim{\mathrel{\rlap{\lower4pt\hbox{\hskip1pt$\sim$}}
    \raise1pt\hbox{$>$}}}         
\def\lsim{\mathrel{\rlap{\lower4pt\hbox{\hskip1pt$\sim$}}
    \raise1pt\hbox{$<$}}}         

\newcommand{\cov}{\mathrm{cov}}

\newcommand{\draft}[1]{}

\newcommand\abs[1]{\left|#1\right|}
\def\beq{\begin{equation}}  
\def\eeq{\end{equation}}  

\def \n0{N_j^{(0)}}

\def\lapprox{\lower .7ex\hbox{$\;\stackrel{\textstyle <}{\sim}\;$}}
\def\gapprox{\lower .7ex\hbox{$\;\stackrel{\textstyle >}{\sim}\;$}}

\begin{document}

\begin{figure}[h]
\epsfig{width=0.35\textwidth,figure=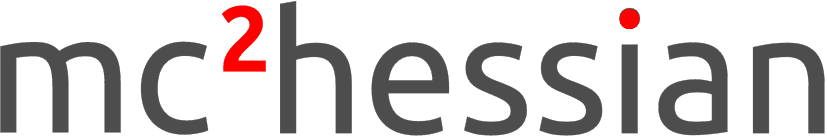}
\end{figure}

\begin{flushright}
TIF-UNIMI-2015-1\\
OUTP-15-04P \\
\end{flushright}

\vspace{0.4cm}

\begin{center}
  {\Large \bf An Unbiased Hessian Representation for Monte Carlo PDFs}
\vspace{.7cm}

Stefano Carrazza$^1$, Stefano Forte$^1$, Zahari Kassabov$^{2,1}$,
Jos\'e Ignacio Latorre$^3$ and Juan Rojo$^4$

\vspace{.3cm}
{\it ~$^1$ TIF Lab, Dipartimento di Fisica, Universit\`a di Milano and
INFN, Sezione di Milano,\\ Via Celoria 16, I-20133 Milano, Italy\\
~$^2$ Dipartimento di Fisica, Universit\`a di Torino and
INFN, Sezione di Torino,\\ Via Pietro Giuria 1, I-10125 Torino, Italy\\
~$^3$ Departament d'Estructura i Constituents de la Mat\`eria, 
Universitat de Barcelona,\\ Diagonal 647, E-08028 Barcelona, Spain\\
~$^4$ Rudolf Peierls Centre for Theoretical Physics, 1 Keble Road,\\ University of Oxford, OX1 3NP Oxford, UK\\}
\end{center}

\vspace{0.1cm}

\begin{center}
{\bf \large Abstract}

\end{center}

We develop a methodology for the construction of a
Hessian representation of
Monte Carlo sets of parton distributions, based on the use of a subset
of the
Monte Carlo PDF replicas as an unbiased linear basis, and of a genetic
algorithm for the determination of the optimal basis.
We validate the methodology by first showing that it faithfully reproduces a
native Monte Carlo PDF set (NNPDF3.0), and then, that if
applied to Hessian PDF set (MMHT14) which was transformed into a Monte Carlo
set, it gives back the starting PDFs with minimal information loss.
We then show that, when applied to a large Monte Carlo PDF set
obtained as combination of several underlying sets, the methodology
leads to a Hessian representation in terms of a rather smaller set of
parameters (MC-H PDFs), 
thereby providing an alternative implementation of the recently
suggested Meta-PDF idea and a Hessian version of the recently
suggested PDF compression algorithm (CMC-PDFs).
The  {\tt mc2hessian} conversion code is made publicly available
together with (through  {\tt LHAPDF6}) a
Hessian representations of the NNPDF3.0 set, and  the MC-H PDF set.

\clearpage

\tableofcontents

\section{Introduction}

The reliable treatment of uncertainties on  the
Parton Distributions  (PDFs) of the proton is currently an
essential ingredient for LHC phenomenology (see for example
Refs.~\cite{Forte:2010dt,Perez:2012um,Forte:2013wc,DeRoeck:2011na,Ball:2012wy} for recent
reviews).
PDF uncertainties are of a peculiar nature, because they are
uncertainties on a space of functions, and two main methods have been used to
provide a representation of them: the Hessian method and the Monte
Carlo (MC) method. 

In the Hessian method (currently used for instance in
the MMHT14~\cite{Harland-Lang:2014zoa} and CT10~\cite{Gao:2013xoa} PDF sets),  a
parametrization based on a fixed functional form is introduced, and a
multigaussian probability distribution is assumed in the space of
parameters. Uncertainties are then given as the inverse of the
covariance matrix of this multigaussian distribution.
This is usually obtained,
assuming linear error propagation and the least-squares method, as
the Hessian matrix with respect to the parameters of a figure of merit
($\chi^{2}$) at its minimum, which is viewed as the best-fit PDF. In
the Monte Carlo method (currently used for instance in the
NNPDF3.0~\cite{Ball:2014uwa} PDF set) PDFs are delivered as an ensemble of
replicas which provide a discrete (Monte Carlo) representation of the underlying
probability distribution: uncertainties are then simply obtained as moments
of this probability distribution.

The Monte Carlo method has the twofold advantage that no Gaussian and
linear error propagation assumption is necessary, and also, that PDFs
can then parametrized with a general-purpose  functional form  with a
large number of parameters (such as neural networks), for which the
least-squares method would fail. The Hessian method, on the other
hand, has the advantage that the (orthogonal) eigenvectors of the Hessian matrix
may be treated as nuisance parameters. This is often a desirable
feature when PDFs are used in experimental analysis,  because other sources of
uncertainty are also represented as nuisance parameters, and also
because through standard methods~\cite{Pumplin:2009nm} it is then
possible to the determine a subset of nuisance parameters which is most
important for a given cross-section or distribution.

Whereas deviations from Gaussianity may be important 
in specific kinematic regions, especially when limited
experimental measurements are available and PDF uncertainties are driven by
theoretical constraints (such as for example the large-$x$ region,
relevant for new physics searches), in most cases, and
specifically when PDF uncertainties are small and driven by abundant
experimental data, the Gaussian approximation is reasonably accurate.
This then raises the question of whether in such case, in which
everything is Gaussian and the Hessian approximation is adequate,
one could have the best of
possible worlds: a Hessian representation with the associate
advantages, but without having to give up the use of a general-purpose
flexible functional form.

It is the purpose of the present paper to achieve this goal. 
We will do this by  using the MC replicas
themselves as the basis of the linear representation of the original
MC sample. Indeed, we will show that if replicas of a very large Monte
Carlo set ($N_{\rm rep}$=1000 replicas) are represented as a linear combination of a
subset of them, not only it is possible to achieve very good accuracy
by using a much smaller subset of replicas as basis functions, but in fact there is an optimal number of
basis replicas, in that  the degeneracy of
replicas is such that larger bases would no longer be linearly
independent. It turns out that this optimal number is quite small, of
the same order of magnitude as 
the typical number of Hessian eigenvectors for standard PDF sets such
as MMHT14 or CT10. All this is true if
the basis replicas are suitably chosen, which we do using a genetic algorithm.
We can then simply construct a  Hessian representation in the space of
these linear expansion coefficients, with essentially  no information
loss or further bias introduced
in comparison to the starting Monte Carlo representation. It is thus
possible to provide a faithful, unbiased  Hessian representation of
any Monte Carlo PDF set, such as those provided by NNPDF.

It is interesting to observe that the inverse problem, namely the
conversion of a Hessian PDF set into a Monte Carlo representation, has
already been considered and solved~\cite{Watt:2012tq}.
An
important advantage of being able to provide a MC
representation of Hessian sets is the construction
of combined PDF sets, which incorporate the information contained in
several individual sets, as required for instance for
Higgs boson coupling extraction or New Physics
searches at the LHC.
Currently, the recommended procedure (the so-called PDF4LHC
recommendation~\cite{Botje:2011sn,Alekhin:2011sk}) is to take an envelope, which has
no clear statistical meaning. However, once converted into a Monte
Carlo representation, PDFs based on a common dataset can be combined
in a simple
way~\cite{Forte:2010dt,Watt:2012tq,Forte:2013wc,Gao:2013bia}.
In this
context, also the problem discussed in this work, namely the conversion from
Monte Carlo to Hessian, has also been handled in the so-called
``Meta-PDF'' approach~\cite{Gao:2013bia}. In this approach, a 
functional form similar to those used in the MMHT14 and CT analyses, the
``meta-parametrization'', is fitted to
the combined Monte Carlo PDF set. This clearly achieves the same
goal as the conversion considered here, but with the further usual bias that a
choice of functional form entails. If appplied to a combined PDF set, 
the methodology presented here
provides thus an unbiased alternative to the Meta-PDF method of
Ref.~\cite{Gao:2013bia}. 

The paper is organized as follows. In Sect.~\ref{sec:methodology} we
describe in the detail our methodology for the Monte Carlo to Hessian
conversion. 
Then, in Sect.~\ref{sec:validation} we first, apply our methodology to a
native Monte Carlo set, NNPDF3.0, benchmark its accuracy, and show
that we end up with an optimal number of Hessian eigenvectors of order
of a hundred. We then apply the methodology to a Monte Carlo set
obtained by applying the Watt-Thorne~\cite{Watt:2012tq} method to a
starting Hessian PDF set, MMHT14. This provides a closure test of the
methodology: we can check explicitly that the starting set is
reproduced very well.
Finally,
in Sect.~\ref{sec:cmcpdfs} we provide a Hessian representation of a
Monte Carlo set obtained by combining several underlying PDF sets
(either native Monte Carlo or converted to Monte Carlo from
Hessian). We end up with a set of eigenvectors, the MC-H PDFs, 
which is of similar
size of the compressed Monte Carlo PDF obtained
recently~\cite{Carrazza:2015hva}  by applying
compression algorithms to the large combined replica set, the so-called
CMC-PDFs.
Therefore,
the PDF set which we obtain in this case provides an alternative to
either the Meta-PDFs of Ref.~\cite{Gao:2013bia}, of which it provides
an unbiased version, or to the CMC-PDFs of
Ref.~\cite{Carrazza:2015hva}, of which it provides a Hessian version.
Details of PDF delivery in {\tt LHAPDF6} are presented  in
Sect.~\ref{sec:delivery}, where conclusions are also drawn.
In Appendix~\ref{sec:appendix} we discuss an alternative strategy to
construct a Hessian representation of MC sets, which is used
to validate our main methodology, and might turn out to be
advantageous for future applications.

\section{Methodology}
\label{sec:methodology}

As discussed in the introduction, the basic idea of our approach is to
construct a linear representation for a set of Monte Carlo PDF
replicas by expressing them as a linear combination of a small subset of
them. Linearized error propagation, which is at the basis of the
Hessian approach, can then be applied to the expansion coefficients,
which immediately provide a representation of the Hessian matrix. It
is important to observe that by ``PDF replica'' we mean the full set
of PDFs at the parametrization scale, i.e., seven PDFs provided as a
function of $x$ for some fixed $Q^2$ value, denoted in the following
by $Q_0^2$.
These are all represented
as a linear combination of the basis replicas with fixed coefficients,
which thus do not depend on either the PDF  or $x$ value. Note
that, because of the linearity of perturbative evolution, once a
replica is expressed as a linear combination at the reference scale,
all PDFs at all scales (including heavy flavors generated  dynamically 
above the corresponding thresholds) are then given by the same linear combination
of basis replicas. This in particular ensures that sum rules are
automatically satisfied.

We will first, describe how the Hessian matrix is constructed, and then,
the optimization of parameters that characterize the procedure,
specifically the choice of basis replicas.

\subsection{Construction of the Hessian matrix}
\label{sec:construction}

We start assuming that we are given a prior set of PDFs represented as
 MC replicas $\{ f^{(k)}_\alpha
\}_{k=1,\ldots,N_{\rm rep}}$ where $\alpha=1,\ldots,N_{\rm pdf}$
denotes the type of
PDF, i.e. $N_{\rm pdf}=2N_f+1$: $N_f$ quarks and antiquarks
and the gluon.
In order to simplify the notation, 
we drop the explicit dependence of the PDFs on $x$ and
$Q^2$.
The central  idea of our strategy consists of finding a subset of replicas, denoted by $\{
\eta^{(i)}_\alpha \}_{i=1,\ldots,N_{\rm eig}} \subset
\{f^{(k)}_\alpha\}$, such that any replica of the prior set,
$f^{(k)}_\alpha$, can be represented as a linear combination
\begin{equation}
  f_\alpha^{(k)} \approx f^{(k)}_{H,\alpha} \equiv f^{(0)}_\alpha +
  \sum_{i=1}^{N_{\rm eig}} a_i^{(k)} (\eta^{(i)}_\alpha -
  f^{(0)}_\alpha)\, , \quad k=1,\ldots,N_{\rm rep} \, ,
  \label{eq:hessian}
\end{equation}
where $f^{(0)}_\alpha$ is the central (average) value of the prior MC set;
$a_i^{(k)}$ are constant coefficients,
independent of $\alpha$, $x$ and $Q^{2}$; and
 $f^{(k)}_{H,\alpha}$ denotes the new Hessian representation
of the original replica $f_\alpha^{(k)}$.
Note that by construction
the central value of the Hessian representation is the same as that
of the original MC set.

In order to determine the parameters $\{a_i^{(k)}\}$ we first define
the covariance matrix in the space of PDFs for the prior set of replicas as
\be
\label{eq:covmat}
{\rm cov}^{\rm pdf}_{ij,\alpha\beta} \equiv \frac{N_{\rm rep}}{N_{\rm
    rep} -1 } \left( \la f^{(k)}_{\alpha}(x_i,Q^2_0)\cdot
f^{(k)}_{\beta}(x_j,Q^2_0)\ra_{\rm rep} -
\la f^{(k)}_{\alpha}(x_i,Q^2_0) \ra_{\rm rep}
\la f^{(k)}_{\beta}(x_j,Q^2_0)\ra_{\rm rep} \right) \, ,
\ee
where the averages are performed over the original set of
$N_{\rm rep}$ replicas. Then, we construct a figure of merit, 
 $\chi^{2(k)}_{\rm pdf}$:
\begin{equation}
\label{eq:chi2def}
\chi^{2(k)}_{\rm pdf} \equiv \sum_{i,j=1}^{N_x}
\sum_{\alpha,\beta=1}^{N_{\rm f}}\Bigg( \lc
f_{H,\alpha}^{(k)}(x_i,Q^2_0) - f_{\alpha}^{(k)}(x_i,Q^2_0)\rc \cdot
\lp {\rm cov^{pdf}}\rp^{-1}_{ij,\alpha\beta} \cdot \lc
f_{H,\beta}^{(k)}(x_j,Q^2_0) - f_{\beta}^{(k)}(x_j,Q^2_0)\rc \Bigg) \,
. 
\end{equation}
Note in Eqs.~(\ref{eq:covmat}) and~(\ref{eq:chi2def}) the use
of the the subscript ``pdf'', to avoid any confusion with the covariance
matrix and the 
$\chi^2$ in the space of experimental data, which do not play any
role here.

The optimal set of expansion coefficients $\{ a_i^{(k)} \}$ for each of the original
$N_{\rm rep}$ replicas is determined by 
 minimization of Eq.~(\ref{eq:chi2def}). This is a convex problem which
can be solved in an efficient way through Singular Value Decomposition
(SVD) techniques. The  problem consists of
finding the vector $\vec{a}$ of dimension $N_{\rm eig}$ that minimizes
the residual of a linear system of dimensions $(N_{x}N_{f})\times
N_{\rm eig}$ for each replica of the original set. The PDF
covariance matrix Eq.~(\ref{eq:covmat}) can be viewed as a
$(N_{x}N_{f})\times(N_{x}N_{f})$ matrix ${\rm cov}^{\rm pdf}_{lm}$,
with indices $l,m$ related to those of the original definition by
$l=N_x(\alpha-1) + i$ and $m=N_x(\beta-1) + j$ . Then we define an
$N_xN_{\rm pdf}\times N_{\rm eig}$ matrix $Y_{mq}$ as 
\be
Y_{mq} =
\eta^{(q)}_\beta(x_j)\,, \ee with the same definition for the index
$m$. We can now lay out the linear system by defining 
\be
\label{eq:matrixes}
A_{lq} =
\sum_{m=1}^{N_{x}N_{\rm pdf}} \left({\rm cov}^{\rm
    pdf}\right)^{-\frac{1}{2}}_{lm} Y_{mq}\,,\quad b_l^{(k)}=
\sum_{m=1}^{N_{x}N_{\rm pdf}} \left({\rm cov}^{\rm
    pdf}\right)^{-\frac{1}{2}}_{lm} f_{\alpha}^{(k)}(x_i)\,, 
\ee
again, with $l=N_x(\alpha-1) + i$. Here $({\rm cov}^{\rm
pdf})^{-\frac{1}{2}}$ stands for a square root of inverse covariance
matrix, i.e., for  a semi-positive definite real matrix
$A$, the matrix such that $(A^{\frac{1}{2}})^tA^{\frac{1}{2}}=A$.  Finally
we can recast the original problem Eq.~(\ref{eq:chi2def}) as that of
finding
$\vec{a}$ that minimizes $\left\Vert A\vec{a}-b^{(k)}\right\Vert$.

If the starting number of MC replicas is large enough, they will not
all be linearly independent. In such case, if the 
 number of eigenvectors $N_{\rm eig}$ is too large, 
the system will be over-determined and the solution will be degenerate
in the space of linear expansion coefficients $\vec{a}$.
In these conditions,
the
correlations between this parameters will be ill-defined, and will
result in a numerically unstable covariance matrix
Eq.~(\ref{eq:covmata}).
On the other hand, if the number of eigenvectors $N_{\rm eig}$ is too small,
it will not be possible to achieve a small value 
of the figure of merit Eq.~(\ref{eq:chi2def}) and the 
Hessian representation of the original covariance matrix will be a
poor approximation.
Therefore,  on quite general grounds one expects
that, if one starts with an extremely large (``infinite'') number of
MC replicas,  there will always be an
optimal value of $N_{\rm eig}$.

In Eq.~(\ref{eq:chi2def}) we
have introduced a sampling in $x$, with a total of $N_x$
points. This immediately raises the issue of choosing both a suitable
spacing and range of the grid of points in $x$. 
Because PDFs are generally quite smooth, 
neighboring points in $x$ are highly correlated, and thus the $x$-grid
cannot be too fine-grained, otherwise the matrix
$\cov^{\rm pdf}$ rapidly becomes ill-conditioned. Furthermore,   
the choice of the $x$-grid range must keep into account not only the
fact that we want the replicas to be especially well-reproduced where
they are accurately known (hence the grid should not be dominated by
points in extrapolation regions), but also that the whole procedure is
meaningful only if the starting probability distribution is at least
approximately Gaussian. The way both issues are handled will be
discussed in detail 
in
Sect.~\ref{sec:numerical} below.

Having determined the expansion coefficients $\{ a_i^{(k)} \}$,
we obtain the  eigenvector directions which describe our original
replica set by  computing their covariance matrix:
\begin{eqnarray}
\label{eq:covmata}
   {\rm cov}^{(a)}_{ij} = \frac{N_{\rm rep}}{N_{\rm rep}-1} \lp \la   a_i^{(k)} a_j^{(k)}\ra_{\rm rep}  -
   \la   a_i^{(k)}\ra_{\rm rep}
   \la  a_j^{(k)}\ra_{\rm rep} \rp \, , \qquad
   i,j=1,\ldots,N_{\rm eig}\, .
\end{eqnarray}
This covariance matrix in the space of the linear expansion parameters
Eq.~(\ref{eq:covmata}) should not be confused with the covariance
matrix in the space of PDFs, defined in Eq.~({\ref{eq:covmat}) (hence
  the different superscripts).
The Hessian matrix is then the inverse of ${\rm cov}^{(a)}_{ij}$,
which we can diagonalize through a rotation matrix $v_{ij}$, thus
obtaining a set of eigenvalues $\lambda_i$ (as in the Meta-PDF method
of Ref.~\cite{Gao:2013bia}).

 We thus obtain the one-sigma uncertainty band associated
to each orthogonal direction by normalizing by $\sqrt{\lambda_i}$.
Therefore the total one-sigma uncertainty will be given by
\be
\sigma^{\rm PDF}_{H,\alpha}(x,Q^2) = \sqrt{ \sum_{i=1}^{N_{\rm
   eig}}\left[\sum_{j=1}^{N_{\rm
   eig}}\frac{v_{ij}}{\sqrt{\lambda_i}}\left(\eta^{(j)}_\alpha(x,Q^2)-f^{(0)}_\alpha(x,Q^2)
\right)\right]^2}\, ,
\label{eq:eigs}
\ee
and our final  Hessian representation of the original Monte
Carlo PDF set is composed by $N_{\rm eig}$ symmetric
eigenvectors, given by
\be
\widetilde{f}_\alpha^{(i)}(x,Q^2)=f^{(0)}_\alpha(x,Q^2)+\sum_{j=1}^{N_{\rm
   eig}}\frac{v_{ij}}{\sqrt{\lambda_i}}\left(\eta^{(j)}_\alpha(x,Q^2)-f^{(0)}_\alpha(x,Q^2)
\right) \, .
\ee
For PDF sets obtained with this Hessian representation, one
should use the symmetric Hessian formula, namely, the one-sigma PDF
uncertainty will be given by 
\be
\label{eq:sigma}
\sigma^{\rm PDF}_{H,\alpha}(x,Q^2) = \sqrt{ \sum_{i=1}^{N_{\rm eig}}
\lp \widetilde{f}_{\alpha}^{(i)}(x,Q^2) -  f_{\alpha}^{(0)}(x,Q^2)   \rp^2} \, ,
\ee
which is the practical recipe for Eq.~(\ref{eq:eigs}).
An analogous expression should be used for the computation of
PDF uncertainties in physical cross-sections.

If the method is successful, Eq.~(\ref{eq:sigma}) should be close to
the original result for the one-sigma PDF uncertainty in the MC representation,
namely
\be
\label{eq:sigmaMC}
\sigma^{\rm PDF}_{\alpha}(x,Q^2) = \sqrt{  \la
  \lp {f}^{(k)}_{\alpha}(x,Q^2)\rp^2\ra_{\rm rep} -  \la
  {f}^{(k)}_{\alpha}(x,Q^2)\ra_{\rm rep}^2 } \, .
\ee
Of course, once the Hessian representation is available, all the
Hessian technology can be used, like the dataset
diagonalization method~\cite{Pumplin:2009nm} or the computation of the
correlation coefficients
between different cross-sections~\cite{Lai:2010nw}.
Likewise, one can now easily include the orthogonal
eigenvectors in a nuisance parameter analysis.

\subsection{Optimization} 
\label{sec:numerical}

We discuss now the determination of the optimal set of parameters
which characterize our procedure, and specifically:
\begin{itemize}
\item the optimal grid of points in $x$ over which the figure of merit
  Eq.~(\ref{eq:chi2def}) is evaluated;
\item the optimal number of eigenvectors $N_{\rm eig}$ and the optimal
  choice of the basis basis replicas.
\end{itemize}
We consider in particular the application of our method to a
prior set of $N_{\rm rep} = 1000$ MC replicas from the 
 NNPDF3.0 NLO set. We then consider also $N_{\rm rep} = 1000$ MC
 replicas  of the   MMHT14 NLO set,  constructed  from the original
Hessian representation using the Hessian to Monte Carlo conversion
methodology of Ref.~\cite{Watt:2012tq}.

\begin{figure}[t]
  \begin{center}
    \includegraphics[width=0.52\textwidth]{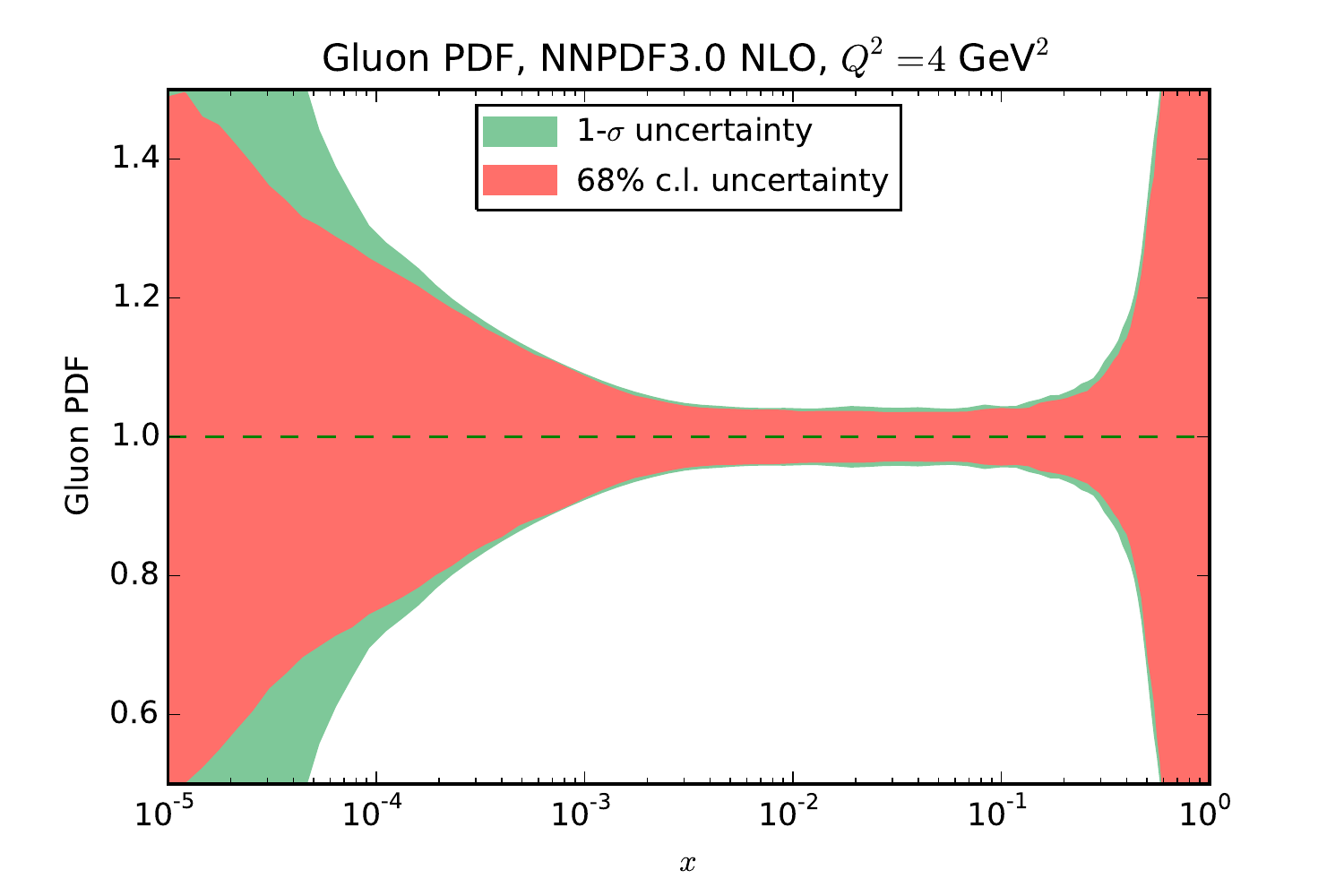}\includegraphics[width=0.52\textwidth]{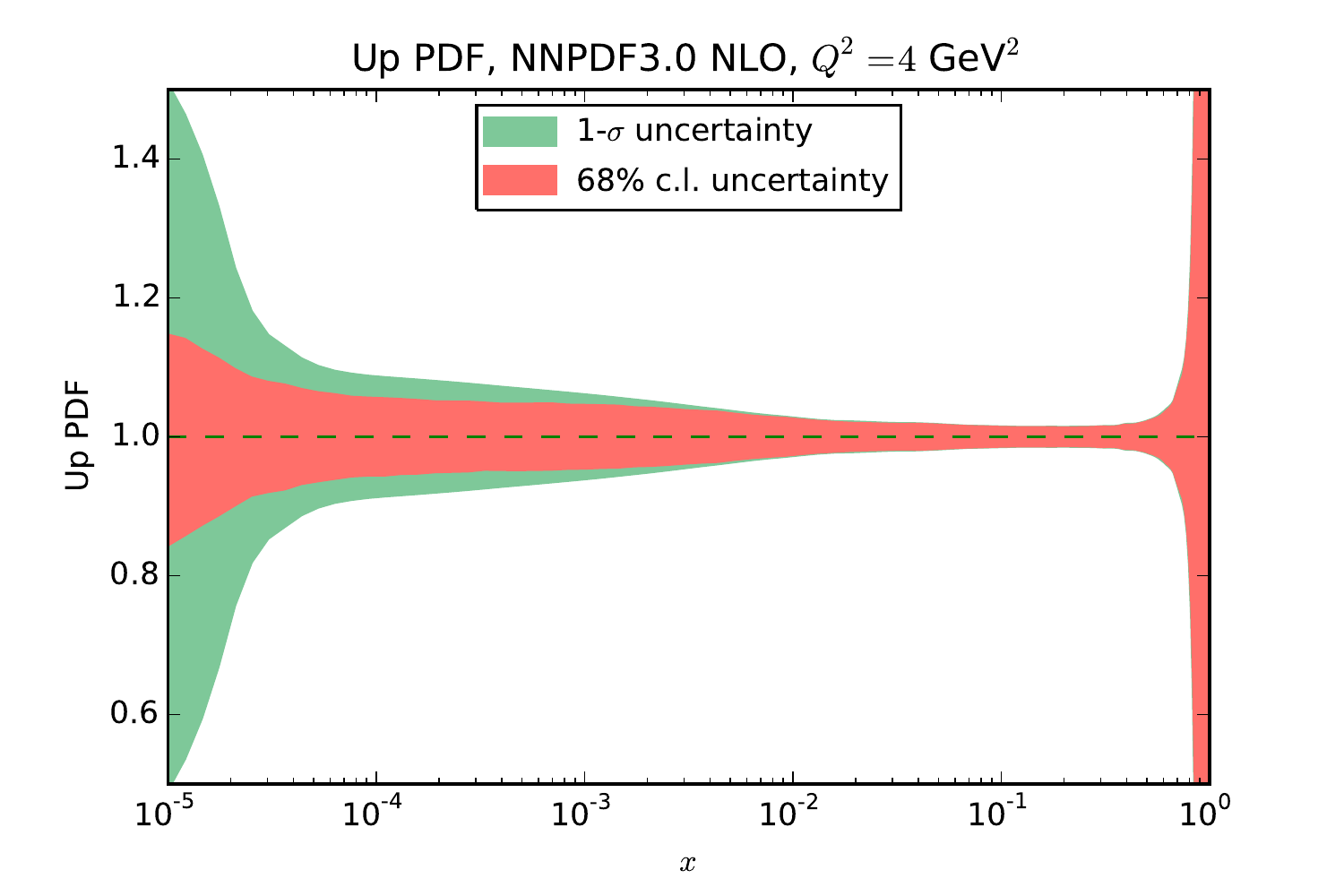}\\
    \includegraphics[width=0.52\textwidth]{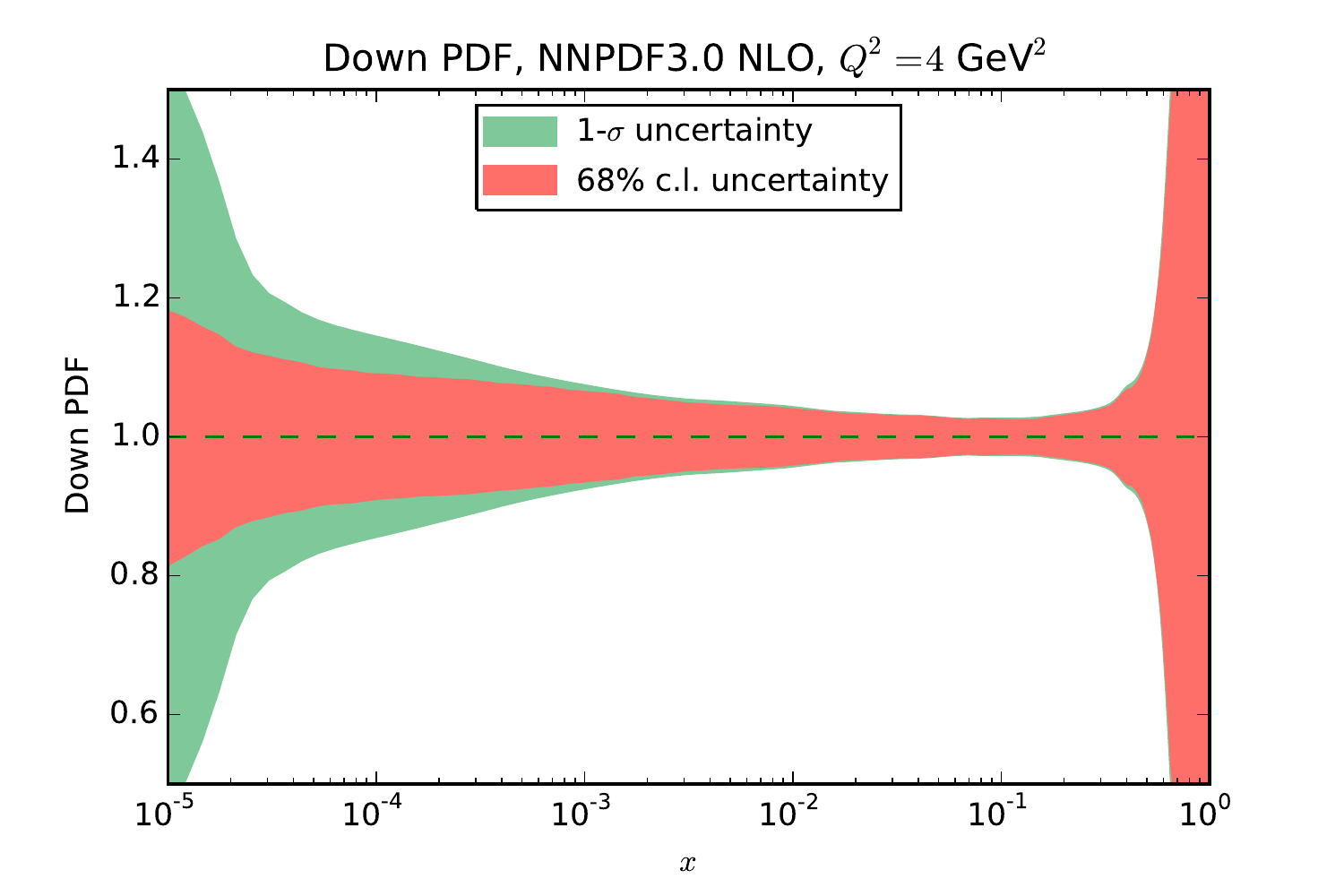}\includegraphics[width=0.52\textwidth]{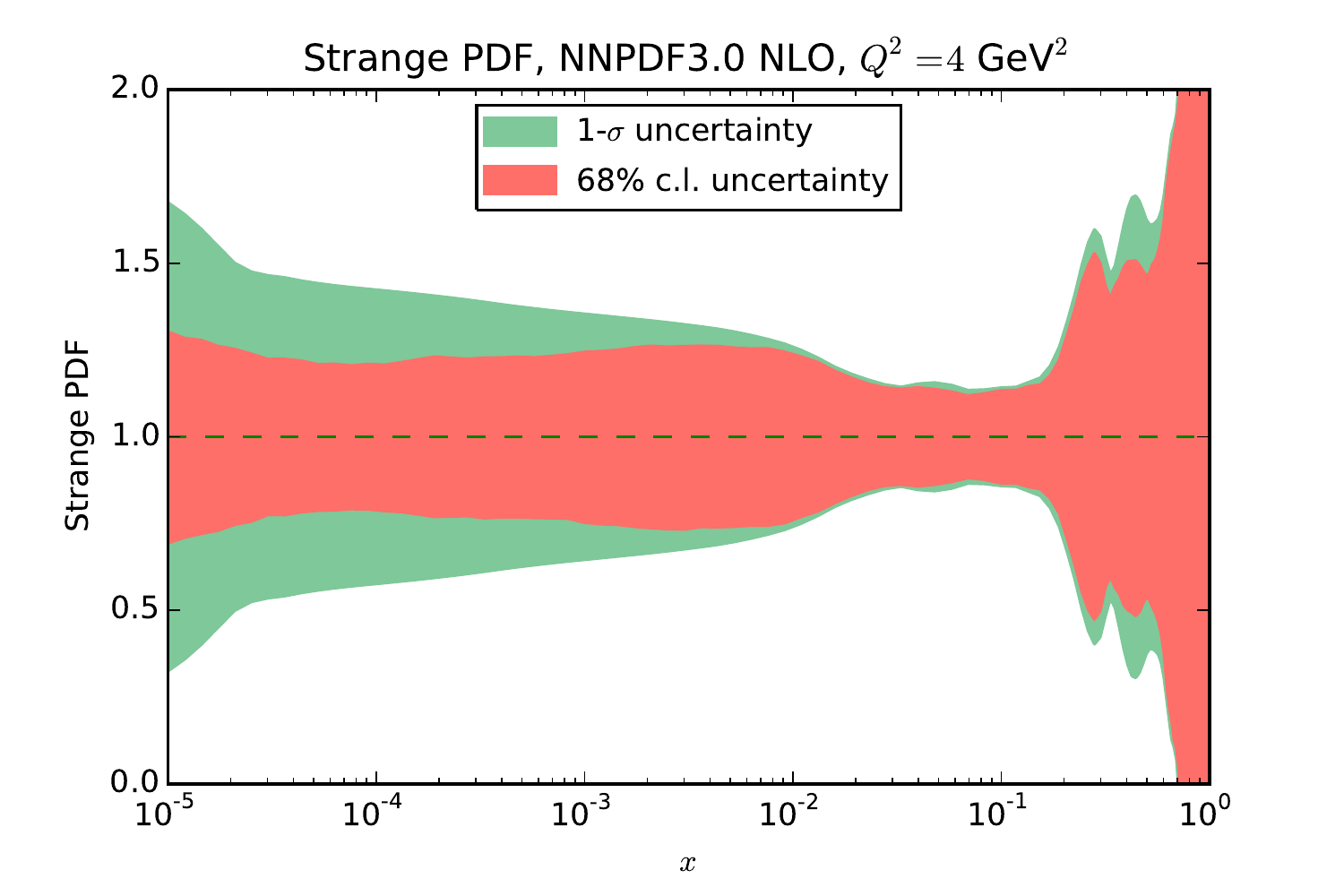}
  \end{center}
  \vspace{-0.3cm}
  \caption{\small \label{fig:gaussian} Comparison of  one-sigma
    and  68\% confidence level intervals for some PDFs from the NNPDF3.0 NLO
    set, determined using a sample of $N_{\rm rep}=1000$ MC replicas, 
at $Q=4$ GeV$^2$. From top to bottom and from left to right the gluon,
  down, up and strange PDFs are shown.}
\end{figure}

A suitable choice for the grid of points
in $x$ is one that ensures that PDFs are well-reproduced in the kinematic region
which is relevant for phenomenology, and that the spacing
of the grid is such that correlations between neighboring points are
not so strong that it becomes impossible to invert the covariance
matrix.
In practice, we proceed as follows: we first  
consider an initial grid of points in $x\in[10^{-5},0.9]$
with $N_{x}=50$ for all PDFs ($N_{\rm pdf}=7$, since heavy flavor
PDFs are generated dynamically), half equally spaced on a logarithmic
scale for  $x\in[10^{-5},10^{-1}]$ and half equally spaced on a linear
scale for  $x\in[0.1,0.9]$. We then determine the eigenvectors of the
ensuing $350\times 350$ ($N_x N_{\rm pdf}\times N_x N_{\rm pdf}$)
covariance matrix, and discard all eigenvectors corresponding to
eigenvalues whose size is smaller than  a factor $10^{-12}$ times
the largest one. This removes points which carry little information
due to large correlations. We then invert the covariance matrix in the
remaining subspace.

A further difficulty arises whenever  the prior
uncertainties are not Gaussian. In such case,
a faithful Hessian representation is (by construction)
impossible, and our procedure, which always leads to a final Hessian
matrix, becomes meaningless. Whenever the starting PDF set has
potentially non-Gaussian uncertainties, it is thus  necessary to quantify the
deviation from gaussianity in order to make sure that the procedure
can be consistently applied.
We do this using the simplest indicator, namely the
second moment of the probability distribution, specifically comparing
the one-sigma and 68\% confidence level intervals~\cite{Ball:2010de},
which for a
Gaussian distribution coincide. 
The comparison is shown  in Fig.~\ref{fig:gaussian} for some PDFs in 
the NNPDF3.0
NLO set  at
$Q^2 = 4$ GeV$^2$. It is clear that deviations from gaussianity can be
significant whenever experimental information is scarce or missing,
specifically at small- and large-$x$, since in these regions the PDF
uncertainty is not determined by gaussianly distributed data but
rather by extrapolation and by theoretical constraints (such as sum
rules and cross-section positivity).

 We thus
define the indicator 
\begin{equation}
  \label{eq:epsestimator}
  \epsilon_\alpha(x_i,Q^2_0) = \frac{\left|\sigma_{\alpha}(x_i,Q^2_0)
      - \sigma^{68}_{\alpha}(x_i,Q^2_0)\right|}{\sigma^{68}_{\alpha}(x_i,Q^2_0)}  \, ,
\end{equation}
where $\sigma_{\alpha}(x_i,Q^2_0)$ and $\sigma_{\alpha}^{68}(x_i,Q^2_0)$
are respectively the one-sigma and 68\% confidence level intervals
for the $\alpha$-th PDF at point $x_i$ and scale $Q^2_0$, computed
from the original MC representation with
$N_{\rm rep}=1000$ MC replicas.
When the
prior set has potentially non-Gaussian uncertainties, first of all we  evaluate
the figure of merit Eq.~(\ref{eq:epsestimator}) on the same grid of
point on which the covariance matrix is computed. We then discard all
points for which the deviation Eq.~(\ref{eq:epsestimator}) exceeds
some threshold value $\epsilon$, i.e. we only include points such that
\begin{equation}\label{eq:threshold}
\epsilon_\alpha(x_i,Q^2_0)< \epsilon.
\end{equation}
We then proceed as above: on the remaining points we compute the
covariance matrix, determine its eigenvectors, and discard
eigenvectors whose size is less than twelve orders of magnitude smaller
than the largest eigenvector.
Needless to day, this additional initial step is not required 
for sets (such as MMHT14) which are obtained from a Monte
Carlo conversion of an original Hessian set, and thus have Gaussian
uncertainties by construction.

We now turn to the  determination of the optimal basis of
replicas for the Hessian representation. This optimization requires
the definition of a statistical estimator which measures the quality
of the Hessian representation. 
Given that the Hessian representation corresponds to a Gaussian distribution,
and that central values are reproduced by construction, the
probability distribution is fully determined by the covariance
matrix. In practice, however, a good assessment of the quality of the
Hessian representation is obtained by simply verifying 
that the diagonal elements of the covariance matrix
are well well reproduced, thanks to the fact that correct correlations are
automatically provided by the use of PDF replicas as a basis, as we
shall explicitly verify below.

We introduce therefore the estimator
\begin{equation}
  \label{eq:estimator}
  \textrm{ERF}_{\sigma} = \sum_{i=1}^{N_x} \sum_{\alpha=1}^{N_{\rm pdf}} \abs{ \frac{\sigma^{\rm PDF}_{H,\alpha}(x_i,Q^2_0)
      - \sigma^{\rm PDF}_{\alpha}(x_i,Q^2_0)}{\sigma^{\rm PDF}_{\alpha}(x_i,Q^2_0)} } \,,
\end{equation}
which compares the  one-sigma standard deviations computed for 
the original MC and their Hessian
representations, as given respectively by Eqs.~(\ref{eq:sigmaMC}) and~(\ref{eq:sigma}).
We then compute the estimator for a given fixed value of $N_{\rm eig}$
basis replicas. The choice of these is random at first, and then 
optimized using a 
Genetic Algorithm (GA). 

The parameters of the GA are chosen based on
the studies of Ref.~\cite{Carrazza:2015hva}, where the related problem
of optimizing the choice of PDF replica set was studied: we find that
a single mutation per iteration of the GA is
sufficient, with the number of mutants chosen to be between one and
four per mutation, with probabilities listed in
Table~\ref{tab:gapars}. It turns out that
$N_{\rm gen}^{\rm max} = 2000$ iterations of the GA are sufficient to
obtain good stability and a sizable improvement of the figure of merit
in comparison to the starting random selection.

\begin{table}[h]
  \begin{center}   
 \begin{tabular}{|c|c|}
      \hline
      \multicolumn{2}{|c|}{{\tt mc2hessian v1.0.0}}    \\ \hline
      \hline 
      $N_{\rm rep}^{\rm mut}$ &   $P_{\rm mut}$ \\
      \hline
      1 & 30\% \\
      2 & 30\% \\
      3 & 10\% \\
      4 & 30\% \\
      \hline
    \end{tabular}
  \end{center}
  \caption{\small Number of mutants per replica 
and respective probabilities for
    each generation of the GA.}
  \label{tab:gapars}
\end{table}

\begin{figure}[t] 
   \begin{center}
      \includegraphics[scale=0.35]{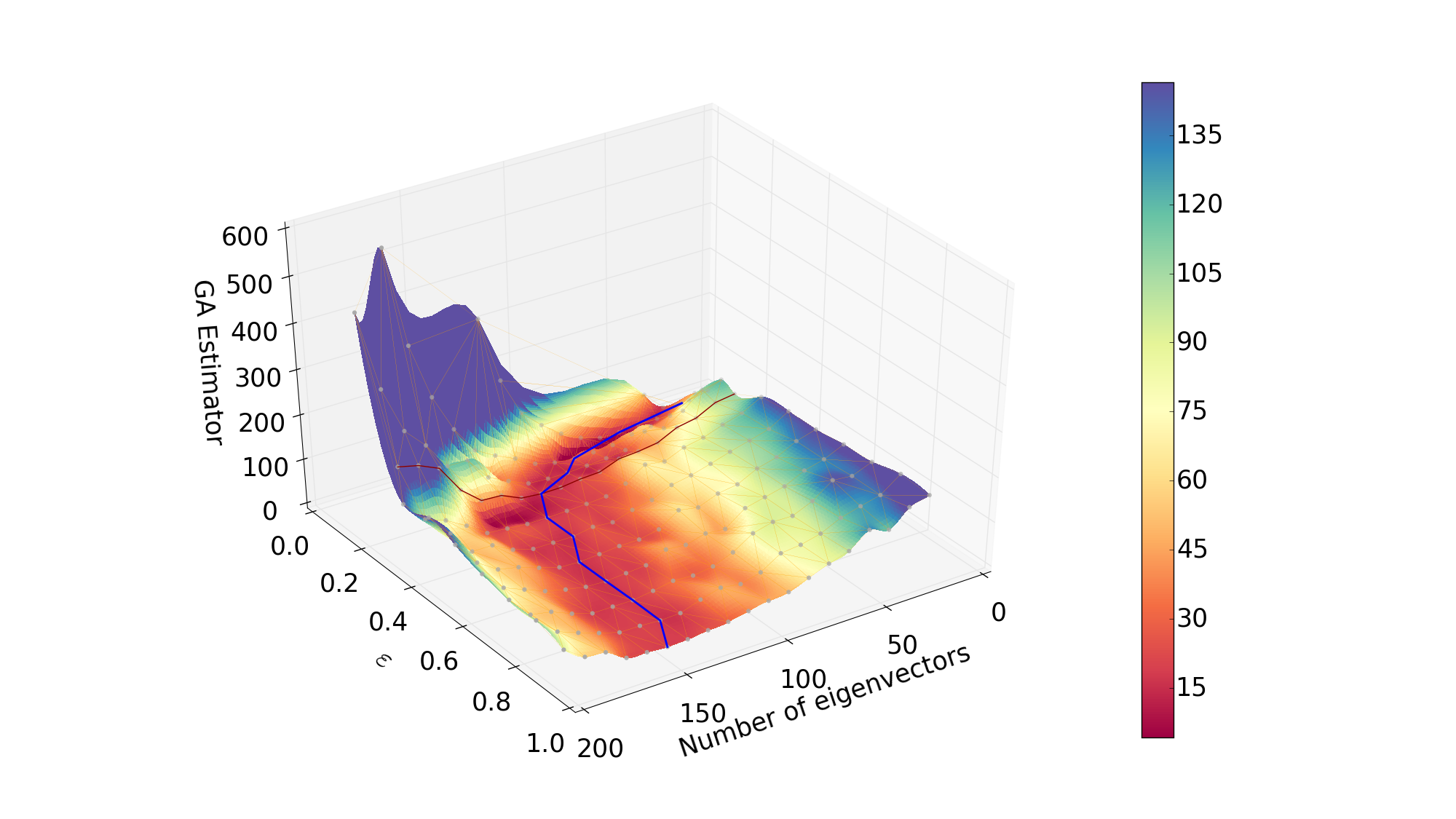} 
   \end{center}
   \vspace{-1.1cm} 
   \caption{\small The figure of merit, 
     Eq.~(\ref{eq:estimator}), for the Hessian
     representation of the NNPDF3.0 NLO set, 
computed for points which satisfy the
     gaussianity criterion Eq.~(\ref{eq:threshold}), plotted versus
     the threshold $\epsilon$  Eq.~(\ref{eq:threshold})  and the
     number of
     eigenvectors $N_{\rm eig}$, after the choice of basis replicas
     has been optimized through a run of the GA with the settings of
     Table.~\ref{tab:gapars}. The value of the
     estimator along the valley of minima, i.e. the
curve determined by finding the value of
     $N_{\rm eig}$ at which the estimator has its absolute minimum for
     each $\epsilon$ is shown (blue curve). The red curve marks the
     value
     $\epsilon=0.25$  which is finally adopted.
     The projection of the valley of minima in the $(\epsilon, N_{\rm eig})$
     plane is shown in Fig.~\ref{fig:2dplot}. 
     \label{fig:3dplot}}
\end{figure}

In Fig.~\ref{fig:3dplot} we plot the value of the 
estimator Eq.~(\ref{eq:estimator}) after GA minimization, 
as a function of the threshold value of $\epsilon$ and
the number $N_{\rm eig}$ of basis replicas.
A valley of minima is clearly seen, and shown in the plot as a curve,
determined by searching for the absolute minimum of the estimator as a
function of $N_{\rm eig}$ for each fixed
 $\epsilon$. 

\begin{figure}[t] 
   \begin{center}
      \includegraphics[scale=0.3]{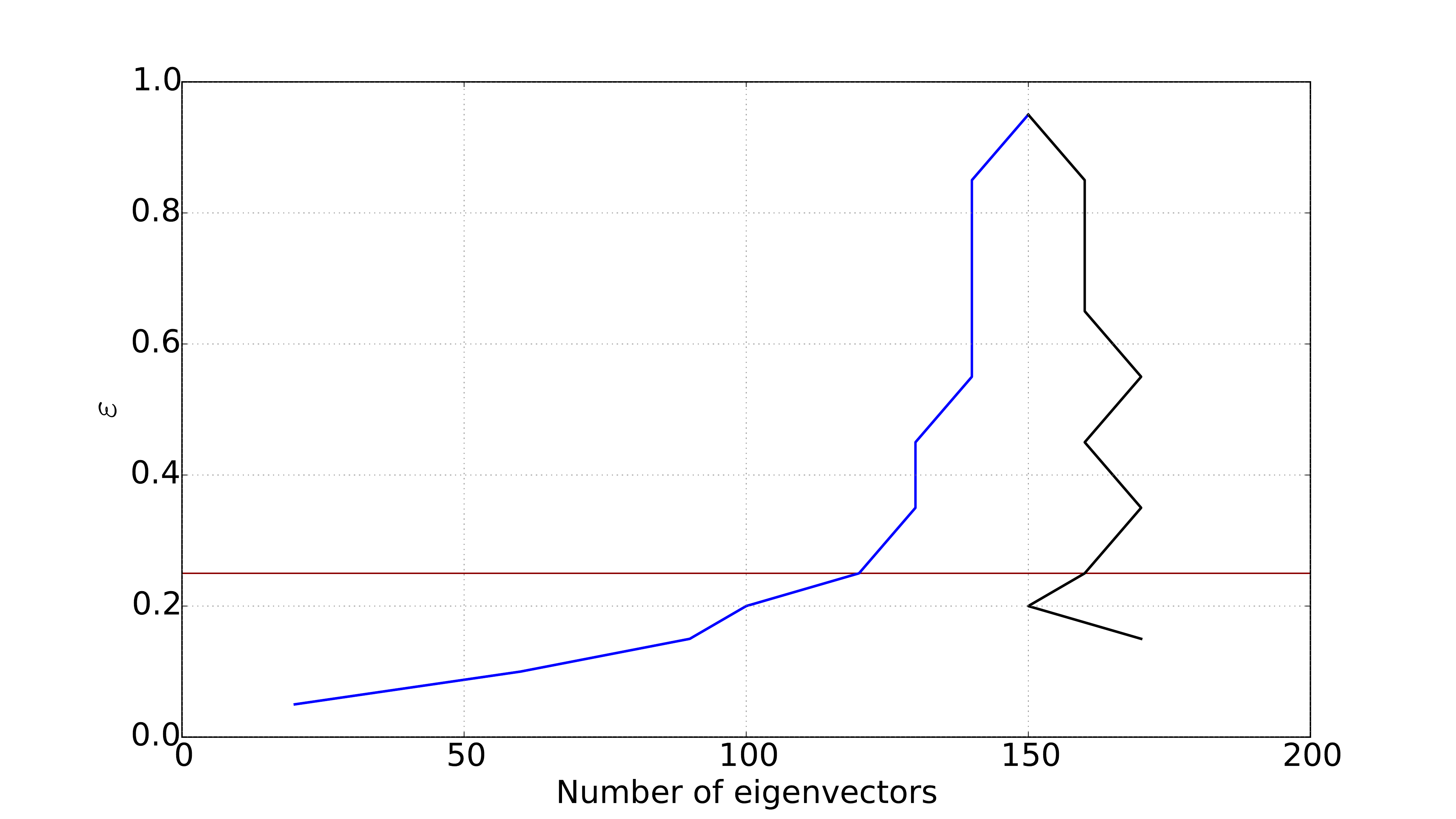} 
   \end{center}
   \vspace{-0.3cm} 
   \caption{\small The valley of minima
     shown in Fig.~\ref{fig:3dplot} (blue curve) shown as a
     projection on the $(\epsilon, N_{\rm eig})$ plane, compared to
     the curve recomputed using the same eigenvector basis but
     including all points in the determination of the figure of merit
     (black curve).
     The red line indicates the
     threshold value
 $\epsilon=0.25$  which is finally adopted.}
   \label{fig:2dplot} 
\end{figure}
An  interesting feature of Fig.~\ref{fig:3dplot} is that for all
values of 
$\epsilon$ the optimal value of  $N_{\rm eig}$ is reasonably small,
and much smaller than the total number of replicas $N_{\rm
  rep}=1000$. As one might expect, when the value of the
threshold  $\epsilon$ is
small, and thus
 a large number of points in $x$ is excluded, the optimal number of
 eigenvectors is also small, rapidly decreasing when $\epsilon\lsim0.2$.
Clearly, however, if $\epsilon$ is too small, only few points will be
retained in the computation of the figure of merit
Eq.~(\ref{eq:chi2def}) and the original MC replicas will not be reproduced
accurately enough. 

In order to determine the optimal value of
$\epsilon$, for each value of $N_{\rm eig}$ and $\epsilon$ shown in
Fig.~\ref{fig:3dplot} we have recomputed the figure of merit
Eq.~(\ref{eq:estimator}) using the same eigenvector basis, 
but now including all points; the valley of minima is then determined
again for this new surface. The dependence on $\epsilon$ is now due to
the fact that the eigenvector basis changes as $\epsilon$ is varied,
though the definition of the figure of merit does not. Therefore, the
difference between the two curves is a measure of how much the
exclusion of nongaussian points by the $\epsilon$ criterion affects
the choice of optimal eigenvector set.
The two curves are compared in
Fig.~\ref{fig:2dplot}: they are seen to diverge
when  $\epsilon\lsim 0.2$. We take this as an indication that, below
this value, the amount of information which is necessary in order to
describe all points starts being significantly different from that
which is sufficient for an accurate description of the points which
have passed the cut, and thus the cut becomes too restrictive. We 
consequently adopt $\epsilon=0.25$ (red curve), as a
 reasonable compromise between only including points for which
 uncertainties are Gaussian, and not loosing too much information.

\begin{figure}[t]
\begin{center}
  \includegraphics[width=0.6\textwidth]{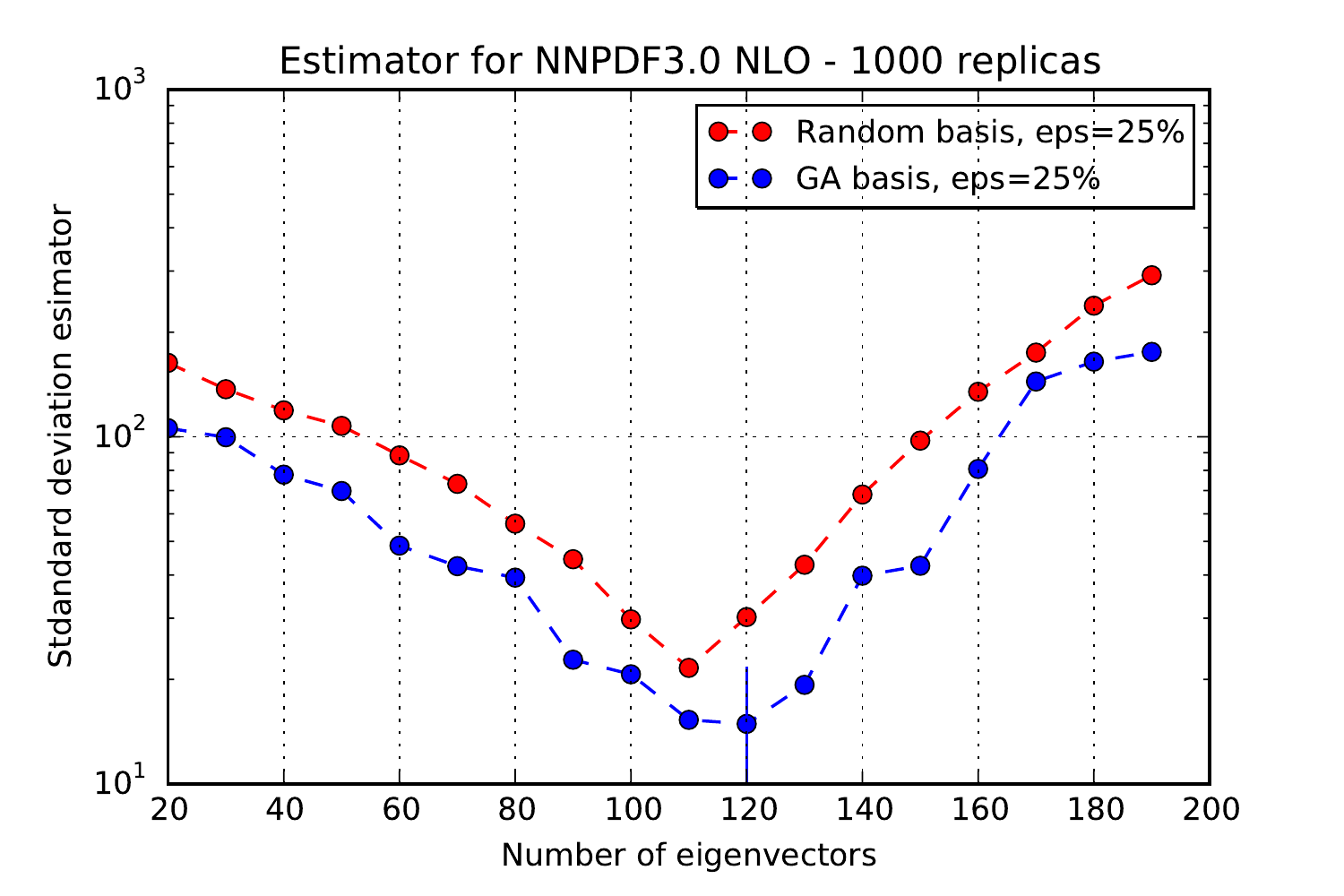}
\end{center}
\vspace{-0.3cm}
\caption{\small \label{fig:estimator1} The figure of merit of
  Fig.~\ref{fig:3dplot}, plotted versus $N_{\rm
    eig}$ for fixed  $\epsilon=0.25$ (blue curve), for
  the NNPDF3.0 NLO set. The optimal value $N_{\rm eig}=120$ is
  denoted by a vertical dash. Results obtained by
  not optimizing the basis through the GA are also shown (red
  curve). 
}
\end{figure}

The profile of the figure of merit with the choice of threshold value 
$\epsilon=0.25$,
is shown  in Fig.~\ref{fig:estimator1}. 
It is seen that the optimal number of eigenvectors is
 $N_{\rm eig} = 120$. The fact that this value is much less than the
starting  $N_{\rm rep}=1000$ means that replicas in the original set
are strongly correlated. This is nicely consistent with the result that it is
possible to construct a ``compressed'' representation of NNPDF3.0, in which the
original probability distribution is reproduced but including a much
smaller, optimized set of Monte Carlo
replicas~\cite{Carrazza:2015hva}. In fact, it turns out that the
optimal number of eigenvectors, and the number of compressed replicas,
are of the same order of magnitude.

\begin{figure}[t]
\begin{center}
  \includegraphics[width=0.6\textwidth]{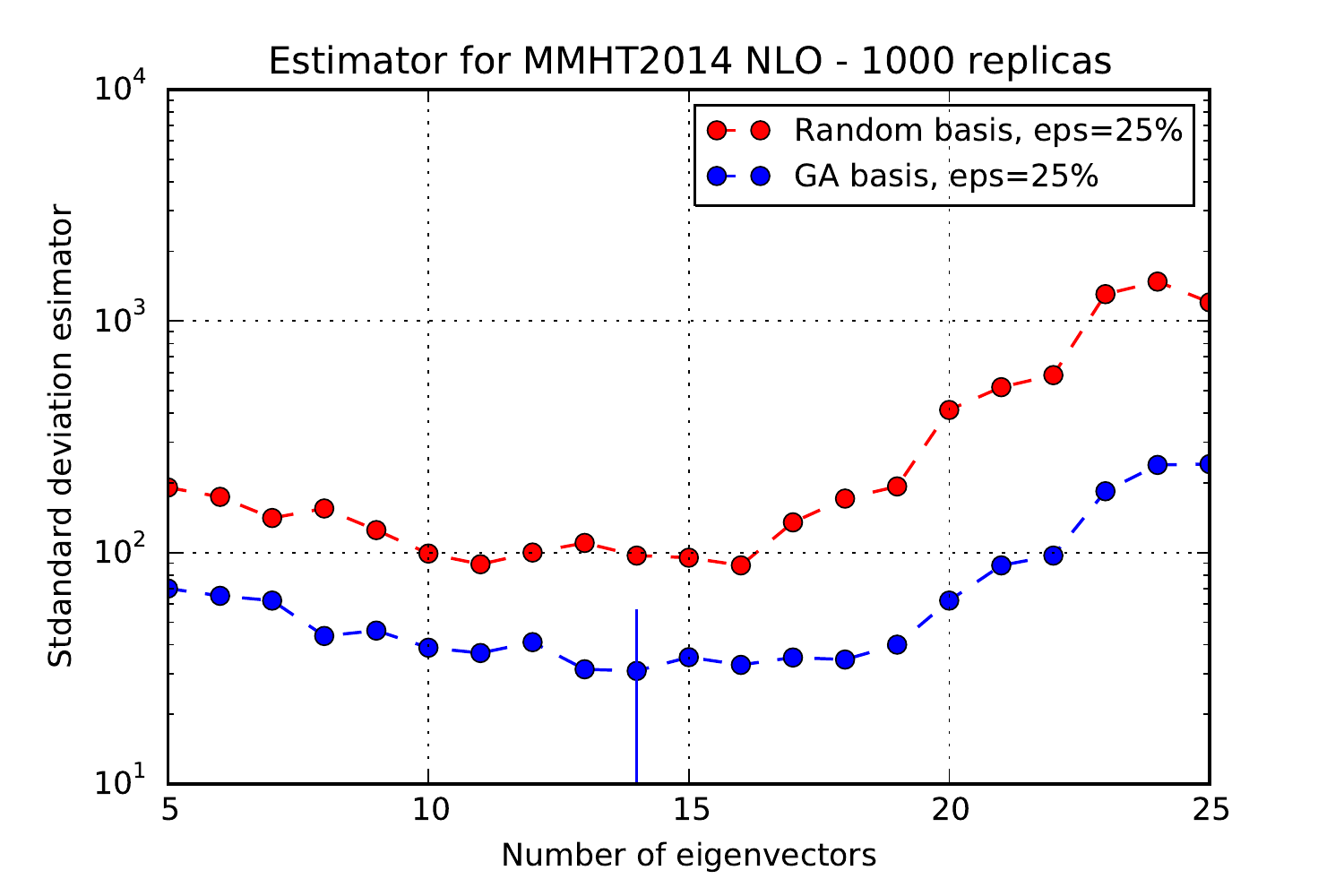}
\end{center}
\vspace{-0.3cm}
\caption{\small \label{fig:estimator2} Same as
  Fig.~\ref{fig:estimator1}, but for the MMHT14 MC NLO PDF set.  In
  this case the gaussianity condition Eq.~(\ref{eq:threshold}) is not
  applied. The optimal value $N_{\rm eig}=14$ is denoted by a
  vertical dash.  }
\end{figure}
In Fig.~\ref{fig:estimator1} we also show the figure of merit when the
basis replicas are chosen randomly, instead of being optimized through
the GA. It is apparent that use of the GA leads to an improvement of
the figure of merit by almost a factor two. It is interesting to
observe that this improvement is in fact achieved 
by modifying only a small fraction of the
initial random selection of replicas: specifically,  only 26 of the
replicas used as initial input for the basis are mutated at the end of
the $N_{\rm gen}^{\rm max}$ iterations of the GA. This suggests that there is still room
for improvement in the selection of the optimal basis, since the GA
only explores combinations that are not to far from the initial basis. 

Finally, we have repeated our construction for the Monte Carlo
representation of the MMHT14 NLO PDF
set. In this case, because the starting PDF set is Hessian, no
 gaussianity requirements are necessary. On the other hand, because
 the underlying PDF set is described by a smaller number of parameters
 than either the NNPDF or the combined set considered previously,
 Monte Carlo replicas tend to be more correlated. As a consequence, it
 is necessary to relax somewhat the criterion for selection of the
 eigenvectors of the covariance matrix: in this case, we keep all
 eigenvectors corresponding to eigenvalues whose size is larger than $10^{-15}$
times that of the largest one (instead of $10^{-12}$ as in the
previous cases). 
We then simply determine the figure
of merit as a function of $N_{\rm eig}$: results are  shown in
Fig.~\ref{fig:estimator2}. Also in this case, the GA leads to an
improvement of the figure of merit by nore than a  factor two. Now,
however, the optimal number of eigenvectors is rather smaller,
$N_{\rm eig} = 14$, as compared to the NNPDF3.0
result.  
Again, this is of the same order as that which is
used when applying the compression algorithm of
Ref.~\cite{Carrazza:2015hva} 
to the MMHT14 PDFs.

\section{Results and validation}
\label{sec:validation}

We now study in detail the results which are obtained when
applying our Monte Carlo to Hessian conversion to NNPDF3.0 NLO  --- a
native Monte Carlo PDF set --- and MMHT14 --- a Hessian PDF set which
has been turned into Monte Carlo using the technique of
Ref.~\cite{Watt:2012tq}. In the former case, we  compare the final
Hessian PDFs to the starting Monte Carlo ones, at the level of central
values, uncertainties, and correlations, thereby validating the
procedure. In the latter case, we  compare the results of the
final Hessian conversion to the original Hessian set, thereby
providing a powerful closure test of the procedure. Finally,
we  compare results before and after Hessian conversion for both
PDF sets  at the level of
physical observables, in order to ascertain the accuracy of our
methodology for realistic applications. PDF comparisons and luminosity
plots
shown in this section  have been
  produced using the {\tt APFEL Web} plotting tool~\cite{Bertone:2013vaa,Carrazza:2014gfa}.

\subsection{Hessian representation of Monte Carlo PDFs}
\label{sec:mch}

We  concentrate on the PDF set obtained starting with $N_{\rm
  rep}=1000$ NNPDF3.0 NLO replicas, and applying our methodology with
the optimal choice of parameters discussed in
Section~\ref{sec:numerical}, namely $\epsilon=0.25$, $N_{\rm eig} = 120$.
In Fig.~\ref{fig:pdfcomp} we compare the original Monte Carlo 
representation to the final Hessian representation of several PDFs at
$Q^2=2$ GeV$^2$:  the excellent
accuracy of the Hessian representation is apparent, with differences 
in the one-sigma
PDF uncertainty bands of the
order 5\% at most (recall that central values coincide by
construction). In Fig.~\ref{fig:pdfcomp-highQ2} the same comparison
is performed at $Q^2=10^4$ GeV$^2$, now shown as a   ratio to 
central values.

\begin{figure}[t]
\begin{center}
  \includegraphics[width=0.48\textwidth]{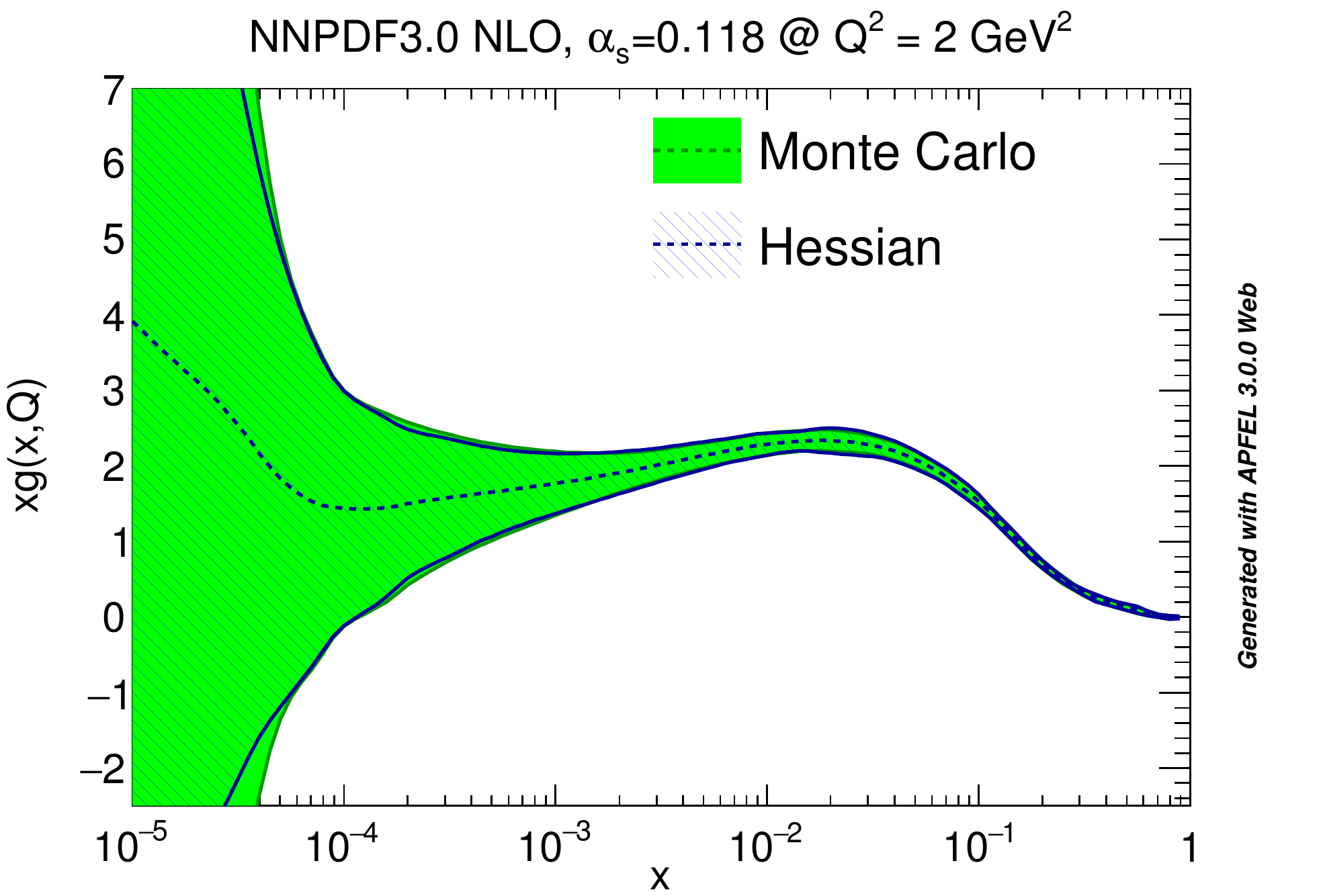}
  \includegraphics[width=0.48\textwidth]{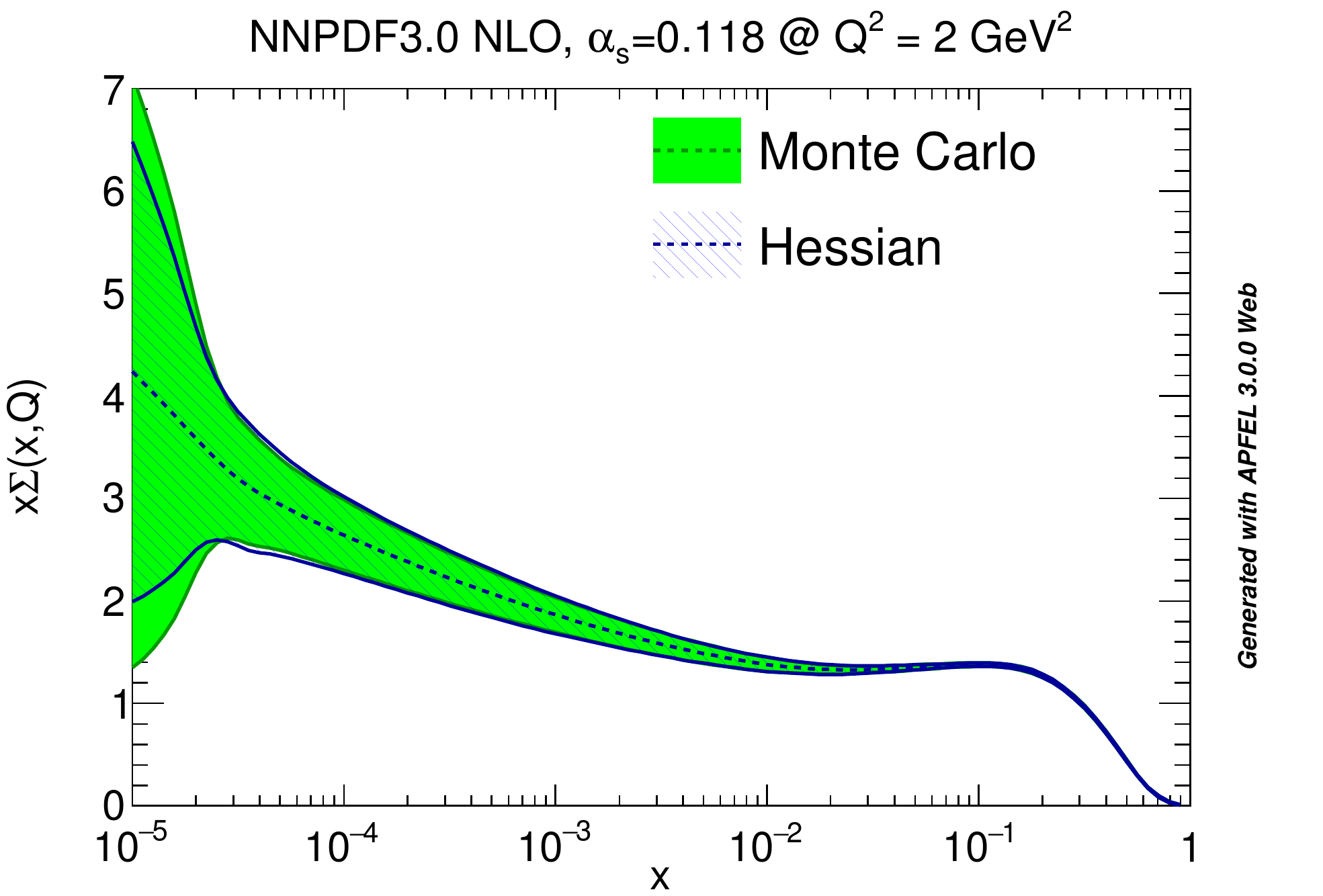}
  \includegraphics[width=0.48\textwidth]{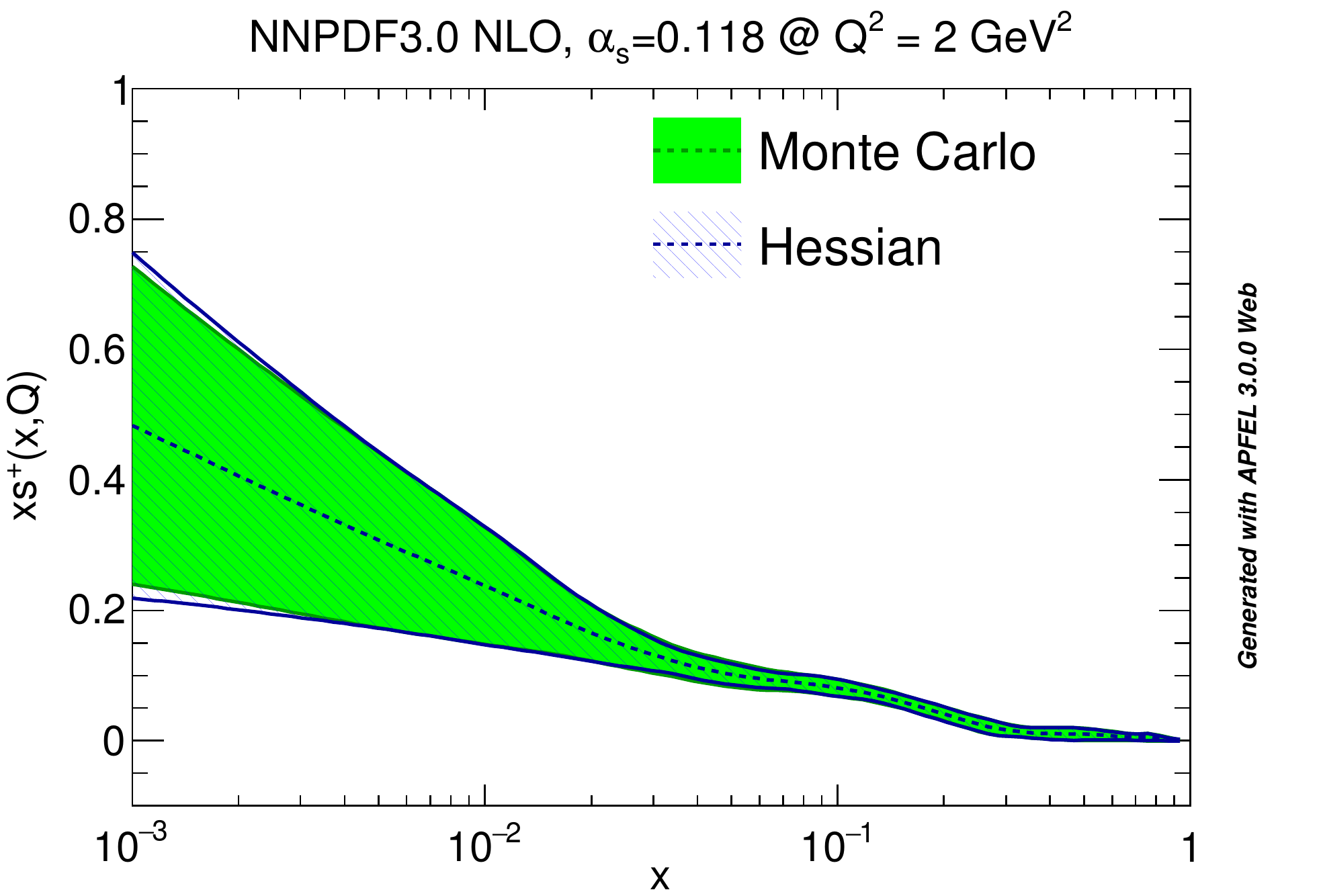}
   \includegraphics[width=0.48\textwidth]{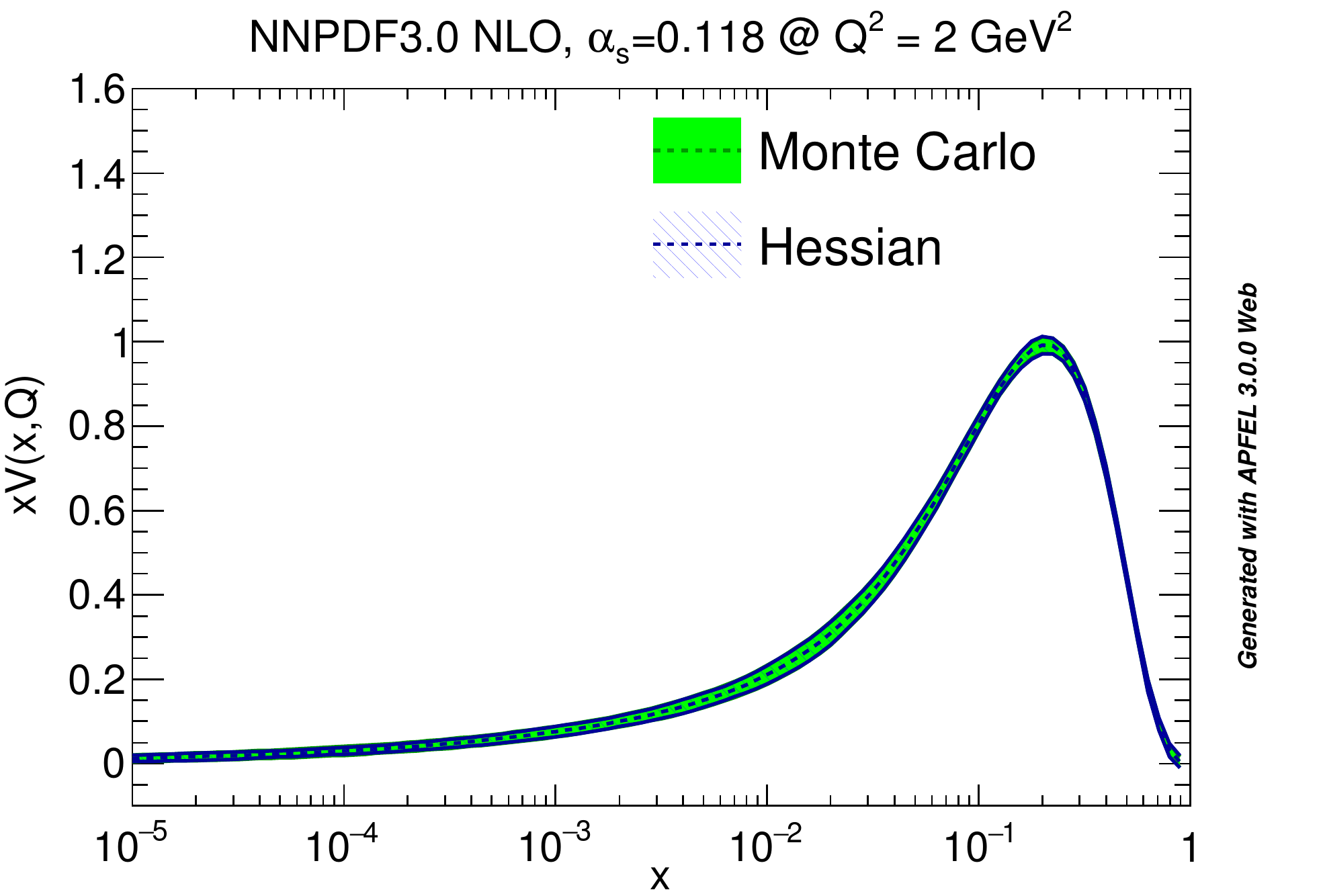}
\end{center}
\vspace{-0.3cm}
\caption{\small \label{fig:pdfcomp}
  Comparison between the starting Monte Carlo representation with $N_{\rm
    rep}=1000$ replicas and the final Hessian representation
with  $N_{\rm eig}=120$ eigenvectors for the 
 for the NNPDF3.0 NLO set with $\alpha_s=0.118$.
  From left to right and from top to bottom we the gluon, 
  total quark singlet,  total strangeness and the total valence
  are plotted vs. $x$ for fixed $Q^2=2$ GeV$^2$.}
\end{figure}

\begin{figure}[t]
\begin{center}
  \includegraphics[width=0.48\textwidth]{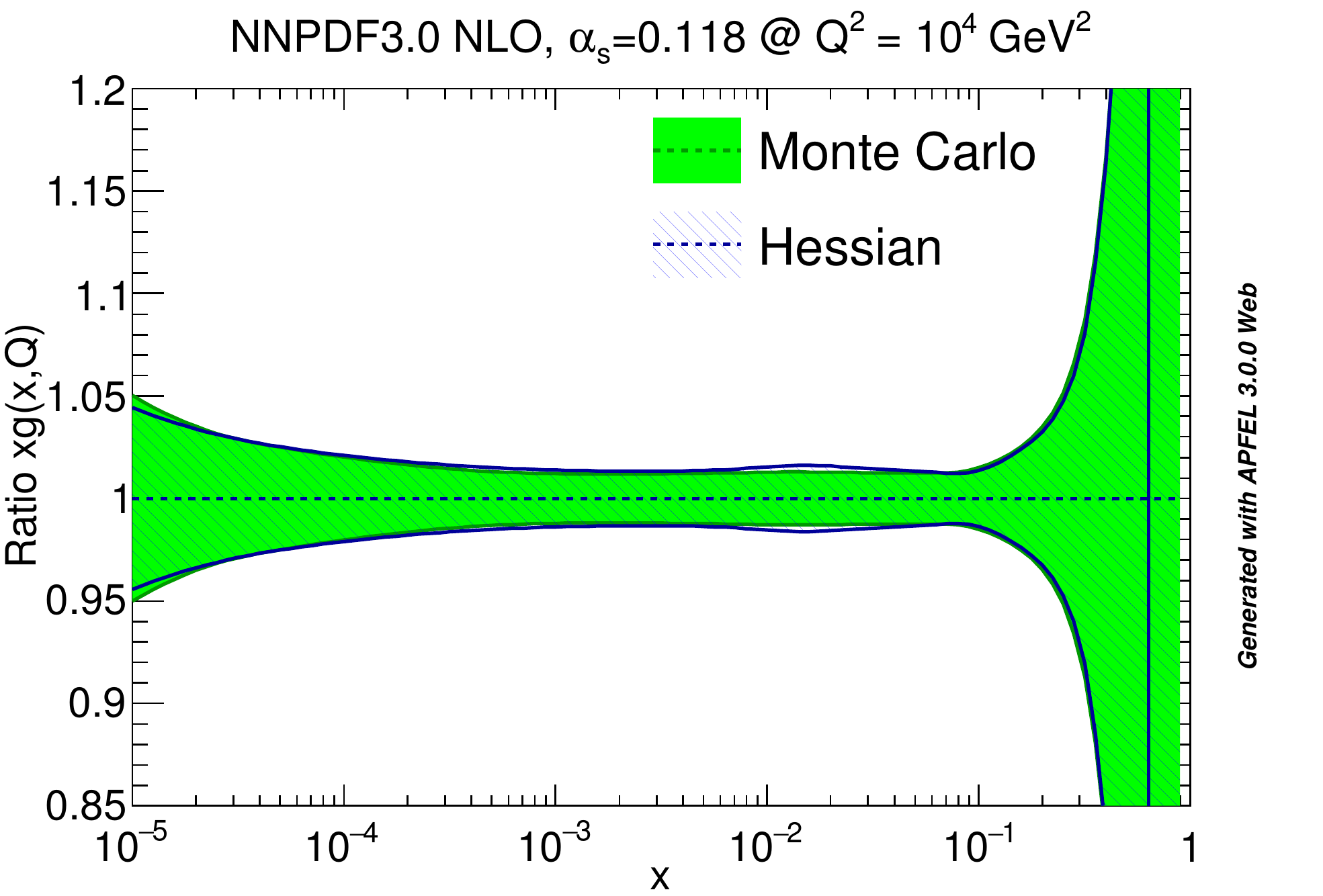}
  \includegraphics[width=0.48\textwidth]{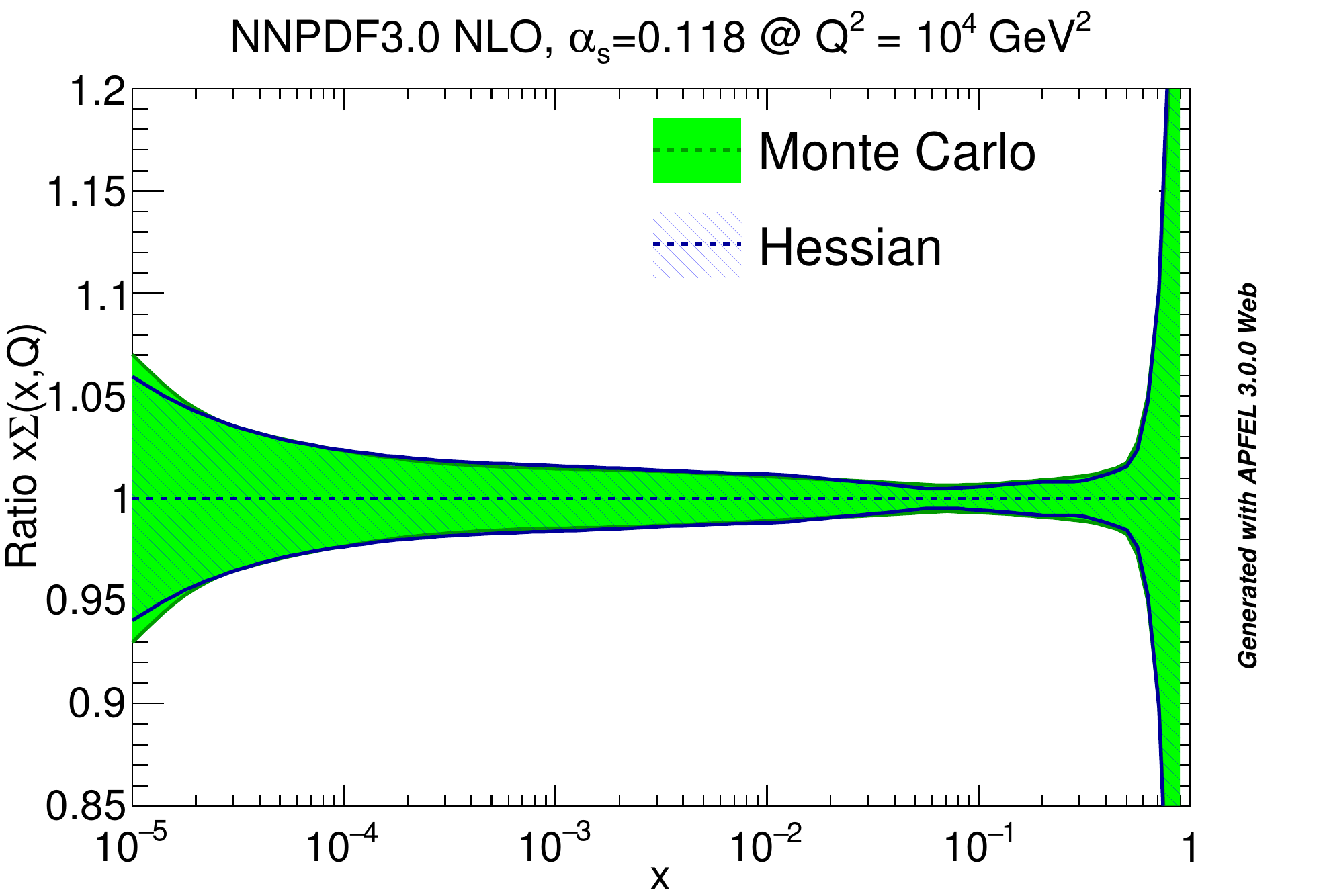}
  \includegraphics[width=0.48\textwidth]{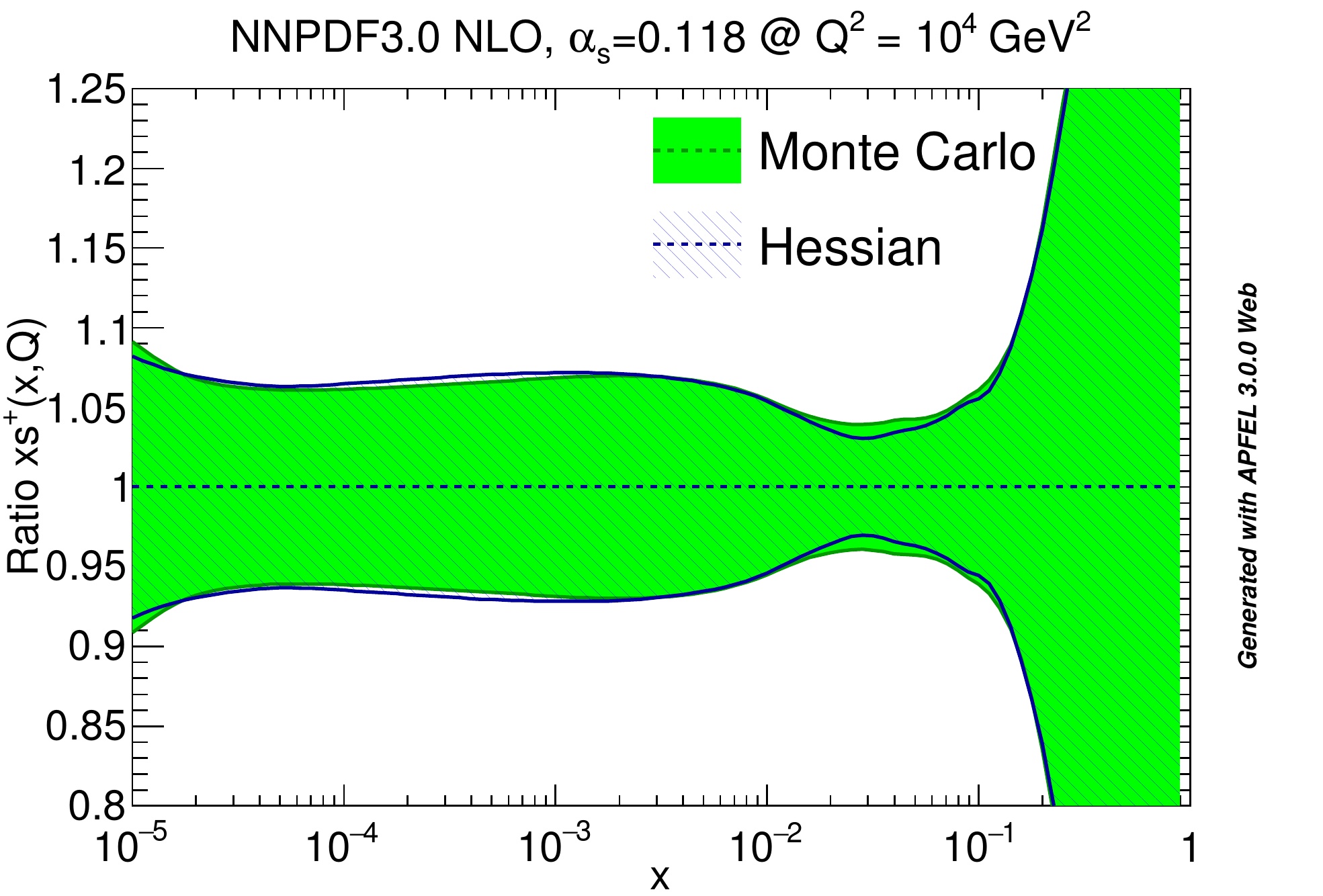}
  \includegraphics[width=0.48\textwidth]{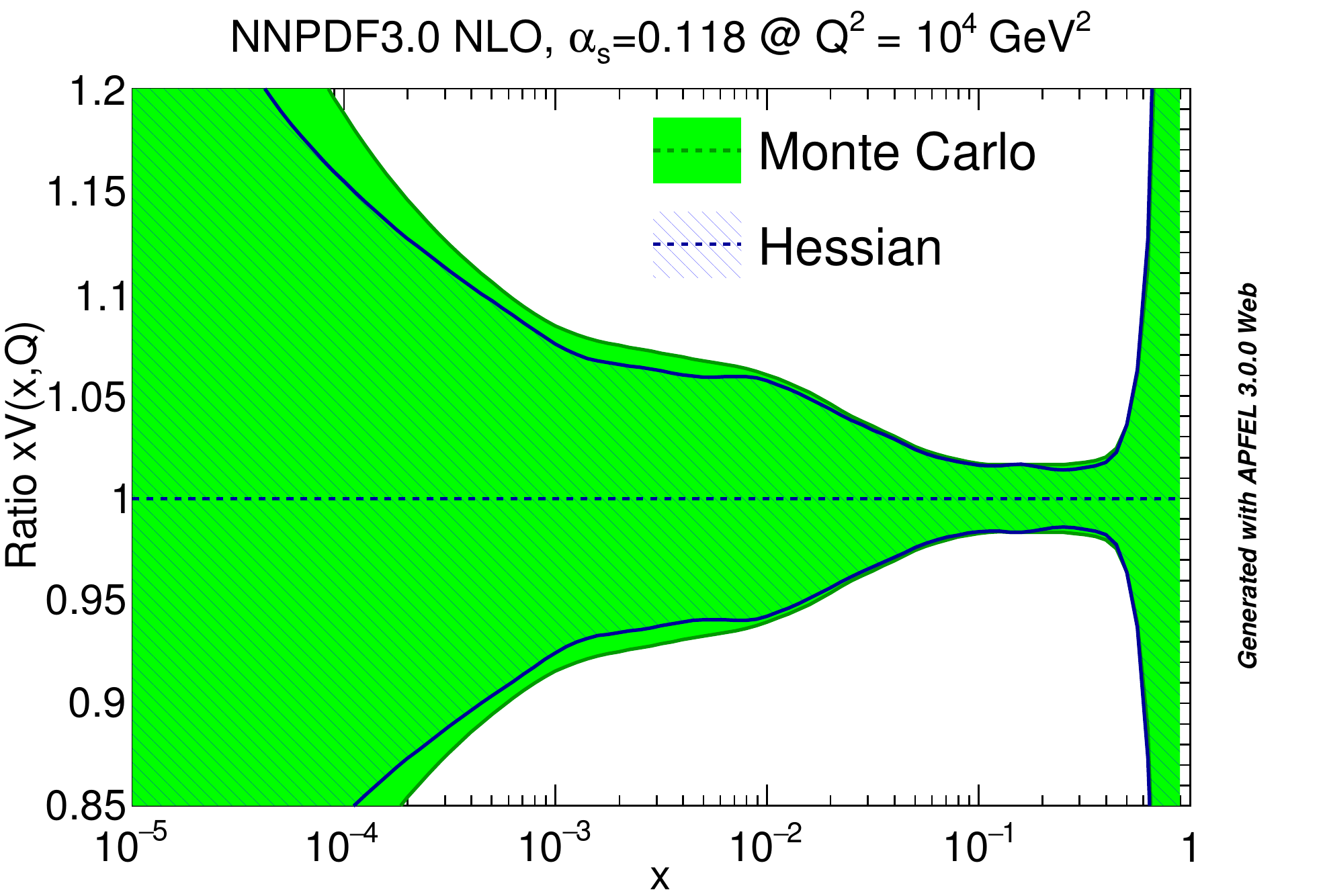}
\end{center}
\vspace{-0.3cm}
\caption{\small \label{fig:pdfcomp-highQ2}
  Same as Fig.~\ref{fig:pdfcomp} but at  $Q^2=10^4$ GeV$^2$, and with
results normalized to the central PDF.  }
\end{figure}

In Fig.~\ref{fig:lumis} we then compare some parton luminosities,
computed  for
proton-proton collisions with a center of
mass energy of 13 TeV, plotted vs.  the invariant mass of the
final state.
Results are shown normalized to the central value of the NNPDF3.0 NLO
Monte Carlo set.  Again, we find excellent agreement, 
 except in the regions of very
small or very large invariant masses (which respectively depend on
small and large-$x$ PDFs).
This is unsurprising given that these are extrapolation regions in
which the Gaussian approximation is less good.

\begin{figure}[t]
\begin{center}
\includegraphics[width=0.49\textwidth]{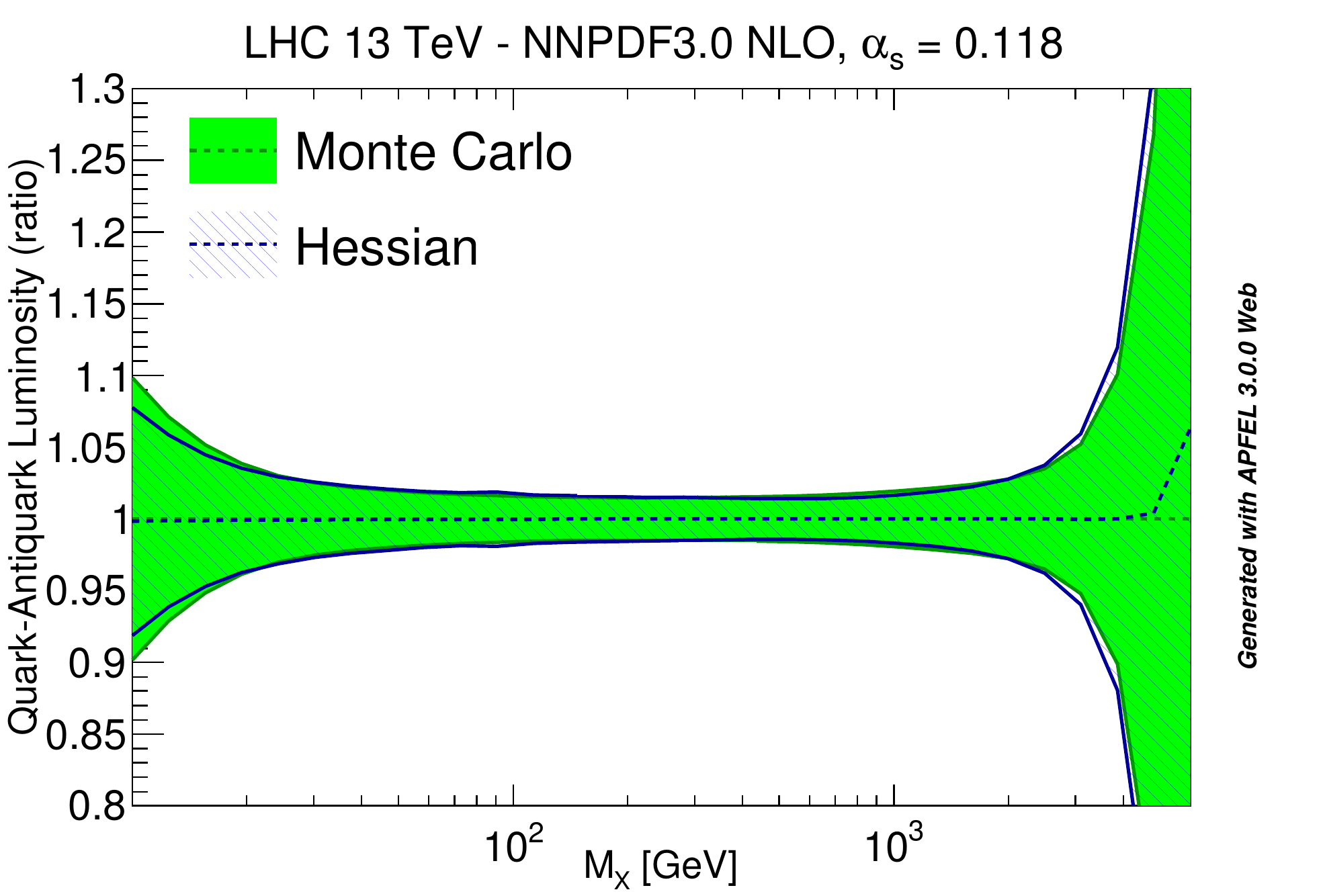}
\includegraphics[width=0.49\textwidth]{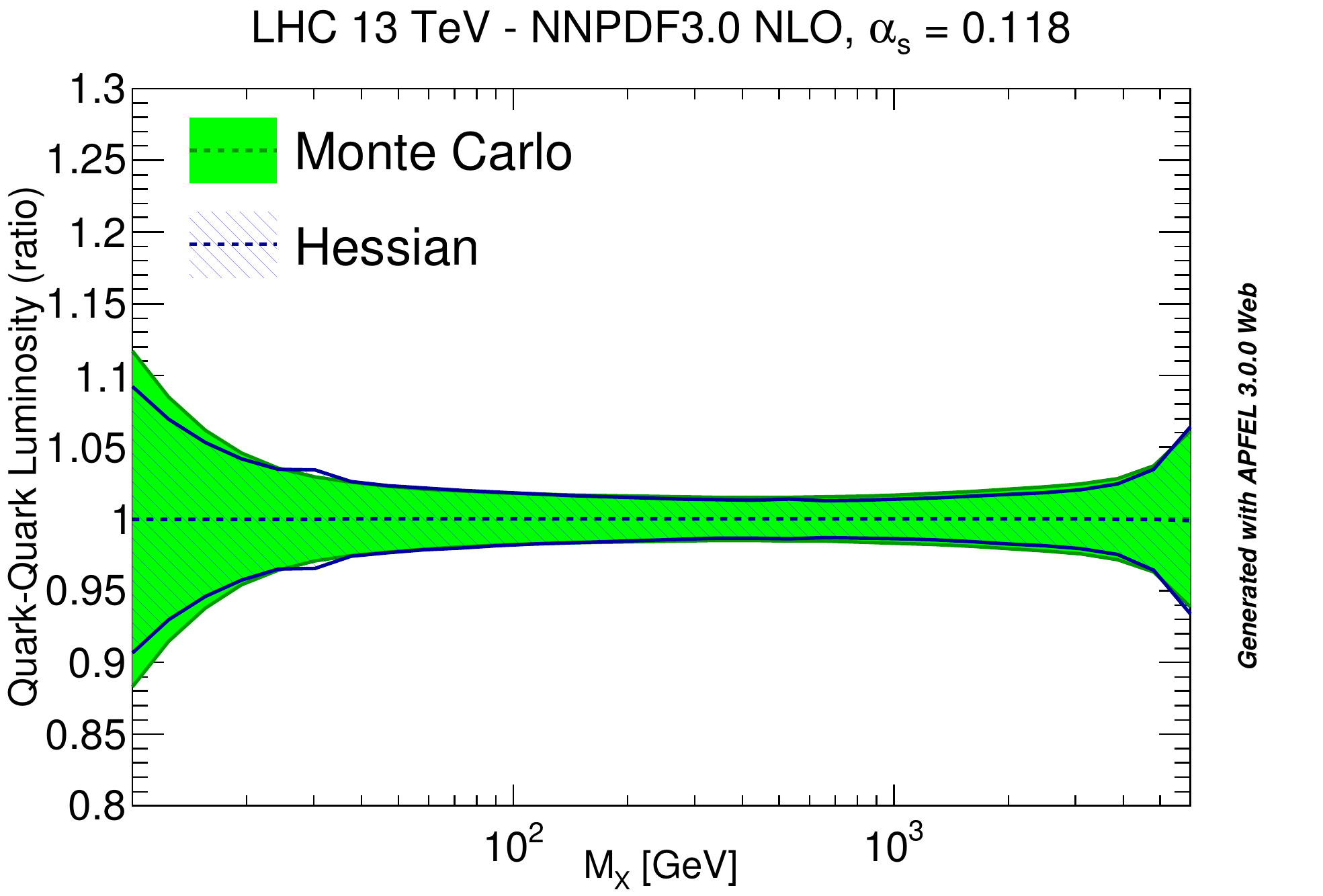}
\includegraphics[width=0.49\textwidth]{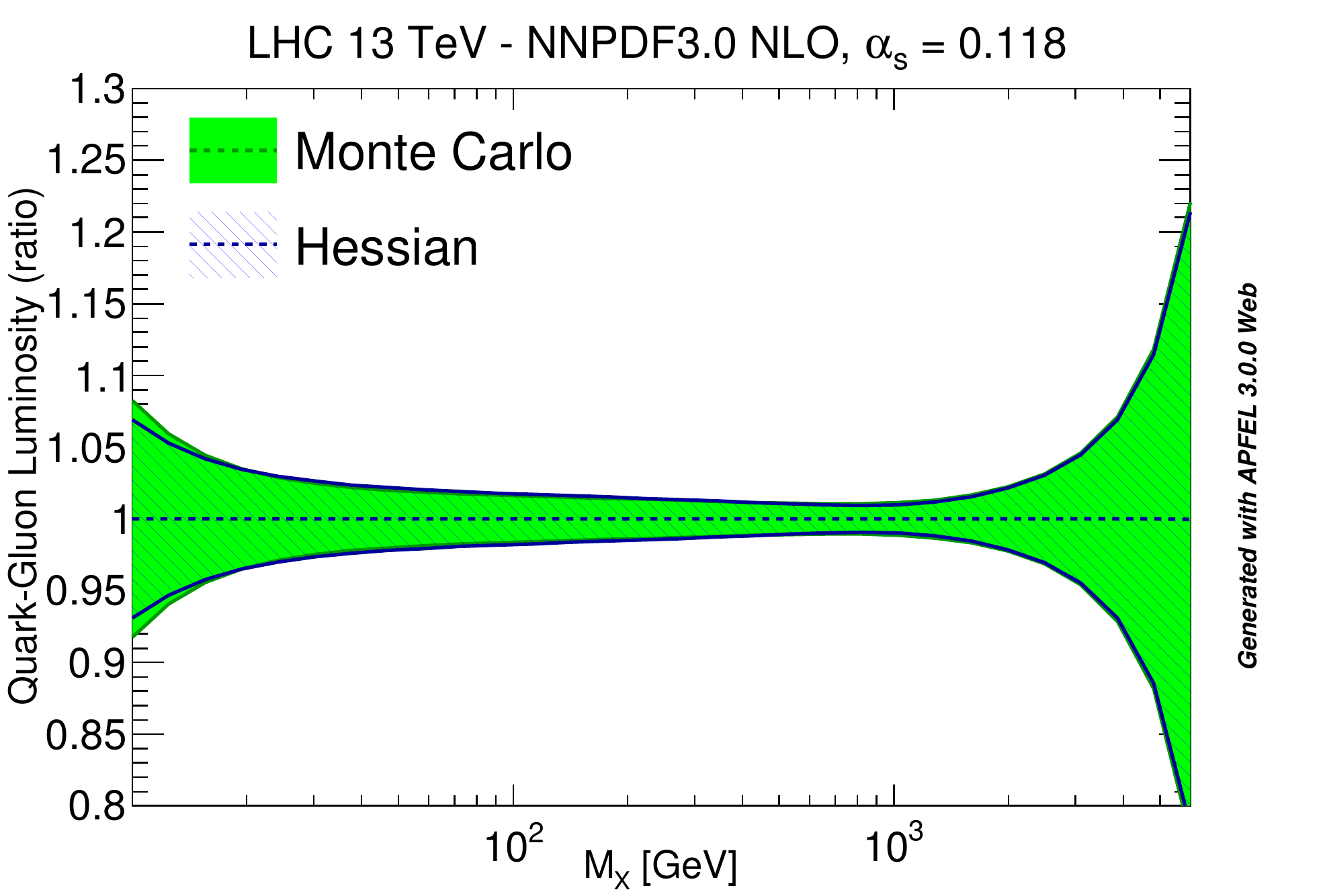}
\includegraphics[width=0.49\textwidth]{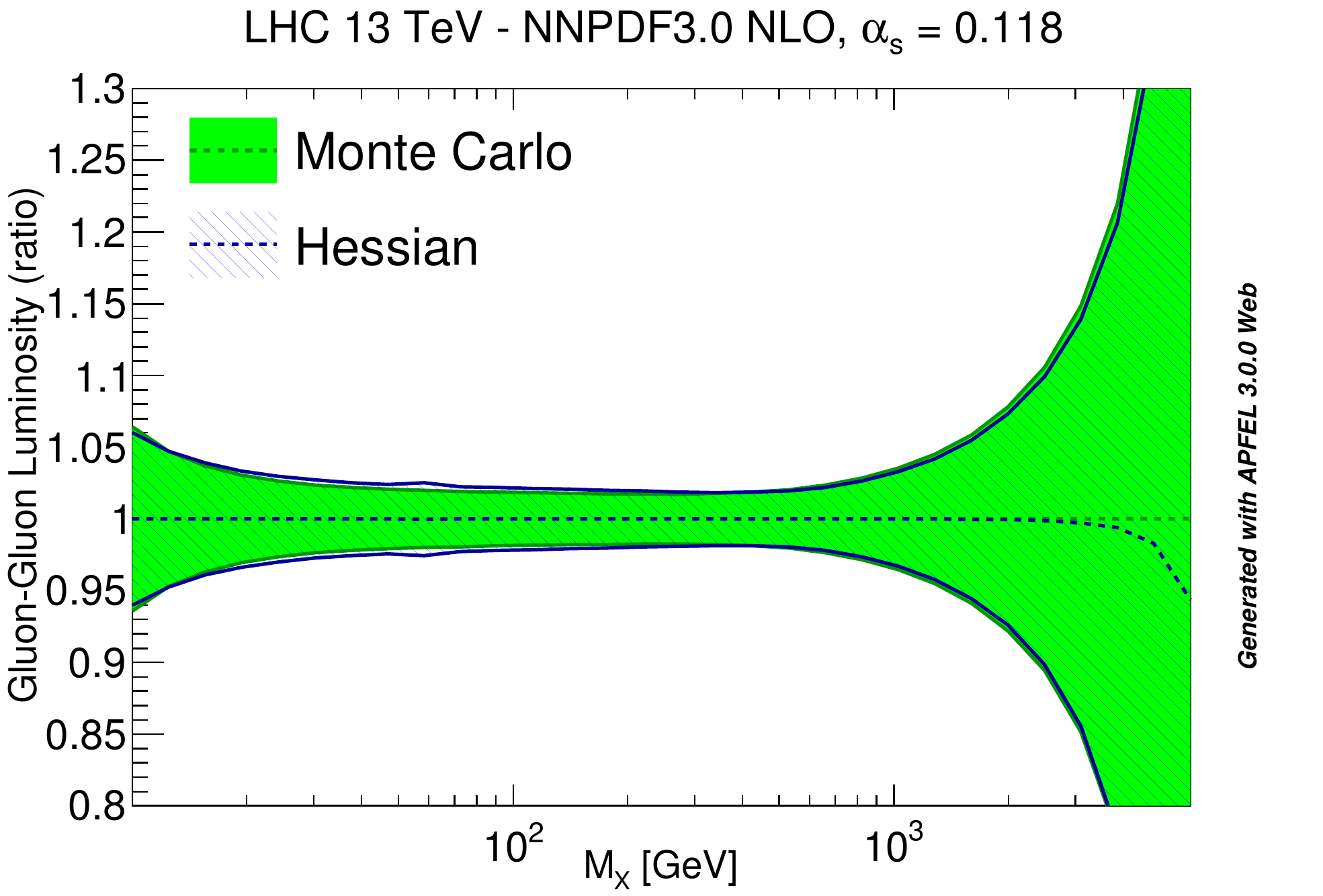}
\end{center}
\vspace{-0.3cm}
\caption{\small \label{fig:lumis}  Same as Fig.~\ref{fig:pdfcomp}, but
  now comparing  parton luminosities for proton-proton collisions
  at~13~TeV, plotted  vs. the invariant mass of the final
  state. Results are shown normalized to the central PDF.}
\end{figure}

As mentioned in Sect.~\ref{sec:numerical}, even though the figure of
merit Eq.~(\ref{eq:estimator}) used for the GA only optimizes the
diagonal elements of the covariance matrix, correlations are
automatically reproduced thanks to the use of the original replicas as
a basis.  We can check this explicitly. The correlation coefficient
between two different PDFs $f_\alpha$ and $f_\beta$, at a given value
of $x$, $Q^2$, in the Monte Carlo representation
is given by~\cite{Demartin:2010er}: 
\be
\label{eq:corr_mc}
\rho^{\alpha\beta}_{\rm MC}(x,Q^2) = \frac{N_{\rm rep}}{N_{\rm rep}-1} \lp
\frac{\la f_\alpha^{(k)}(x,Q^2)f_\beta^{(k)}(x,Q^2)\ra_{\rm rep}- \la
  f_\alpha^{(k)}(x,Q^2)\ra_{\rm rep}\la f_\beta^{(k)}(x,Q^2)\ra_{\rm
    rep}}{\sigma^{\rm PDF}_\alpha(x,Q^2) \cdot \sigma^{\rm PDF}_\beta(x,Q^2) } \rp \, , 
\ee where
averages are taken over the $N_{\rm rep}=1000$ replicas of the sample,
and $\sigma_\alpha(x,Q^2)$ and $\sigma_\beta(x,Q^2)$ are the
standard deviations Eq.~(\ref{eq:sigmaMC}). In the Hessian
representation the same quantity  is given by
\be
\label{eq:corr_hessian}
\rho^{\alpha\beta}_{\rm hessian}(x,Q^2) = \frac{\sum_{k=1}^{N_{\rm eig}}\lc \lp
   \widetilde{f}_\alpha^{(k)}(x,Q^2) - f_\alpha^{(0)}(x,Q^2)\rp\lp
   \widetilde{f}_\beta^{(k)}(x,Q^2) -
  f_\beta^{(0)}(x,Q^2)\rp \rc}{\sqrt{\sum_{k=1}^{N_{\rm eig}}\lp
     \widetilde{f}_\alpha^{(k)}(x,Q^2) - f_\alpha^{(0)}(x,Q^2) \rp^2} \sqrt{\sum_{k=1}^{N_{\rm
  eig}}\lp \widetilde{f}_\beta^{(k)}(x,Q^2) - f_\beta^{(0)}(x,Q^2) \rp^2}} \, , 
\ee
where now the sum is performed over the $N_{\rm eig}$ eigenvectors,
and $f_\alpha^{(0)},f_\beta^{(0)}$ are the respective central sets, which
coincide with the MC average values.

The correlation coefficients
Eqs.~(\ref{eq:corr_mc}-\ref{eq:corr_hessian}) before and after Hessian
conversion  are compared in Fig.~\ref{correlations}: again, very good
agreement is seen, with differences compatible with the uncertainty on
the Monte Carlo representation. We have checked explicitly that
a  similar level of agreement is found at the level
of
correlations  between a number of
LHC cross-sections and differential distributions.

\begin{figure}[t]
\begin{center}
  \includegraphics[width=0.52\textwidth]{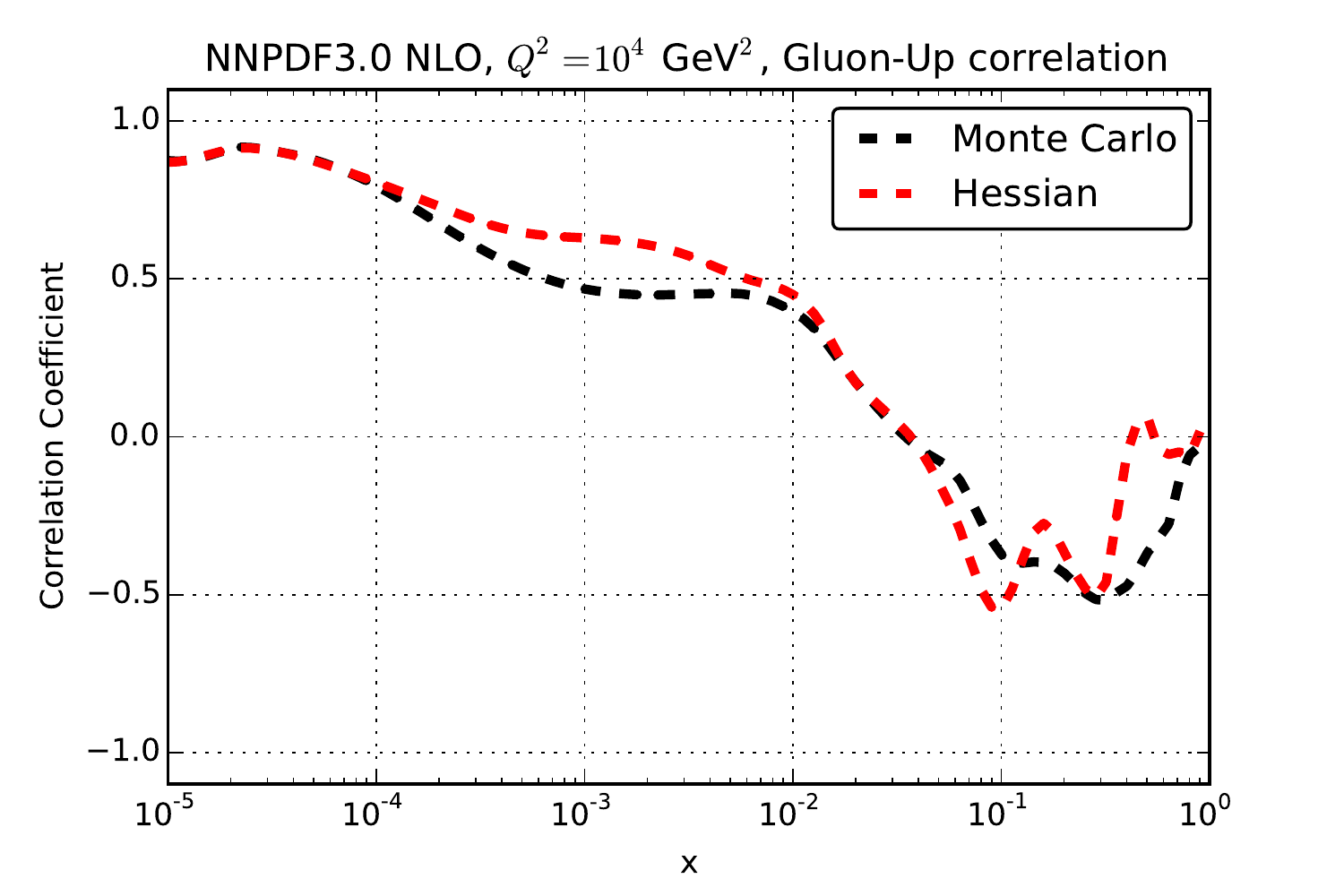}\includegraphics[width=0.52\textwidth]{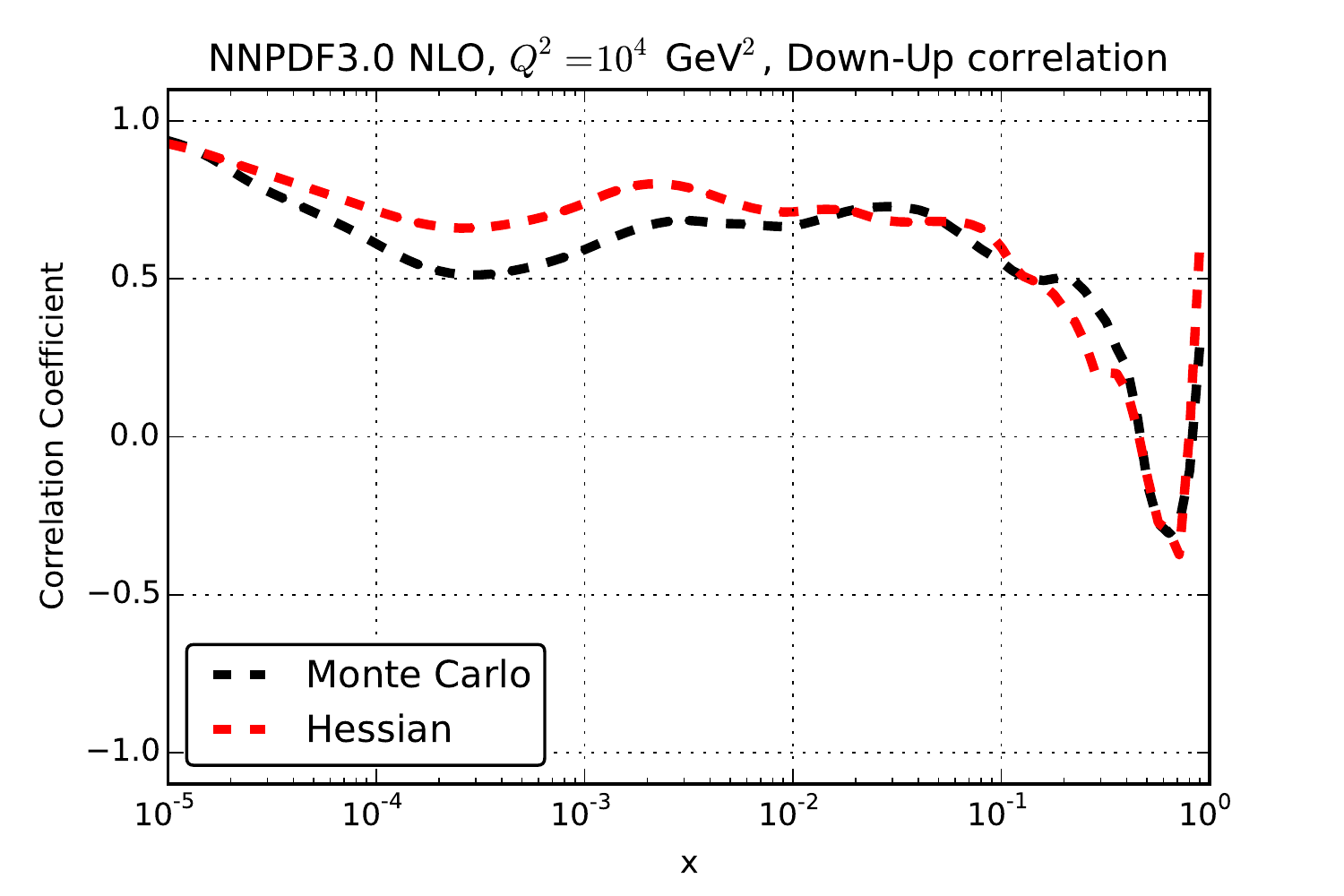}\\
  \includegraphics[width=0.52\textwidth]{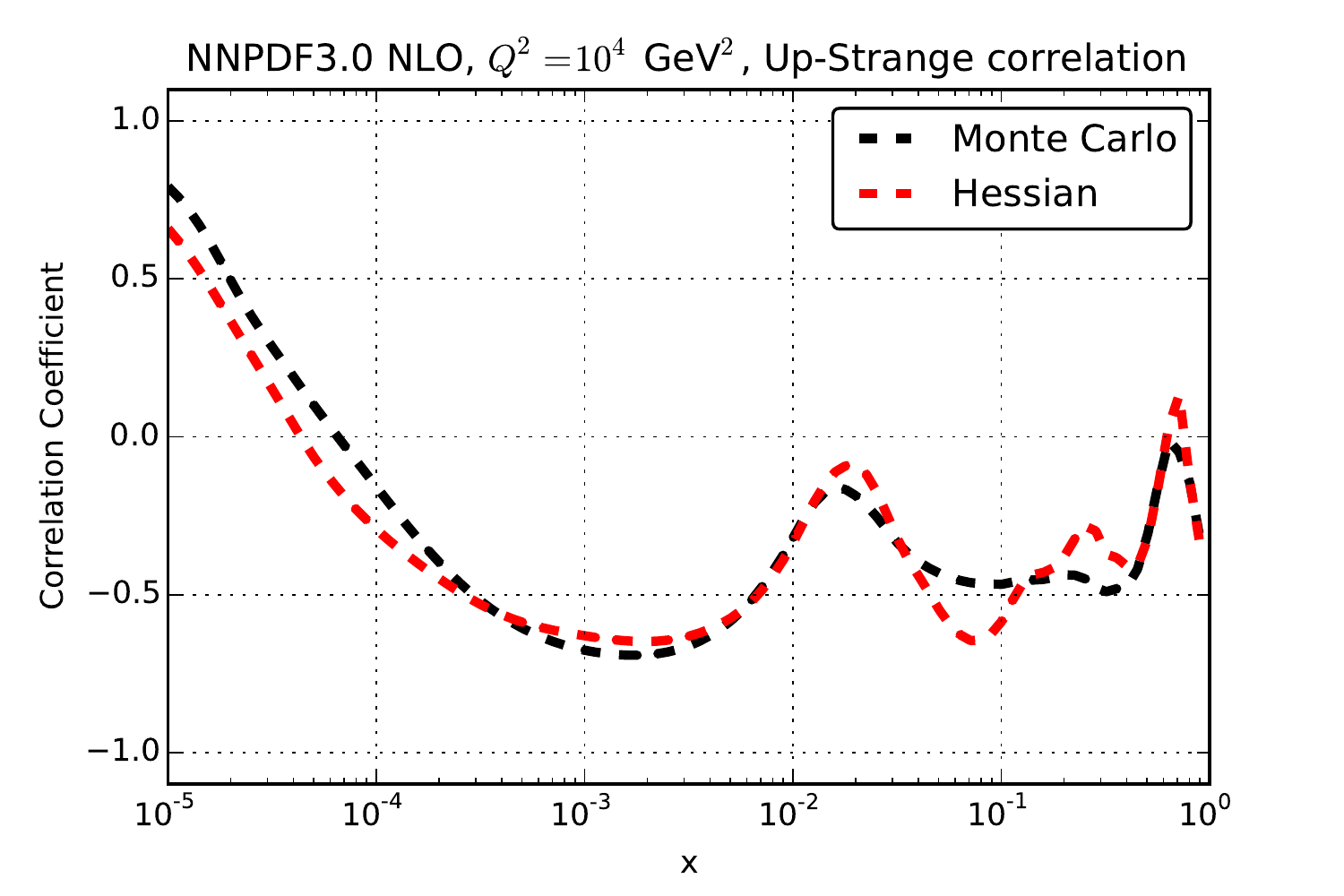}\includegraphics[width=0.52\textwidth]{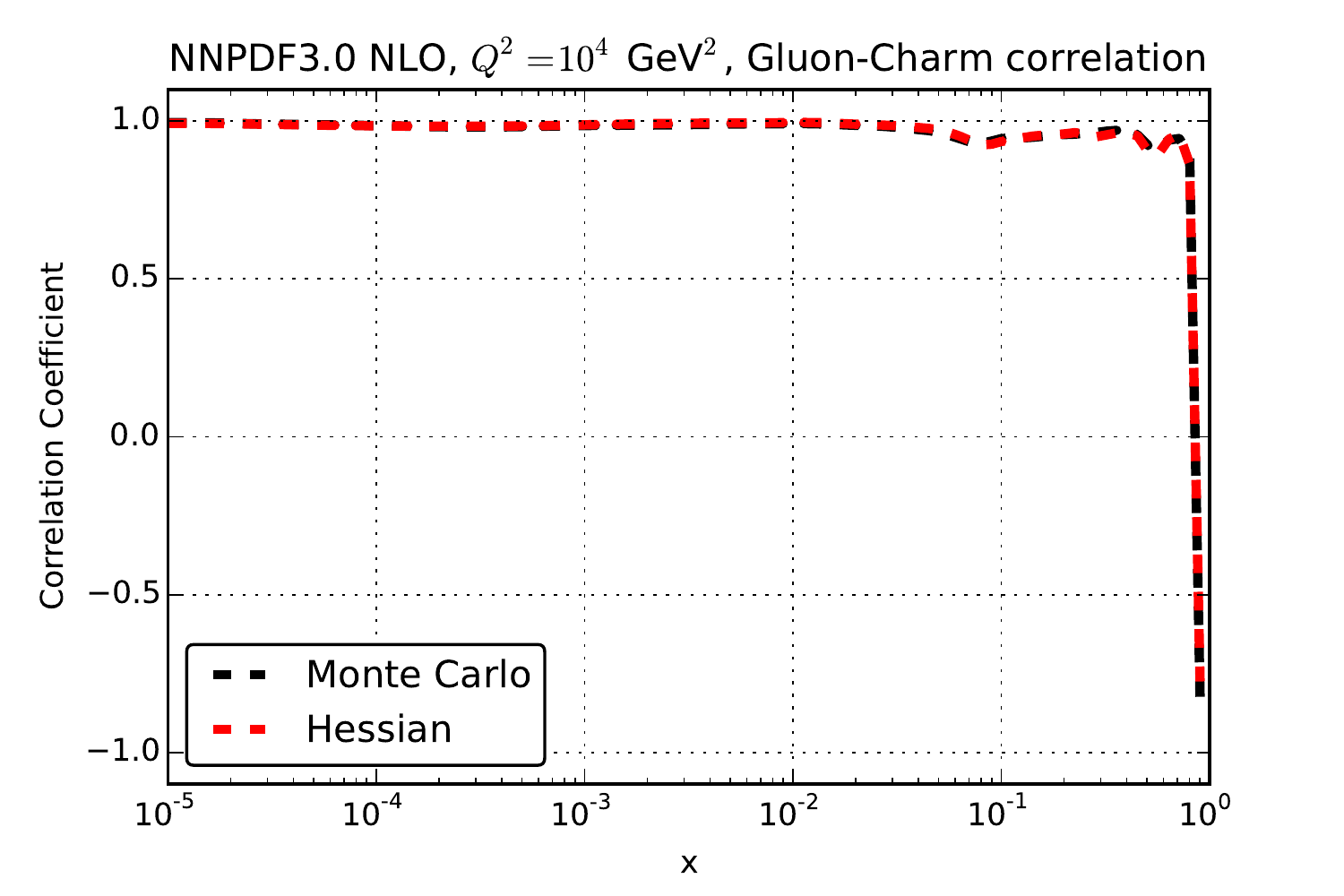}
\end{center}
\vspace{-0.3cm}
\caption{\small \label{correlations}
  Correlation coefficients  between pairs of PDFs at a common value of
  $x$ and $Q^2$ versus $x$ for  $Q^2=10^4$ GeV$^2$ before and after
  Hessian conversion.
}
\end{figure}

%
We conclude that at the level of second moments the Hessian
representation of the starting Monte Carlo probability distribution is
very accurate. Of course, to the extent that higher moments deviate
from Gaussian behaviour they will be accordingly not so well 
reproduced.

\subsection{A closure test: Hessian representation of Hessian PDFs}

We now consider the MMHT14 NLO PDF set. In this case, the Monte Carlo
representation which is converted into Hessian is in turn obtained by
starting from an initial Hessian representation, using the methodology
of Ref.~\cite{Watt:2012tq}. We can then compare the starting and
final Hessian sets, thereby obtaining a closure test. 
This provides a powerful test of the basis-independence of our
procedure, in that the starting Hessian is defined in the space of
parameters of a specific functional form, which is then turned into
Hessian by running a Monte Carlo in parameter space, while our final
Hessian representation uses the ensuing Monte Carlo replicas as basis
functions. 

Again, we adopt the optimal choice of parameters  discussed in
Section~\ref{sec:numerical}, namely $N_{\rm eig} = 14$. Note that this
was obtained by relazing somewhat the criterion for keeping
eigenvectors of the covariance matrix, due to the greater correlation
of MMHT PDF replicas: indeed, we have verified that use of the same
criterion as for the sets we considered previously would need to a
smaller optimal number of eigenvectors ($N_{\rm eig} = 12$ instead of
$N_{\rm eig} = 14$), and a considerable loss of accuracy.

The starting Hessian representation and our final Hessian conversion
are compared 
in Fig.~\ref{fig:pdfcomp-mmht14} at $Q^2=2$ GeV$^2$.
Again we find agreements of the uncertainty to better than 5\%.
It is interesting to observe that the original Hessian representation had
$N_{\rm eig}=25$
asymmetric eigenvectors (corresponding to 50 error sets),
 while our final Hessian conversion only needs
$N_{\rm eig}=14$ symmetric
eigenvectors. This means that our algorithm has managed to achieve a
compression of the 
information in the native Hessian representation, thanks to the use of
replicas as a basis, with minimal information loss.

\begin{figure}[t]
\begin{center}
  \includegraphics[width=0.48\textwidth]{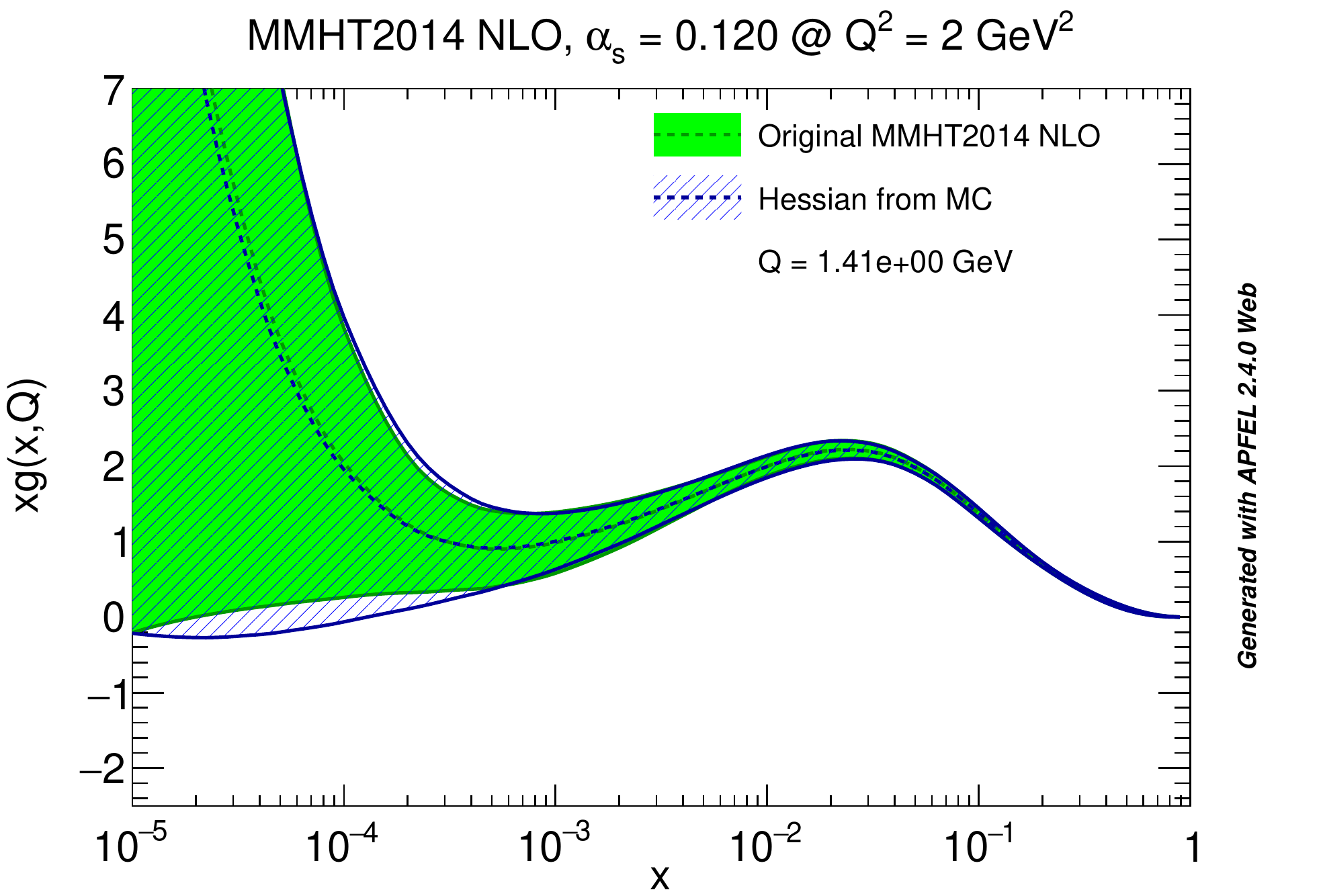}
  \includegraphics[width=0.48\textwidth]{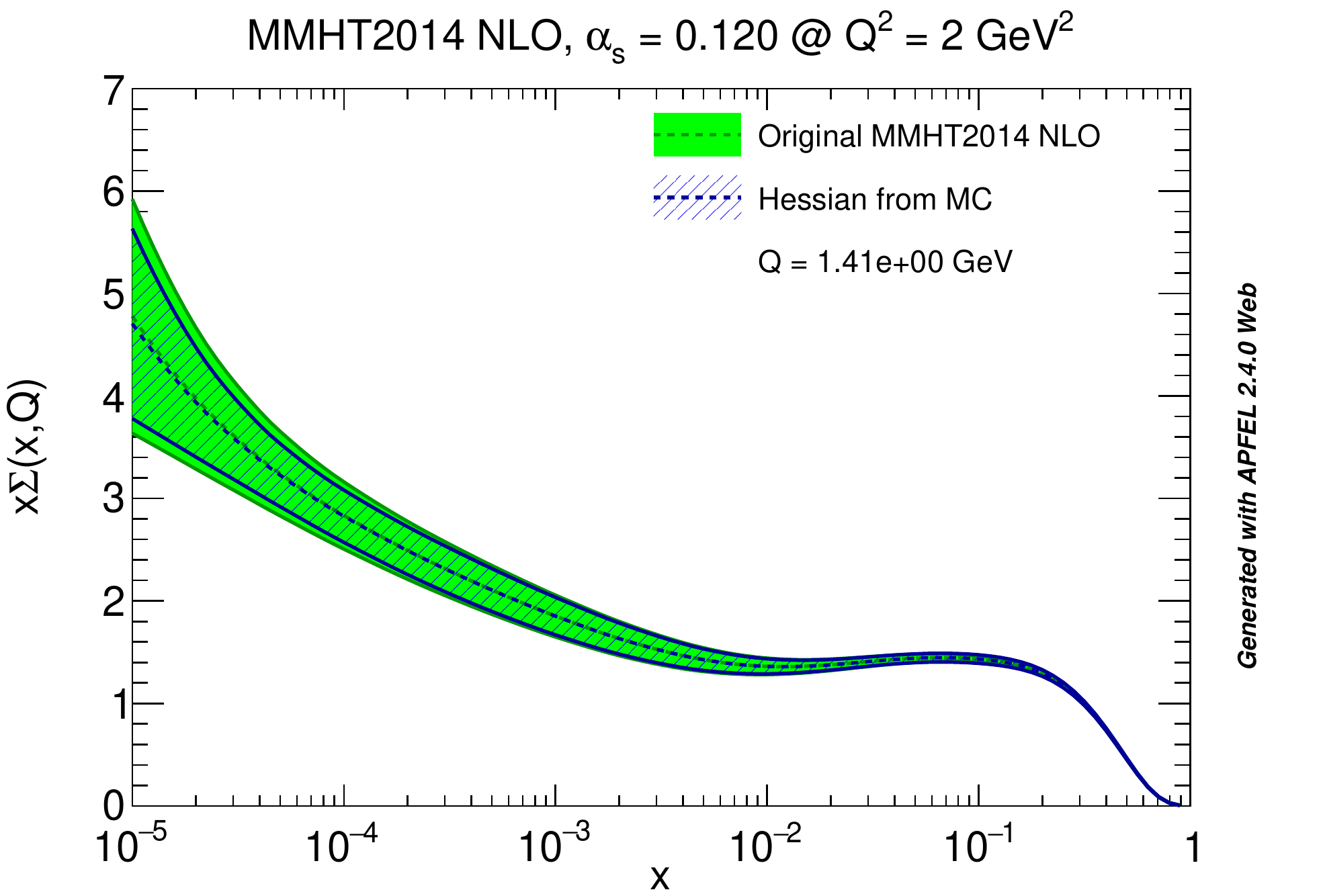}
  \includegraphics[width=0.48\textwidth]{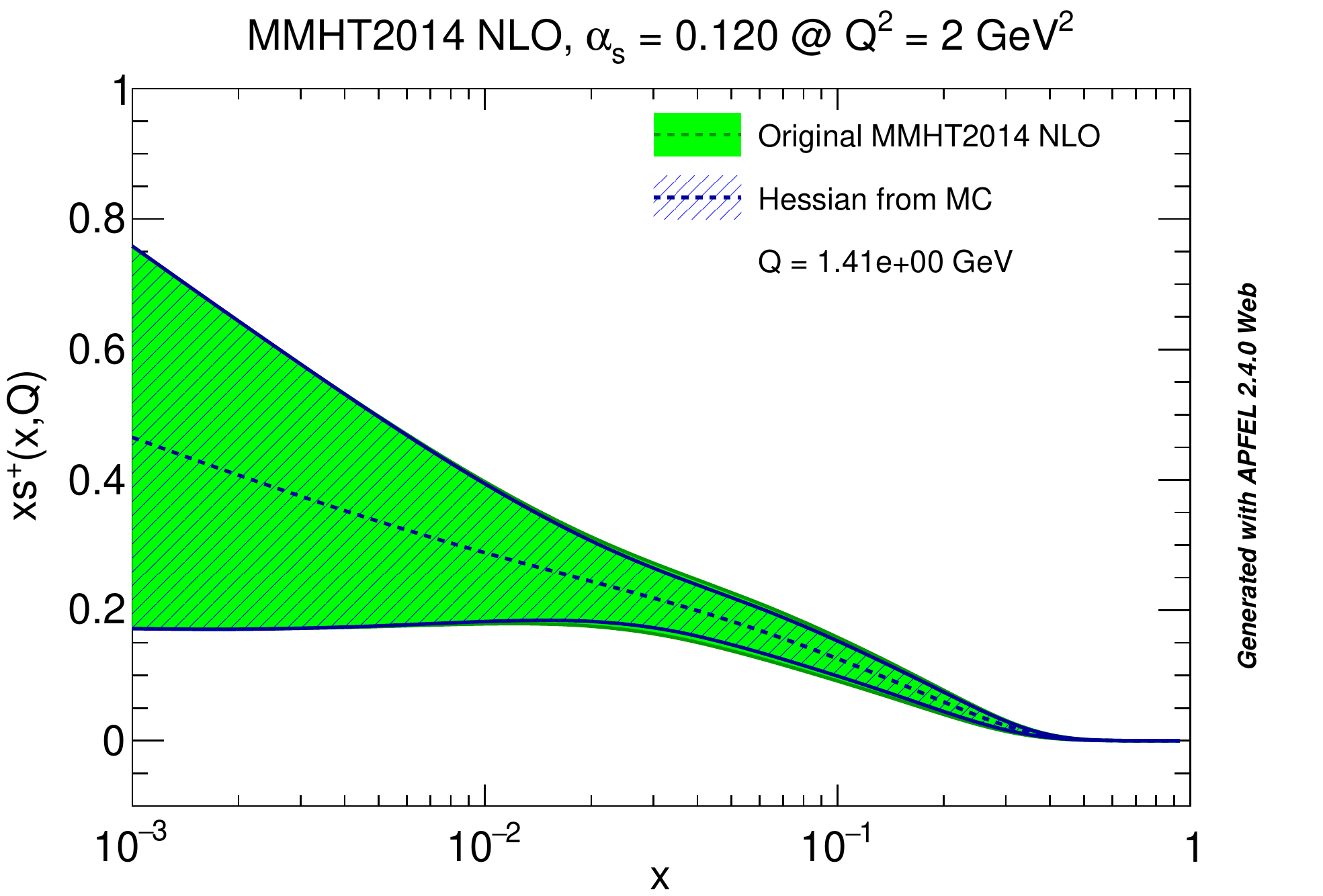}
   \includegraphics[width=0.48\textwidth]{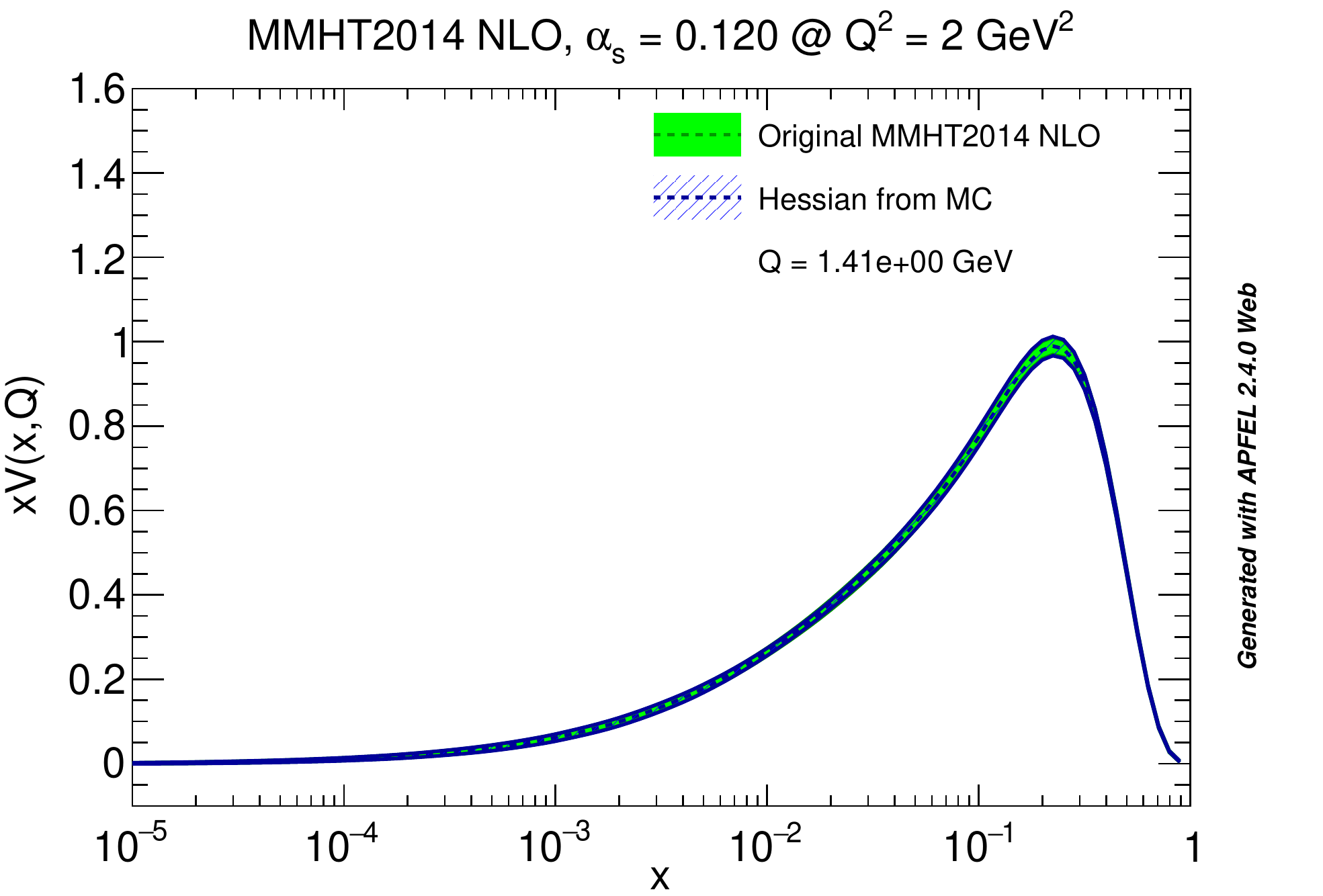}
\end{center}
\vspace{-0.3cm}
\caption{\small \label{fig:pdfcomp-mmht14} Same as Fig.~\ref{fig:pdfcomp}, but
  for MMHT14 NLO PDFs, with $N_{\rm eig}=25$ asymmetric eigenvectors in
  the starting Hessian set and  $N_{\rm eig}=14$ symmetric
  eigenvectors in our final Hessian conversion.}
\end{figure}

\subsection{LHC phenomenology}
\label{sec:lhcpheno}

We finally validate our Hessian conversion at the level of physical
observables:  standard candle total cross-sections
and differential distributions. For simplicity, we perform all
comparisons at NLO, given that, clearly, the accuracy of the Hessian
approximation is essentially independent of the perturbative order.

In  Fig.~\ref{fig:lhcinc} we compare results obtained
using the Monte Carlo and Hessian NNPDF3.0 representations,
and the original and final Hessian representation of MMHT14 PDFs, for the
total cross-sections  for Higgs
production in gluon fusion obtained using the {\tt
  ggHiggs} code~\cite{Ball:2013bra}, top quark pair production
obtained with
{\tt top++}~\cite{Czakon:2011xx}, and inclusive $W$ and $Z$
production obtained with {\tt VRAP}~\cite{Anastasiou:2003ds}.
Results are always shown normalized to the 
value of the original Monte Carlo set.
For NNPDF3.0, the agreement is very good, with PDF uncertainties
consistent with 10\% differences at most.
Somewhat larger differences are found for MMHT14.

\begin{figure}[t]
\begin{center}
  \includegraphics[width=0.49\textwidth]{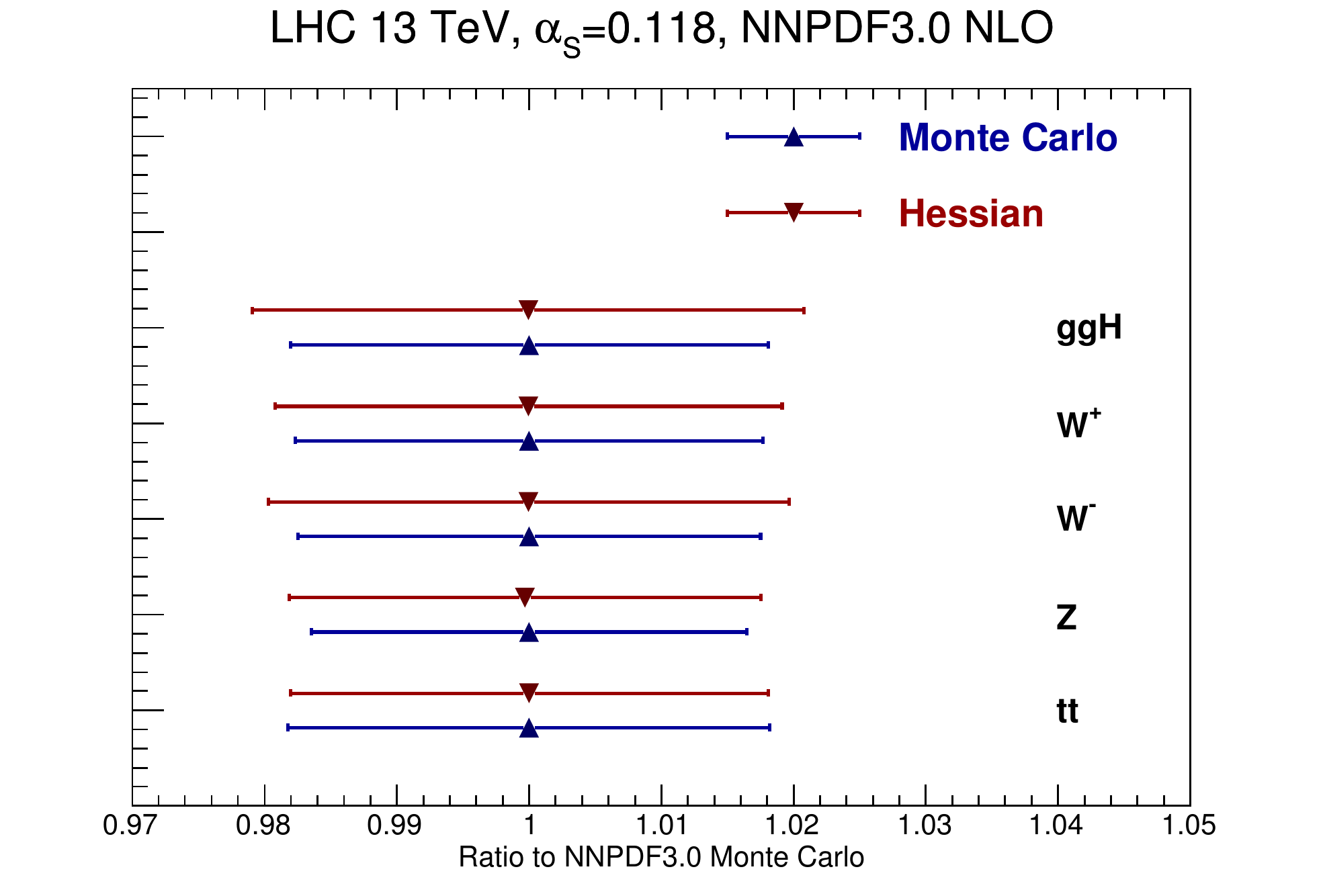}
  \includegraphics[width=0.49\textwidth]{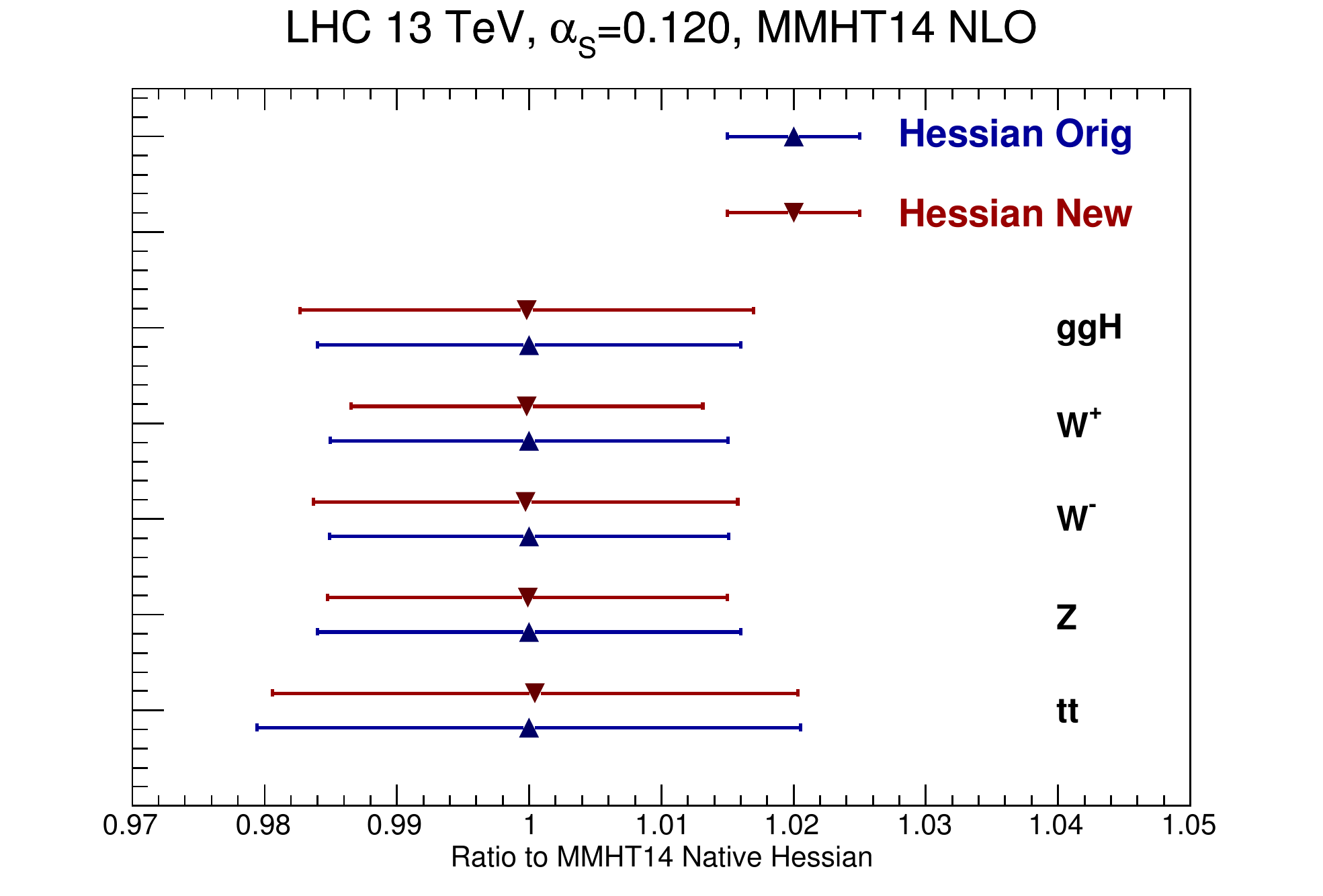}
\end{center}
\vspace{-0.3cm}
\caption{\small \label{fig:lhcinc} Comparison of
NLO inclusive cross-sections at the LHC, computed using PDFs before
and after Hessian conversion: for NNPDF3.0 (left) the Hessian
representation is compared to the original Monte Carlo, while for MMHT14
(right) the final Hessian conversion is compared to the original Hessian.
}
\end{figure}

%

\begin{figure}[t]
\begin{center}
  \includegraphics[width=0.42\textwidth]{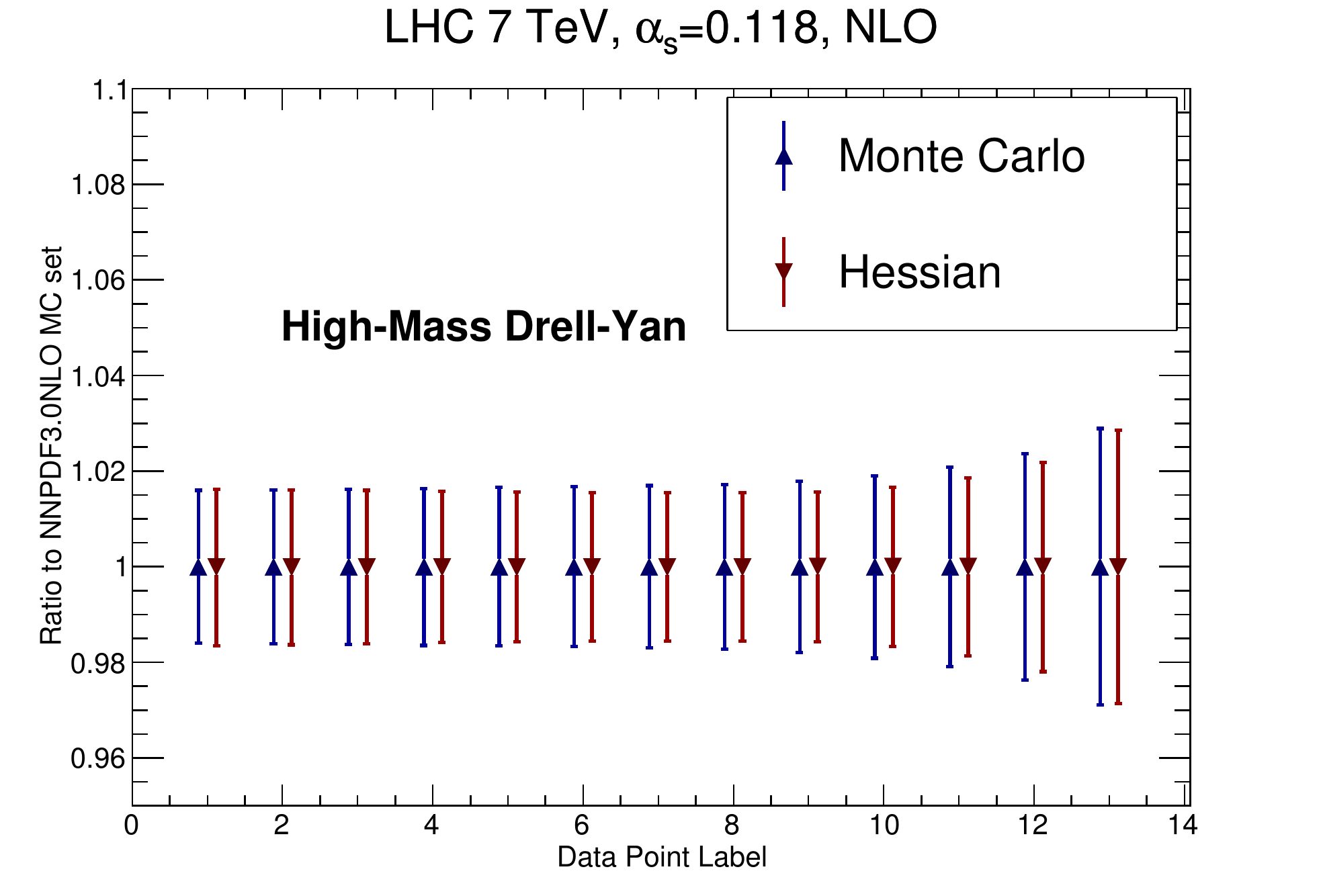}
  \includegraphics[width=0.42\textwidth]{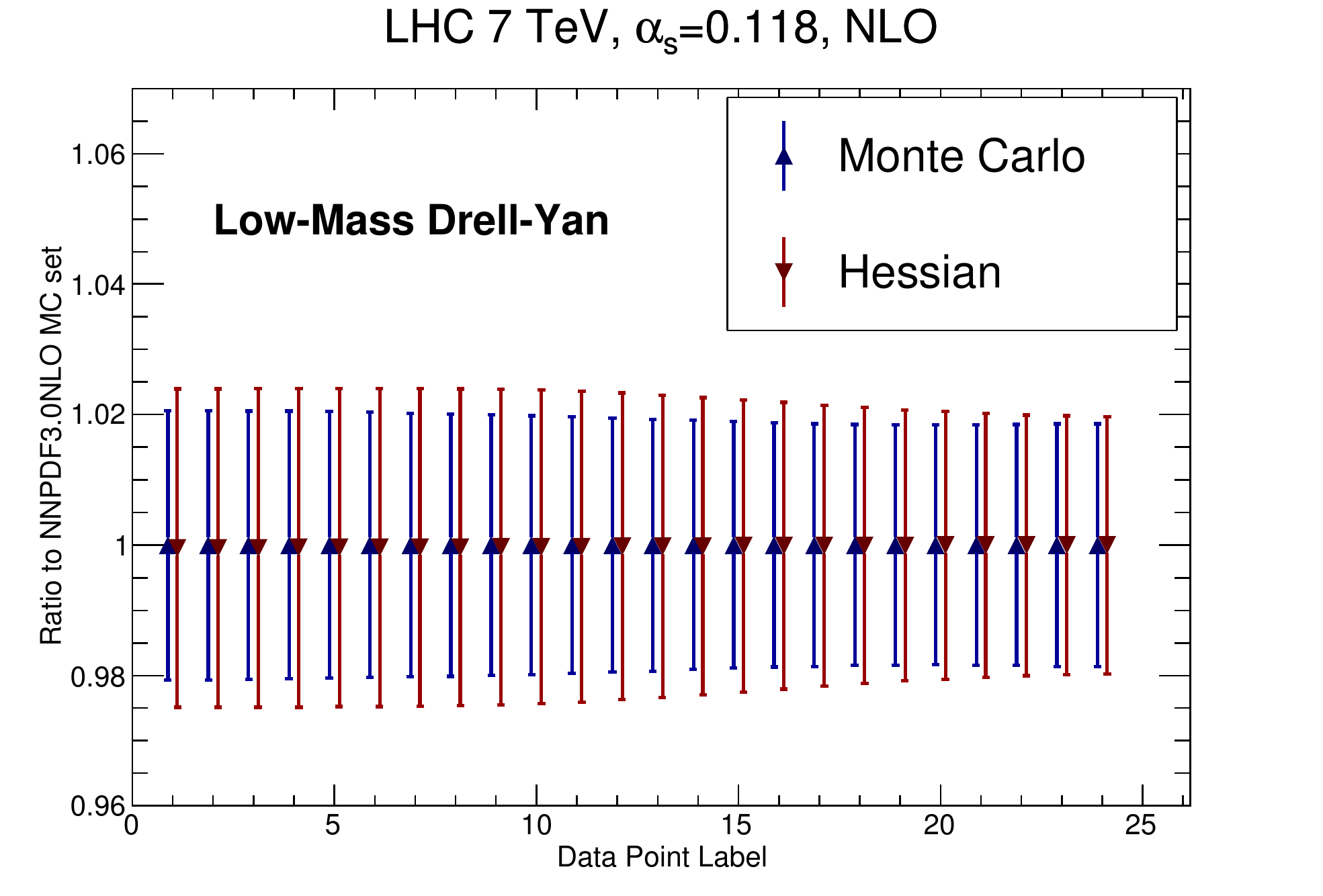}
  \includegraphics[width=0.42\textwidth]{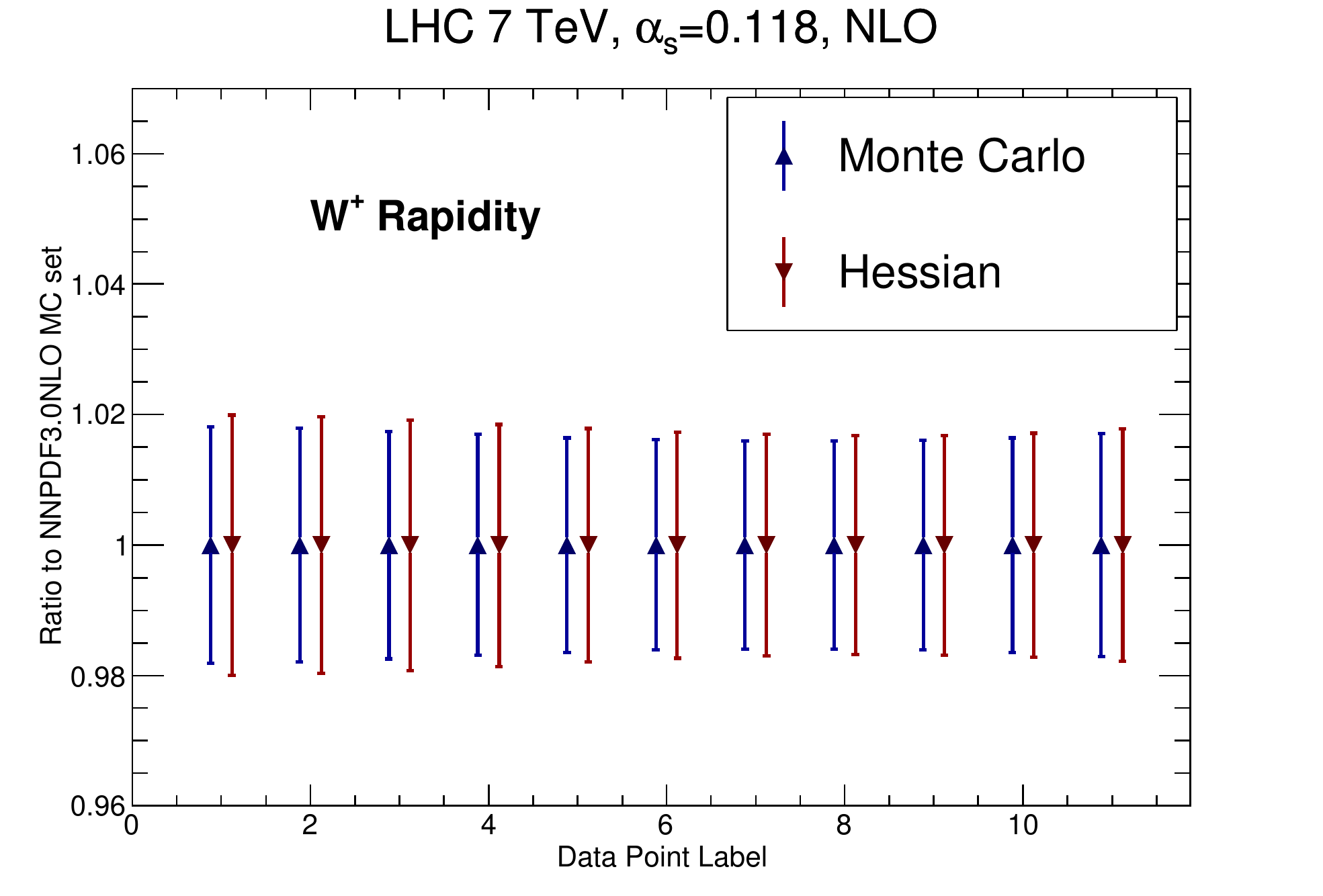}
  \includegraphics[width=0.42\textwidth]{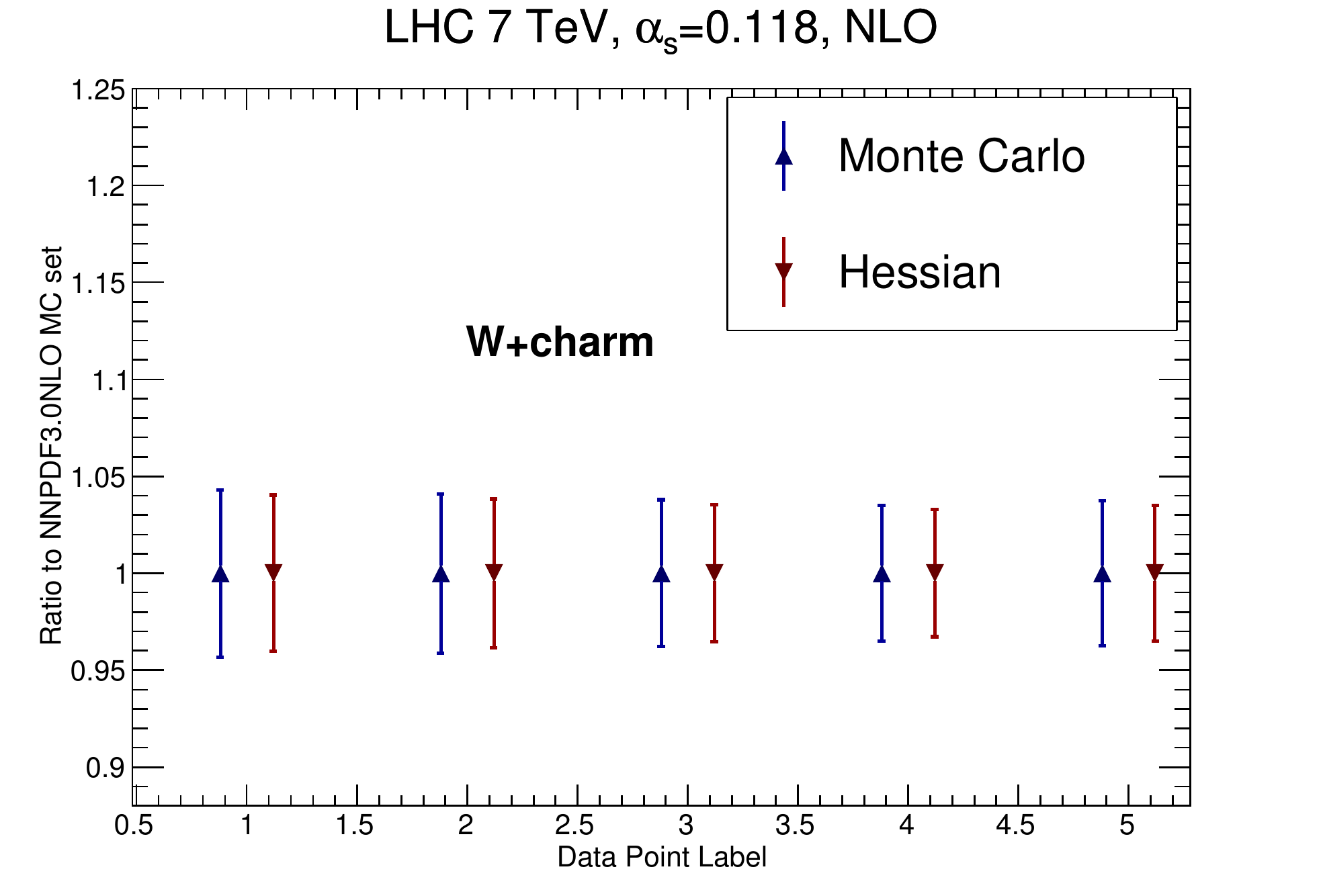}
  \includegraphics[width=0.42\textwidth]{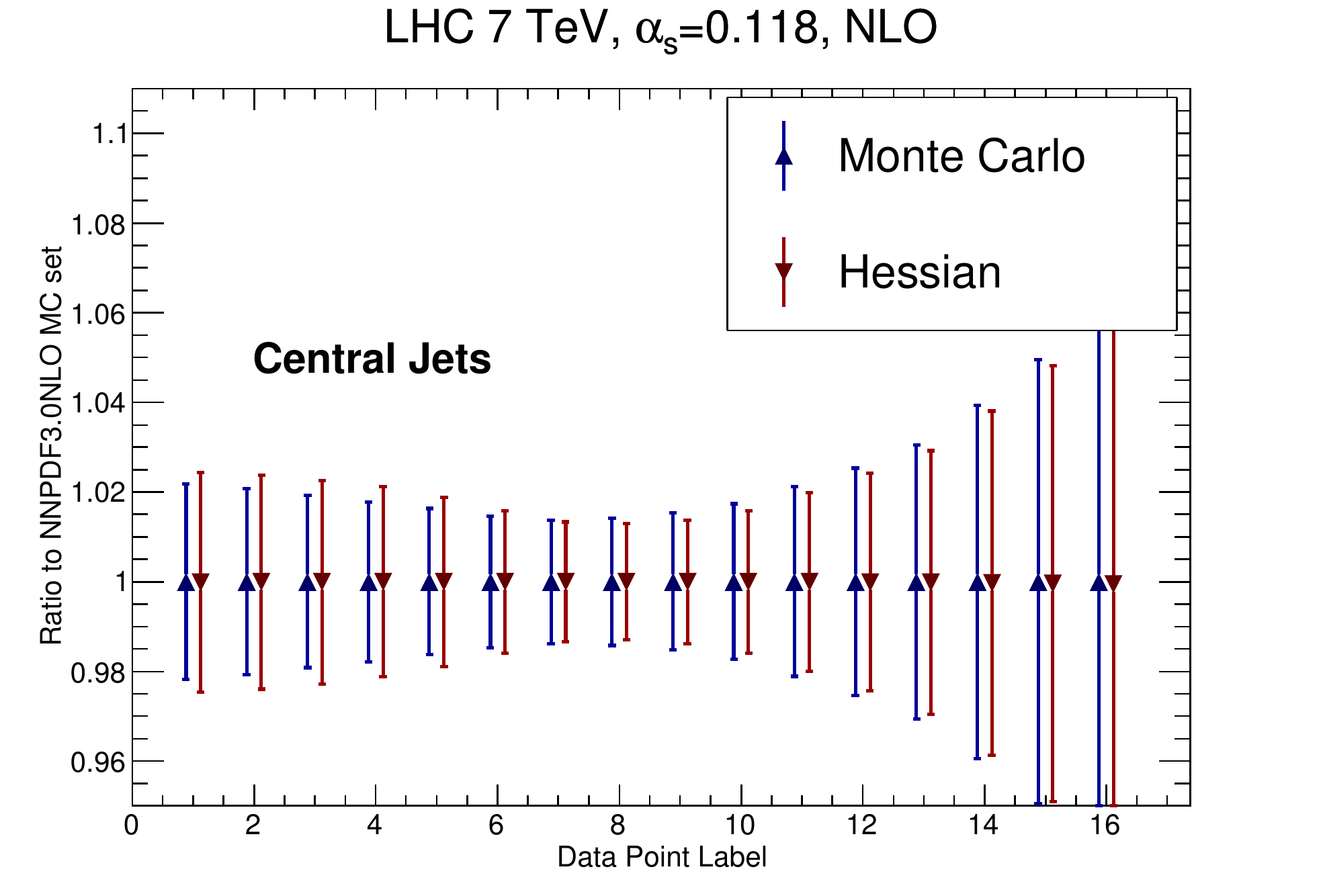}
  \includegraphics[width=0.42\textwidth]{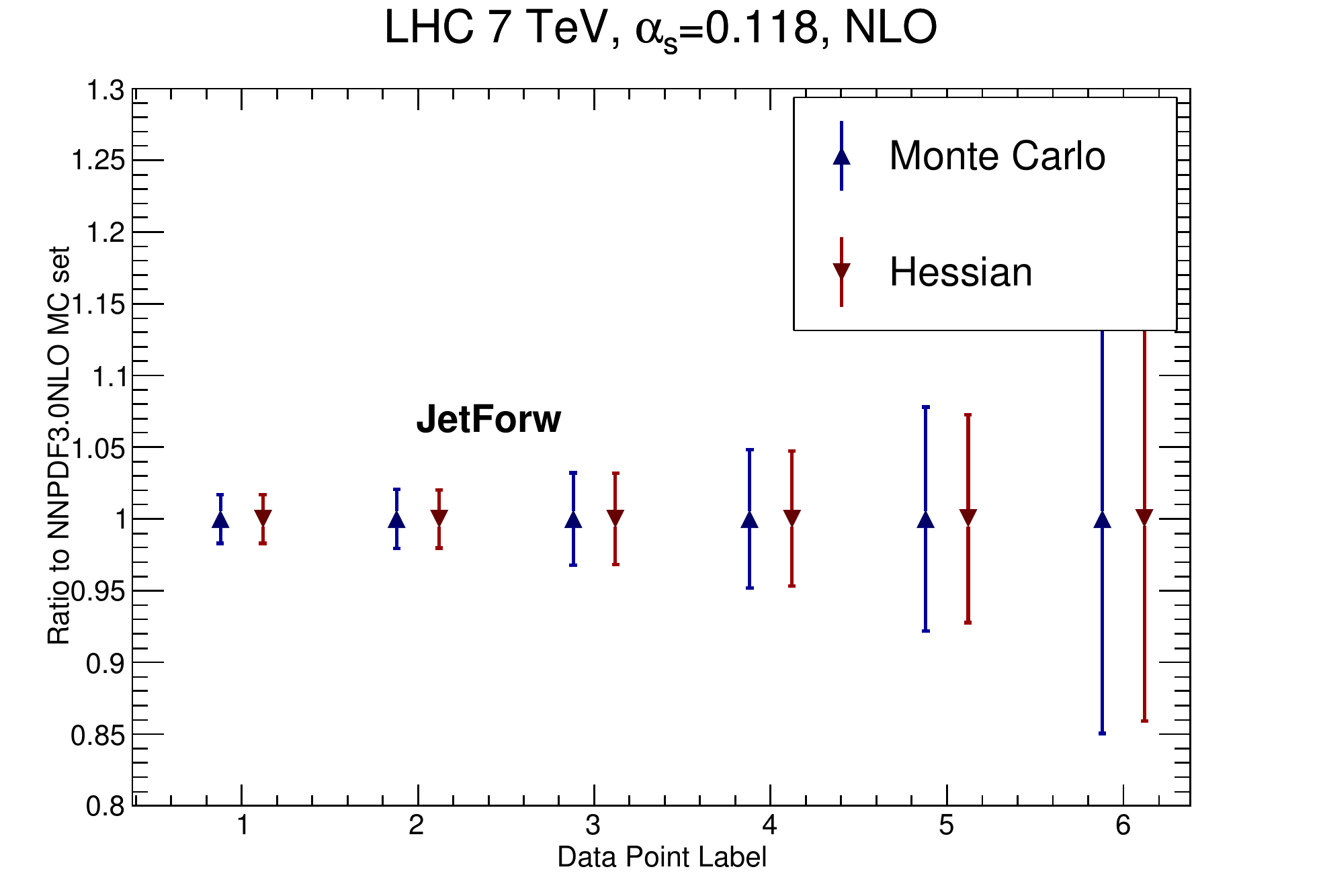}
\end{center}
\vspace{-0.3cm}
\caption{\small \label{fig:lhcapplgrid1} Comparison 
  of the original Monte Carlo representation and the  new
  Hessian representation of NNPDF3.0 NLO for a number
  of differential distributions at the LHC 7 TeV.
  The error band corresponds to the one-sigma PDF uncertainty in each bin.
  Results are shown normalized to the central value of the NNPDF3.0 NLO
  Monte Carlo set.
 See text for more details.
}
\end{figure}

\begin{figure}[t]
\begin{center}
  \includegraphics[width=0.42\textwidth]{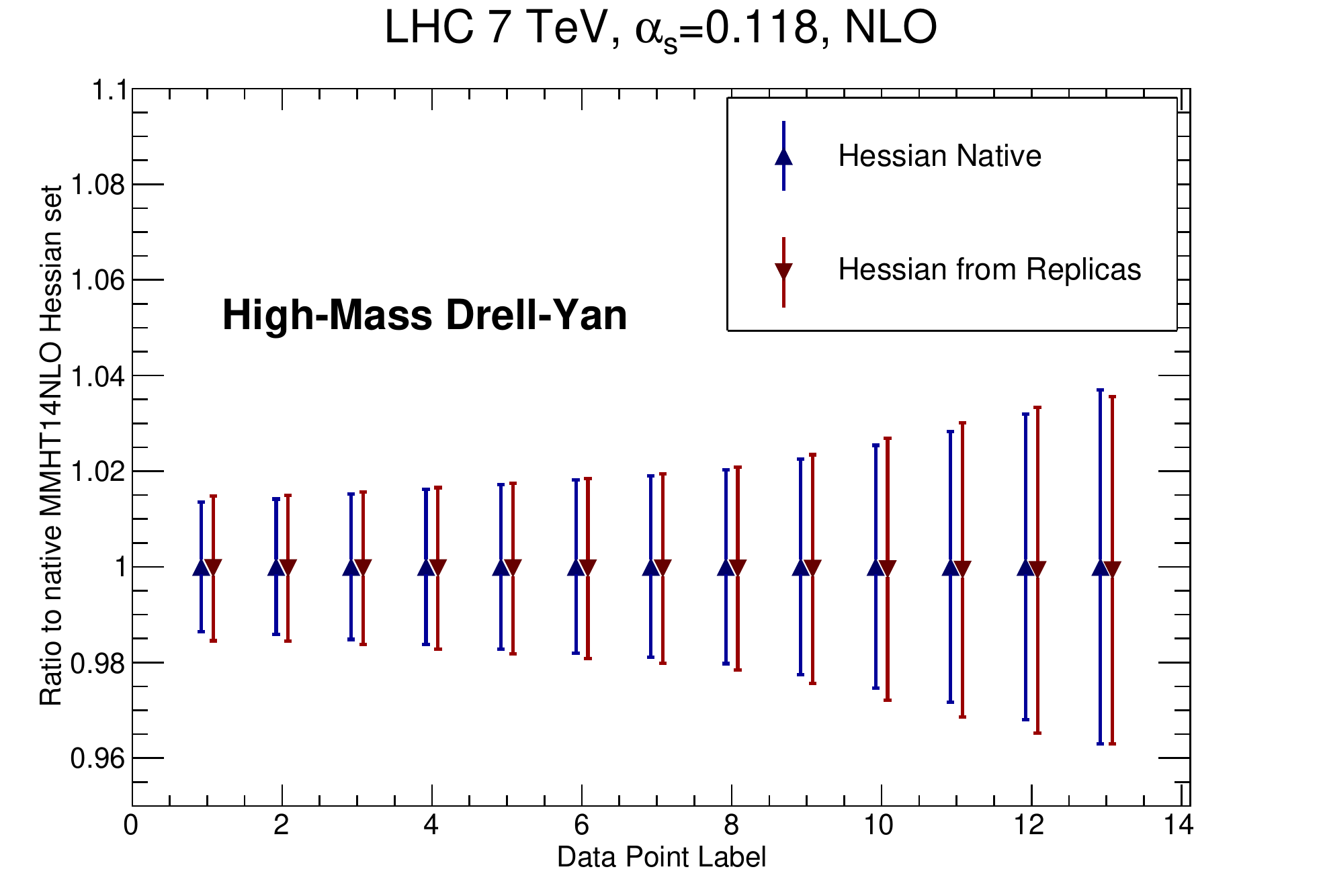}
  \includegraphics[width=0.42\textwidth]{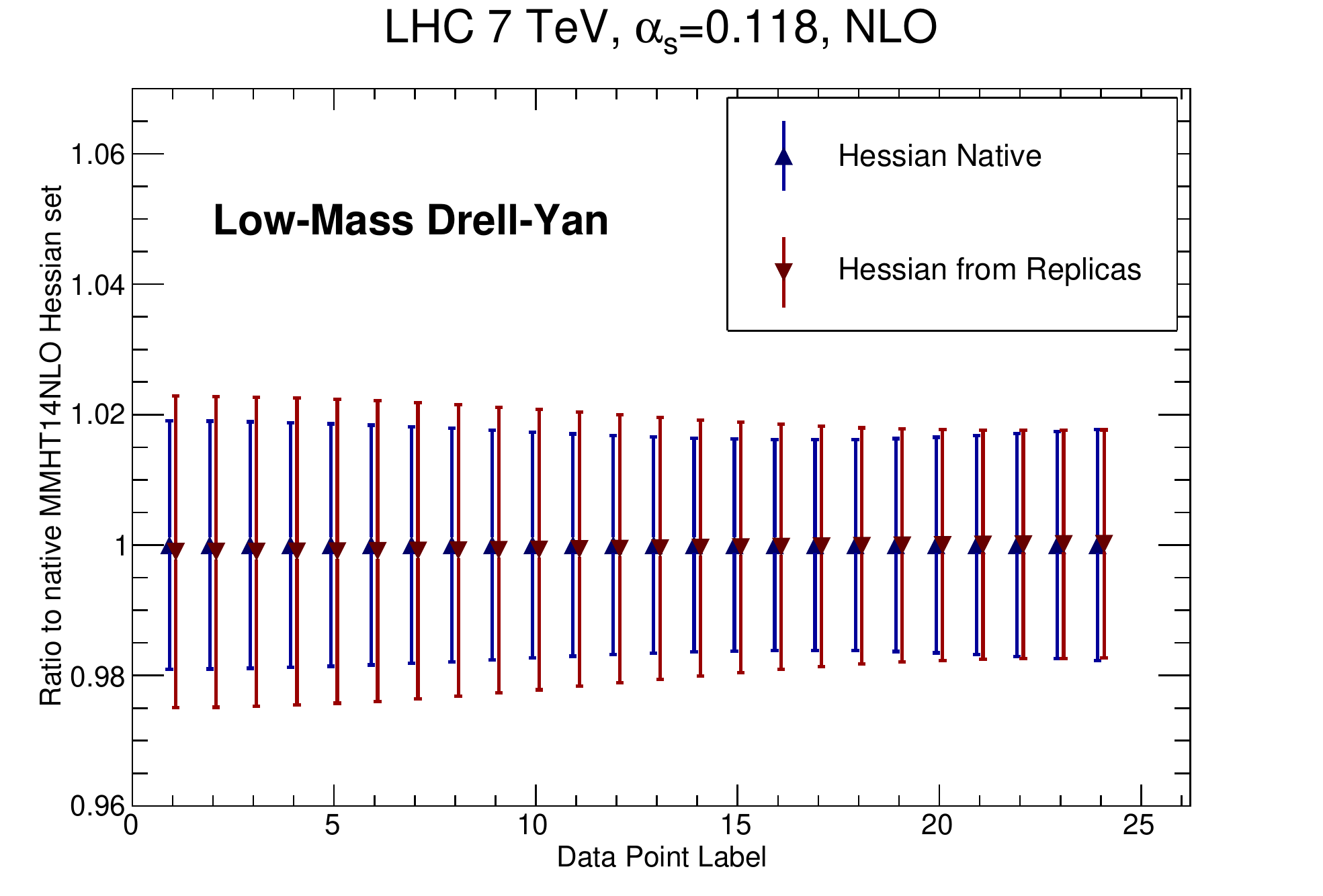}
  \includegraphics[width=0.42\textwidth]{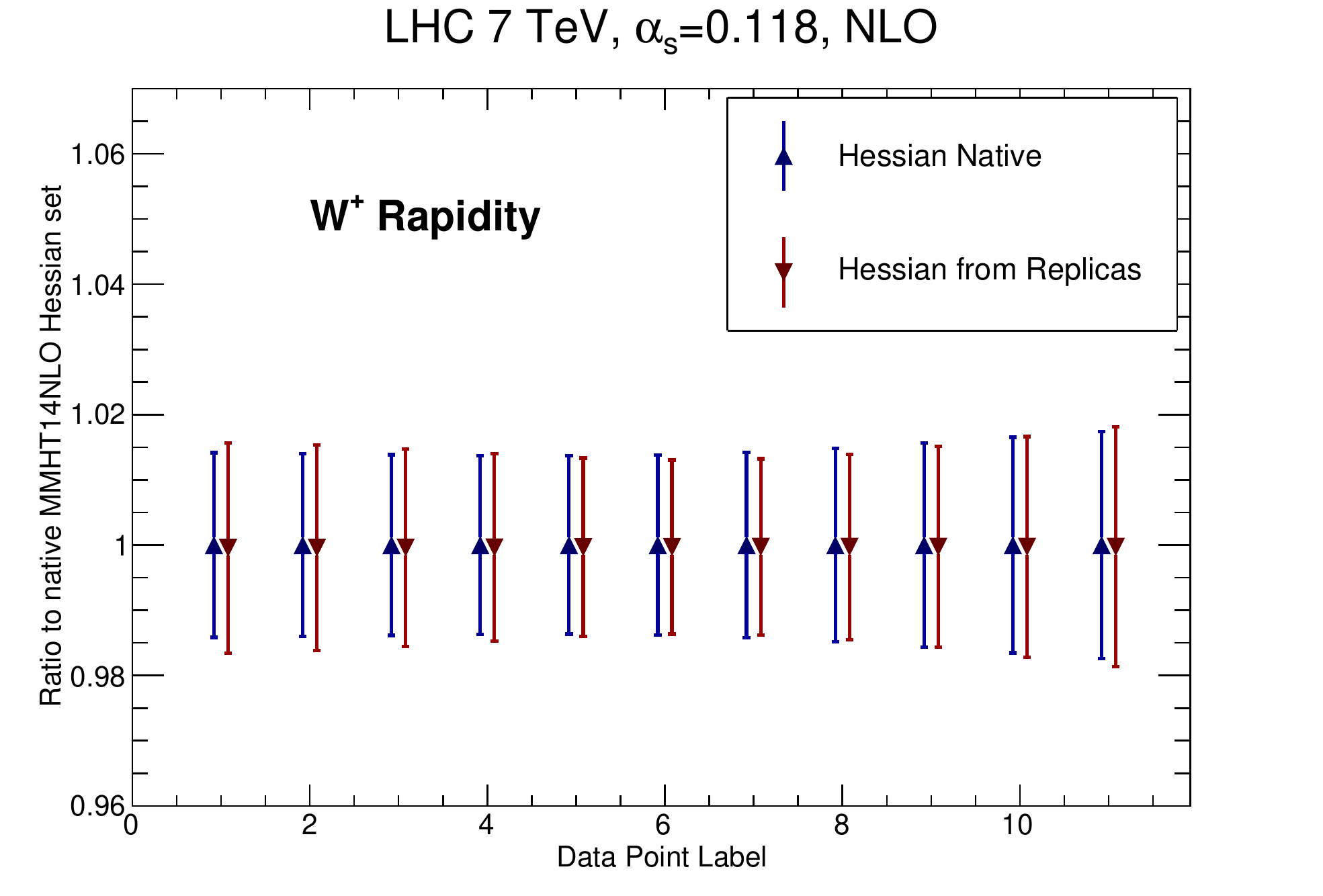}
  \includegraphics[width=0.42\textwidth]{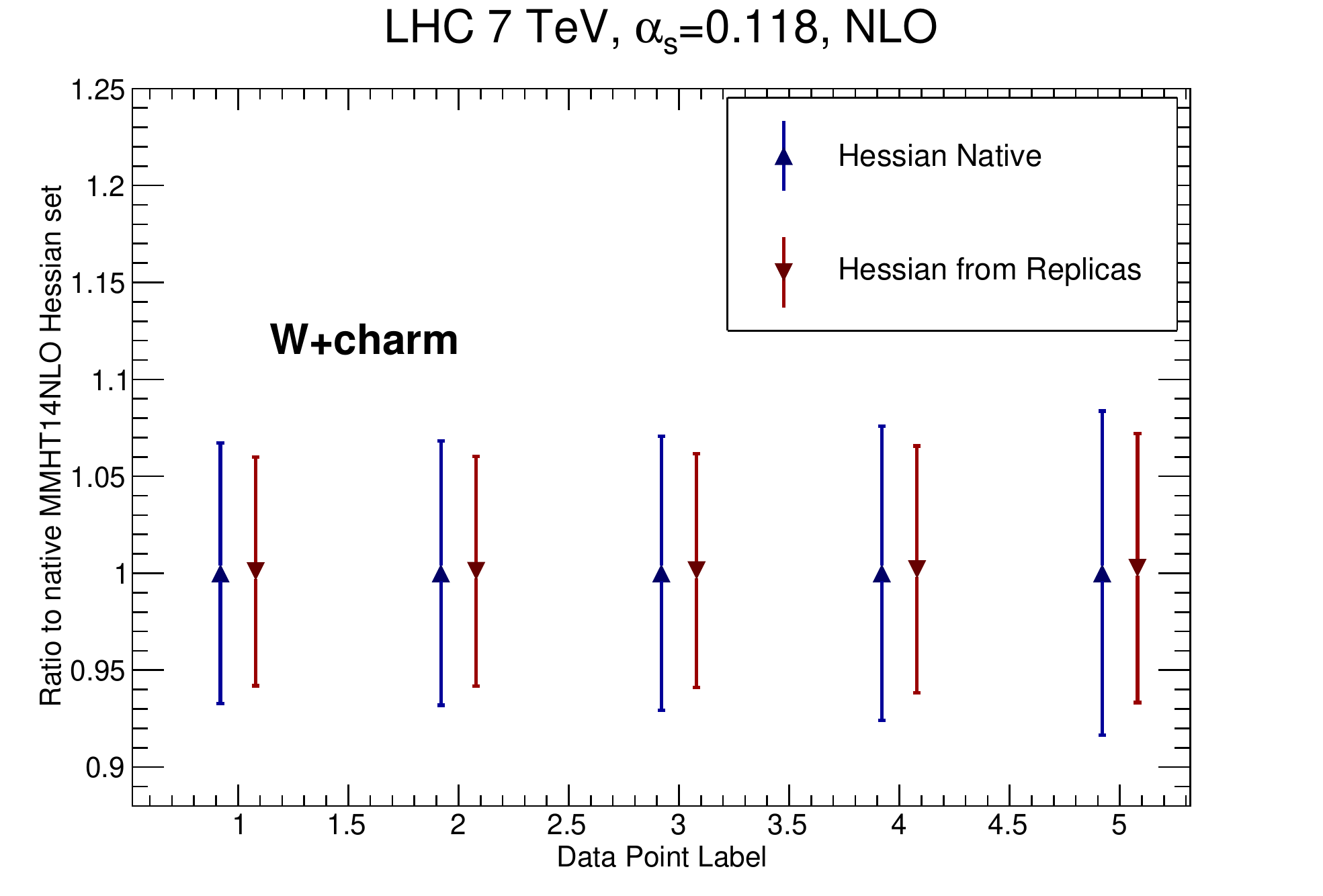}
  \includegraphics[width=0.42\textwidth]{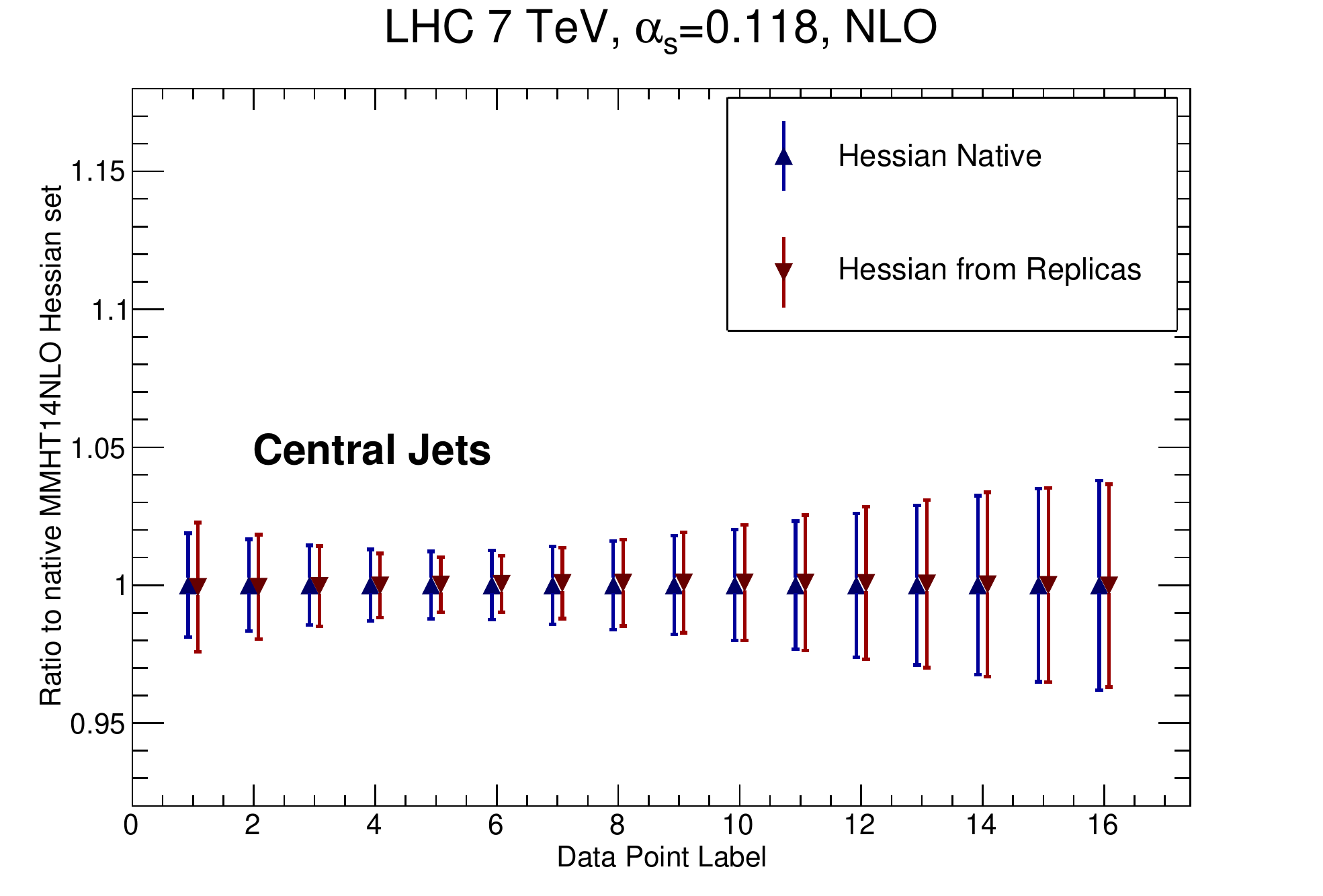}
  \includegraphics[width=0.42\textwidth]{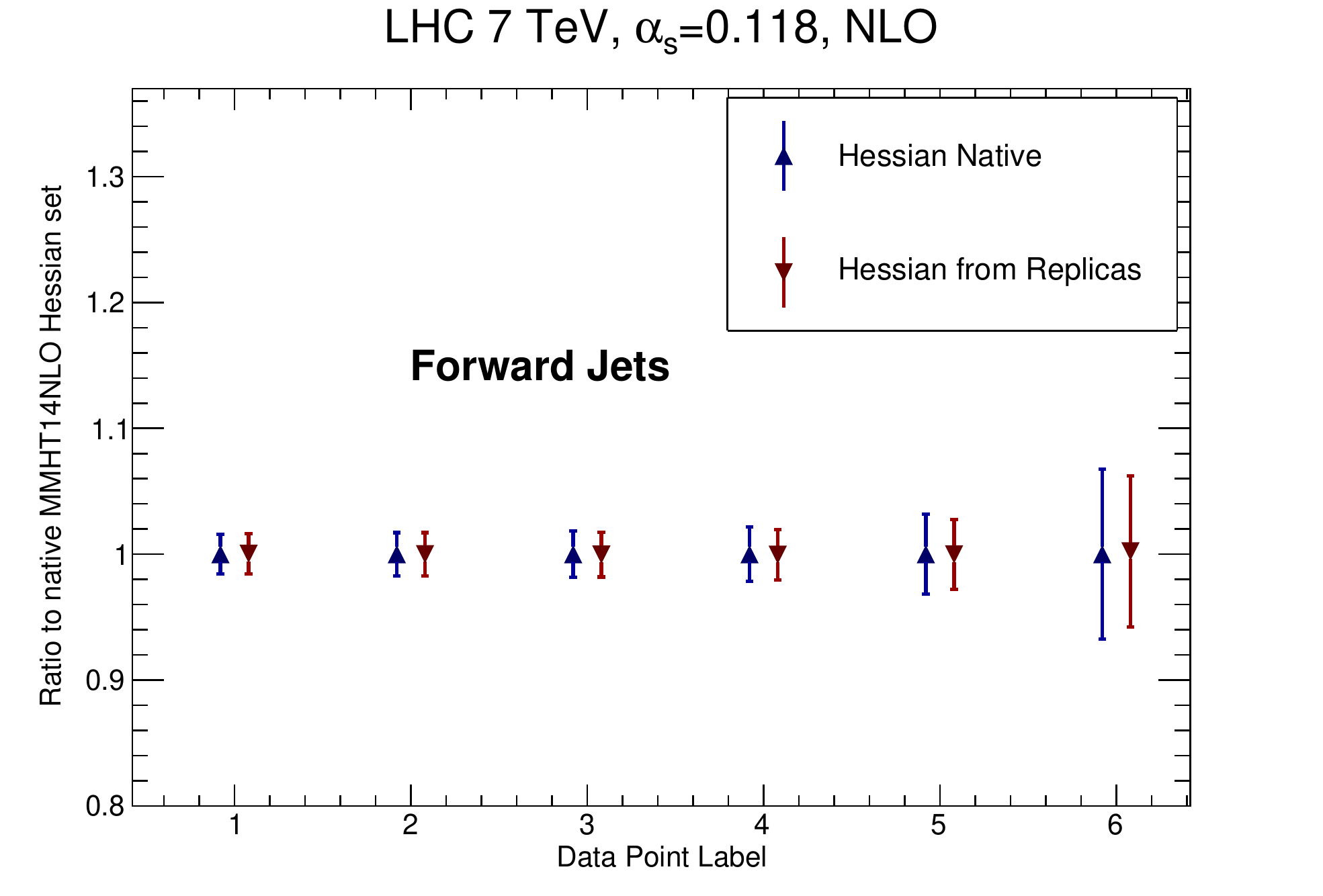}
\end{center}
\vspace{-0.3cm}
\caption{\small \label{fig:lhcapplgrid2}
  Same as Fig.~\ref{fig:lhcapplgrid1}, but for  MMHT14 NLO PDFs. The
  final Hessian set obtained after conversion of the Monte Carlo
  replicas is compared to the starting Hessian representation.
}
\end{figure}

We then compare several differential
distributions, chosen among those which have been used in the NNPDF3.0
PDF determination, and for which fast interfaces are  available, either {\tt
  APPLgrid}~\cite{Carli:2010rw}, {\tt
  FastNLO}~\cite{Wobisch:2011ij} or {\tt
  aMCfast}~\cite{amcfast}. 
Again, results are always shown 
normalized to the central value of either NNPDF3.0 NLO
Monte Carlo set or the MMHT14 NLO native Hessian.
In particular we consider:
\begin{itemize}
\item The ATLAS high-mass Drell-Yan measurement~\cite{Aad:2013iua},
  integrated over rapidity $|y_{ll}|\le 2.1$, and binned as a function
  of the di-lepton invariant mass pair $M_{ll}$,
\item the CMS double differential Drell-Yan measurement~\cite{CMSDY}
  in the low-mass region, $20 \ge M_{ll} \ge 30$ GeV, as a function of
  the di-lepton rapidity $y_{ll}$,
\item The CMS $W^+$ lepton rapidity
  distribution~\cite{Chatrchyan:2013mza},
\item The CMS measurement of $W^+$ production in association with charm quarks
  measurement~\cite{Chatrchyan:2013uja}, as a function of the lepton
  rapidity $y_l$,
\item The ATLAS inclusive jet production measurement~\cite{Aad:2011fc}
  in the central rapidity region, $|y_{\rm jet}|\le 0.3$, as a
  function of the jet $p_T$, and
\item The same ATLAS inclusive jet production
  measurement~\cite{Aad:2011fc} now in the forward rapidity region,
  $4.0 \le |y_{\rm jet}|\le 4.4$, as a function of the jet $p_T$.
\end{itemize}
These observables probe a wide range of PDF combinations, from light
quarks and anti-quarks (Drell-Yan) and strangeness
($W+c$) to the gluon (jets) in a wide range of
Bjorken-$x$ and momentum transfers $Q^2$.

Results are shown 
in Fig.~\ref{fig:lhcapplgrid1} for NNPDF, and in
Fig.~\ref{fig:lhcapplgrid2} 
for MMHT14. Again, there is a good
agreement between the original Monte Carlo and the new Hessian
representations, with differences smaller than 10\%.

\section{A Hessian representation of combined MC sets: MC-H PDFs}
\label{sec:cmcpdfs}

As discussed in the introduction, the  Monte Carlo representation of
PDFs offers the possibility of constructing combined PDF sets which
incorporate information from different PDF determinations, and thus
provide an alternative~\cite{Forte:2010dt,Watt:2012tq,Forte:2013wc,Gao:2013bia} to the current
PDF4LHC recommendation~\cite{Botje:2011sn} for the combination of
predictions obtained using different PDF sets, which is less than
ideal from a statistical point of view. 

An obvious shortcoming of a
combined Monte Carlo set is that it contains generally a large number
of replicas, which can be cumbersome to handle, and which is
computationally very
intensive.
This difficulty has
been handled in Ref.~\cite{Carrazza:2015hva} by developing a
compression algorithm, whereby the number of replicas in a Monte Carlo
set is optimized by means of a GA without significant
loss of information. This has enabled the construction of sets of less
than $N_{\rm rep}=50$ replicas which reproduce most of the information
contained in a starting $N_{\rm rep}=300$ replica set.

Combined Monte Carlo set are generally non-Gaussian even when obtained
by combining individually Gaussian PDF sets. However, once again the Gaussian
approximation may often be adequate in practice, and then a Hessian
representation may be useful for applications as repeatedly
mentioned. The so-called Meta-PDF method~\cite{Gao:2013bia} 
has been proposed as a way of 
dealing with this problem: it consists of re-fitting a fixed functional
form to the final combined Monte Carlo set, and thus it has the usual
shortcomings related to a fixed choice of functional form.
We now show how by applying our Monte Carlo to Hessian conversion to
a combined Monte Carlo set we directly obtained a Hessian
representation with a small number of eigenvector, therefore obtaining
a compressed Hessian representation, which we call MC-H PDFs.

\begin{figure}[t]
\begin{center}
\includegraphics[width=0.49\textwidth]{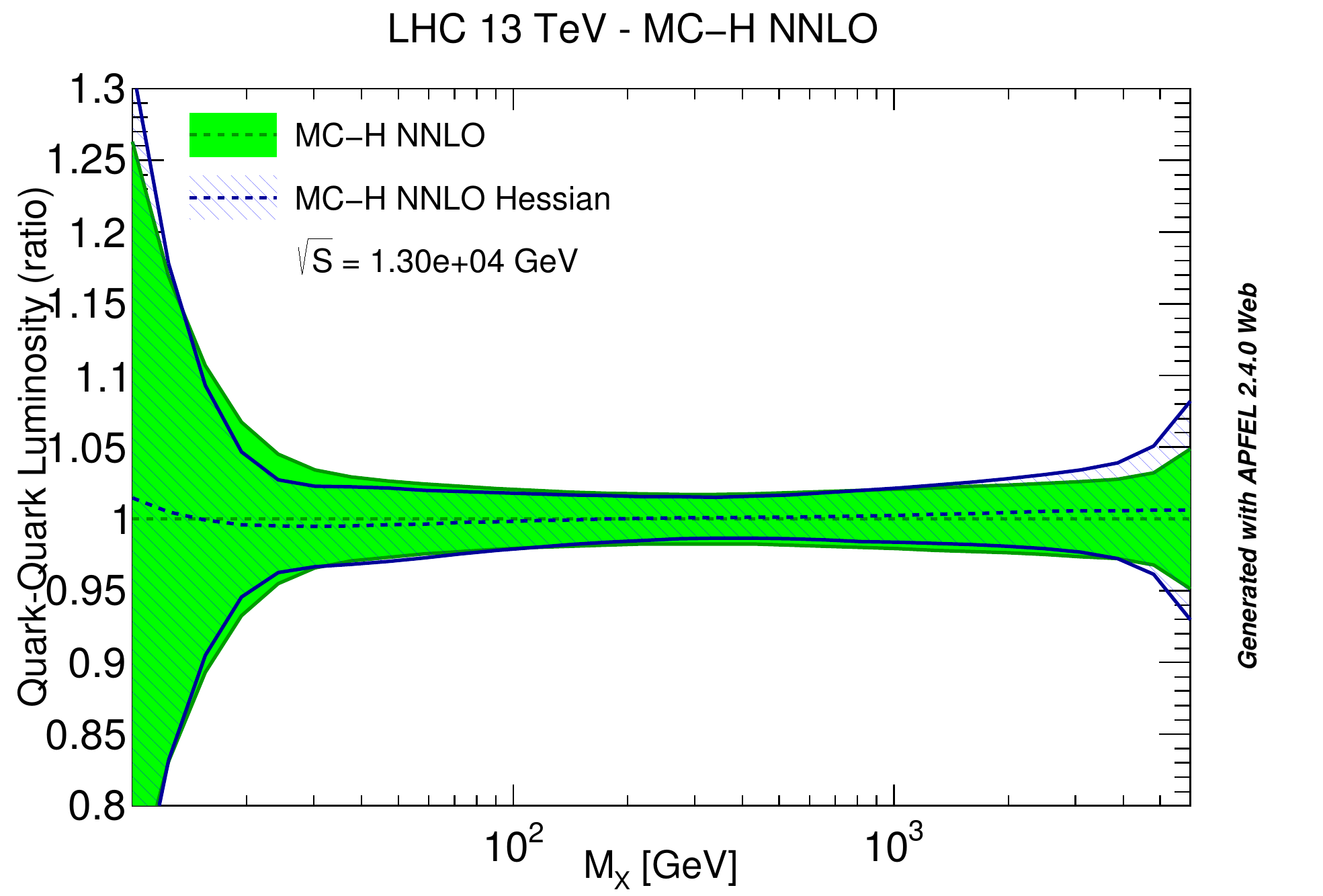}
\includegraphics[width=0.49\textwidth]{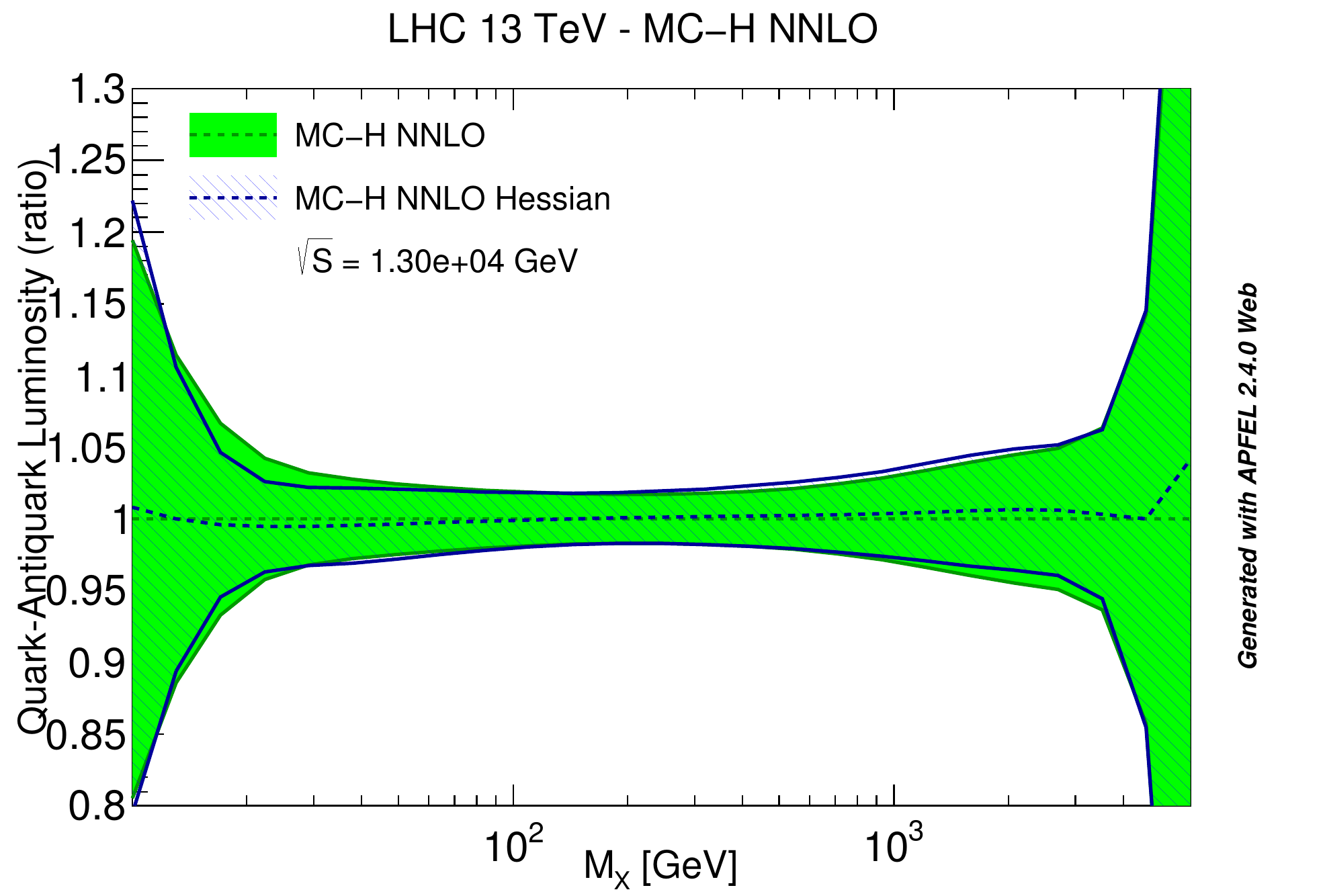}
\includegraphics[width=0.49\textwidth]{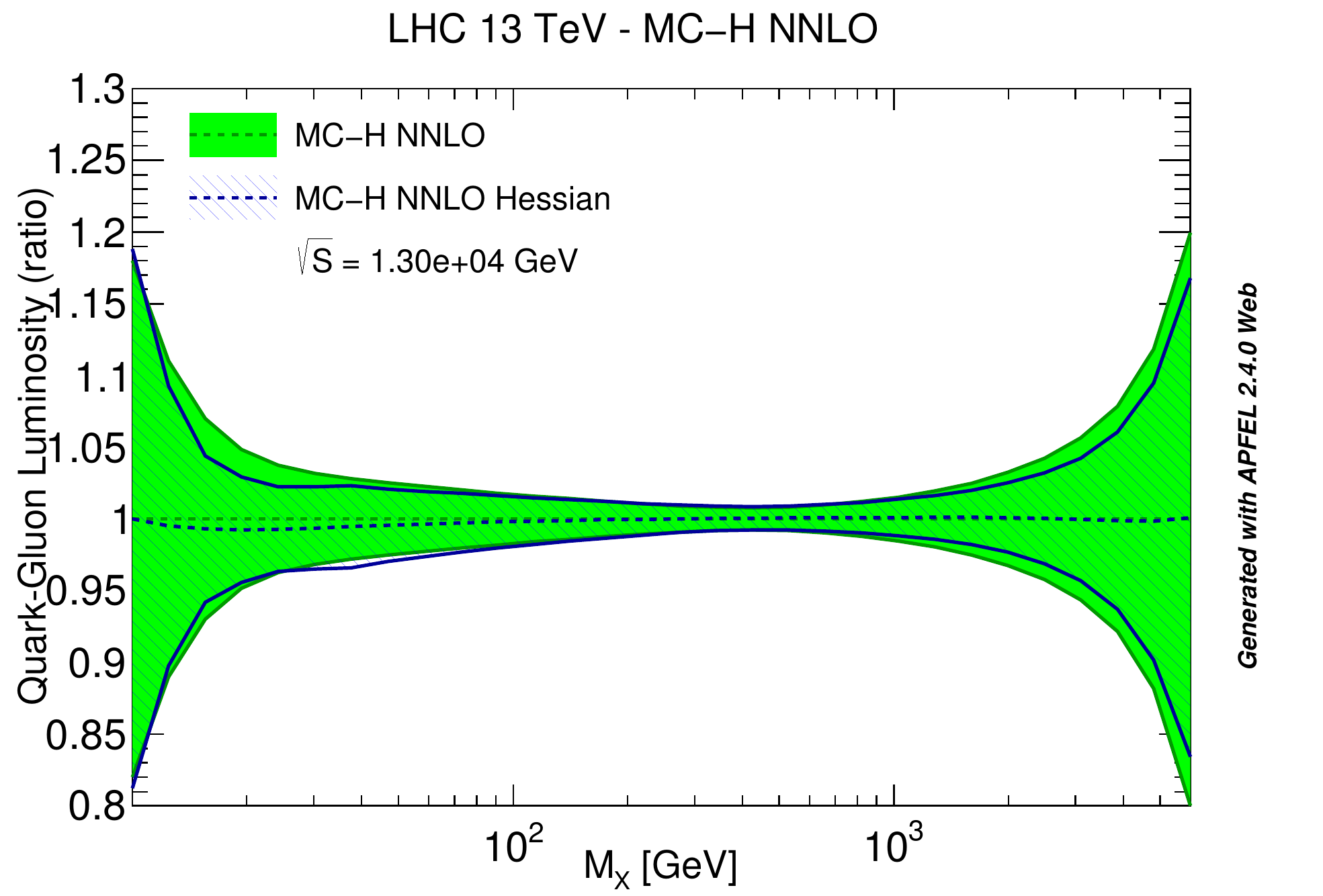}
\includegraphics[width=0.49\textwidth]{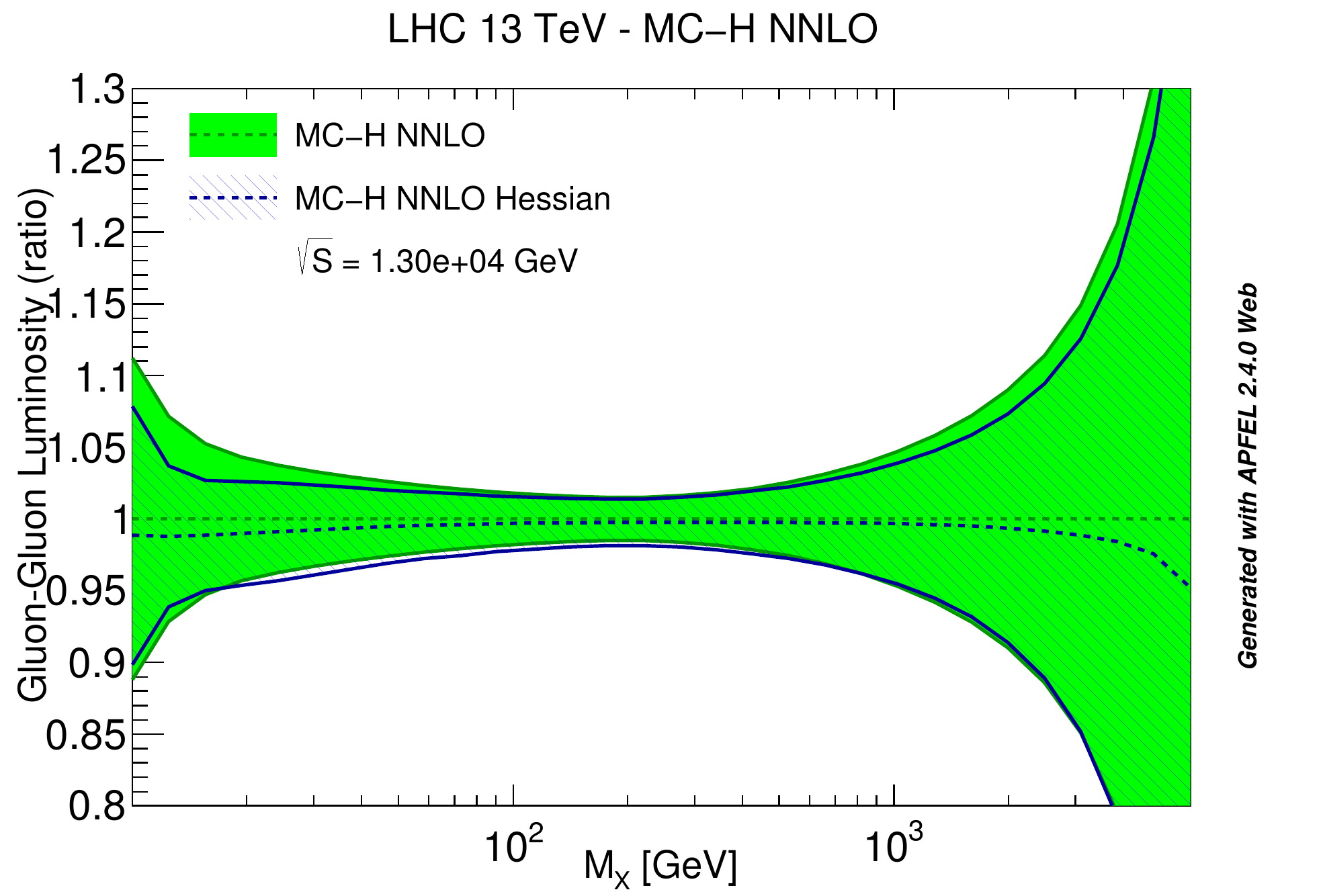}
\end{center}
\vspace{-0.3cm}
\caption{\small \label{fig:lumiscmc}
  Same as Fig.~\ref{fig:lumis}, but now comparing a combined MC PDF
  set with with $N_{\rm rep}=300$ replicas to its MC-H Hessian
  representation with $N_{\rm eig}=90$ eigenvector.
  The starting set is  obtained from the combination of 
 Monte Carlo representations of
    the 
  NNPDF3.0, CT14 and MMHT14 NNLO PDF sets containing $N_{\rm rep}=100$ replicas each.
   }
\end{figure}


We start with a Monte Carlo combination of
the  NNPDF3.0, CT14 and MMHT14 NNLO PDF sets, with
$\alpha_s(M_Z)=0.118$. This is the starting point of the construction
of the compressed sets of  Ref.~\cite{Carrazza:2015hva}, where further
details are given, and it contains $N_{\rm rep} = 300$ replicas.\footnote{Note
  that the CT14 PDF set included in this combination is still preliminary.
We thank Pavel Nadolsky for providing this preliminary version of CT14.
}
We could in principle then first, run the compression algorithm
of Ref.~\cite{Carrazza:2015hva}, and then perform a Monte Carlo to
Hessian conversion of the ensuing compressed set of replicas. However,
each of these two steps entails potential information loss, and thus
it is more advantageous to perform directly a Hessian conversion of
the starting set of  $N_{\rm rep} = 300$ MC replicas. 

We thus apply our Monte Carlo to Hessian conversion to the
combined prior with $N_{\rm rep}=300$, following  the methodology
presented in
Sect.~\ref{sec:methodology}.  It actually turns out that significant
deviations from Gaussian behavior are observed for PDFs  for which
direct experimental information is scarce, and thus theoretical bias
or constraints play some role, such as the strange PDF.
Once a
final combined set is made available for phenomenology, a choice will
have to be made in order to decide whether a Hessian approximation is
viable. For the time being, given the preliminary nature of the
existing set, we simply choose $\epsilon=0.25$ as in
Sect.~\ref{sec:numerical} as a threshold for discarding non-Gaussian points.
 We then end up with an optimal conversion  
with $N_{\rm eig} = 90$ eigenvectors, somewhat larger though of the
same order than the number
of compressed replicas of the CMC set ($N_{\rm rep}\sim 40$).
%

\begin{figure}[t]
\begin{center}
  \includegraphics[width=0.42\textwidth]{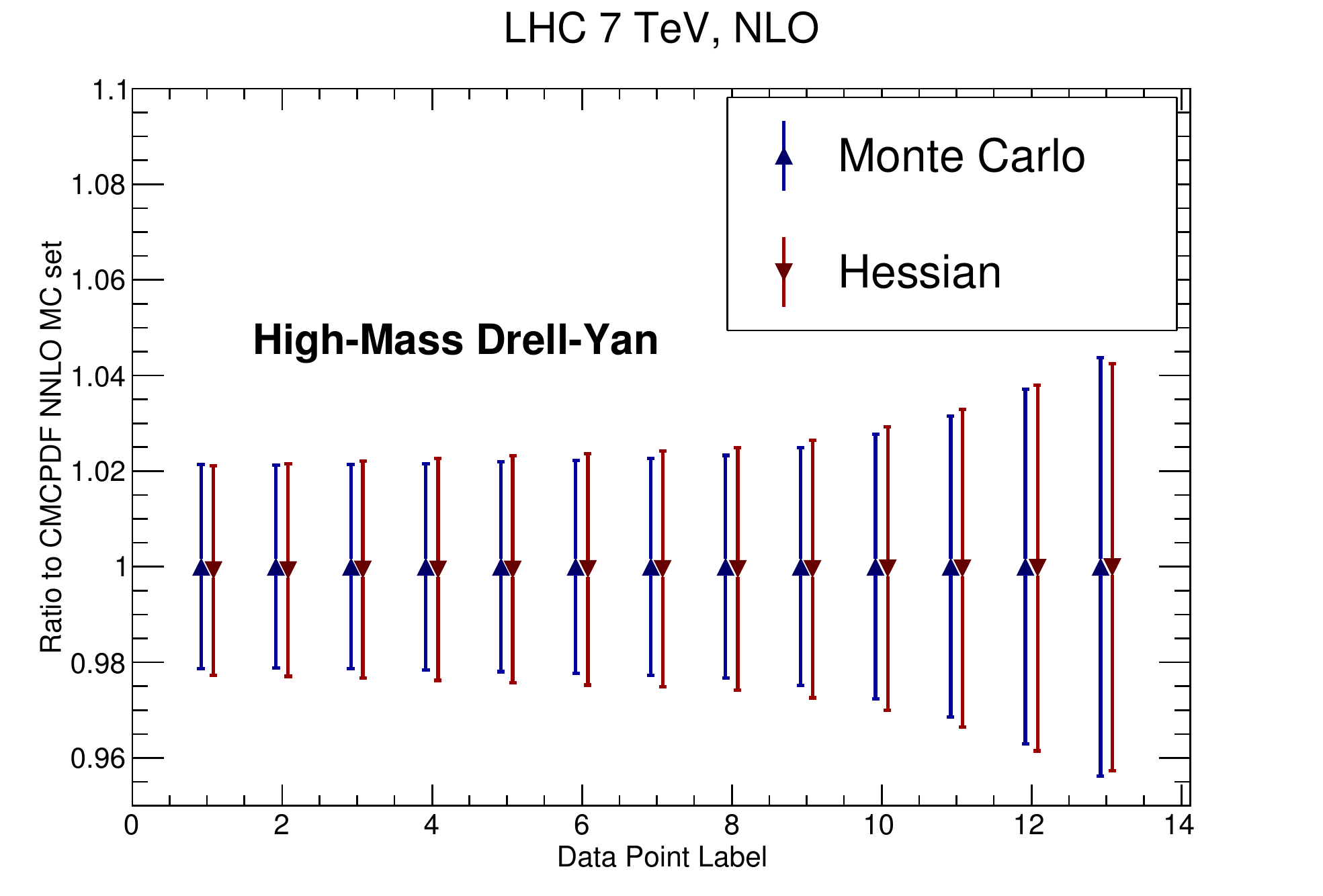}
  \includegraphics[width=0.42\textwidth]{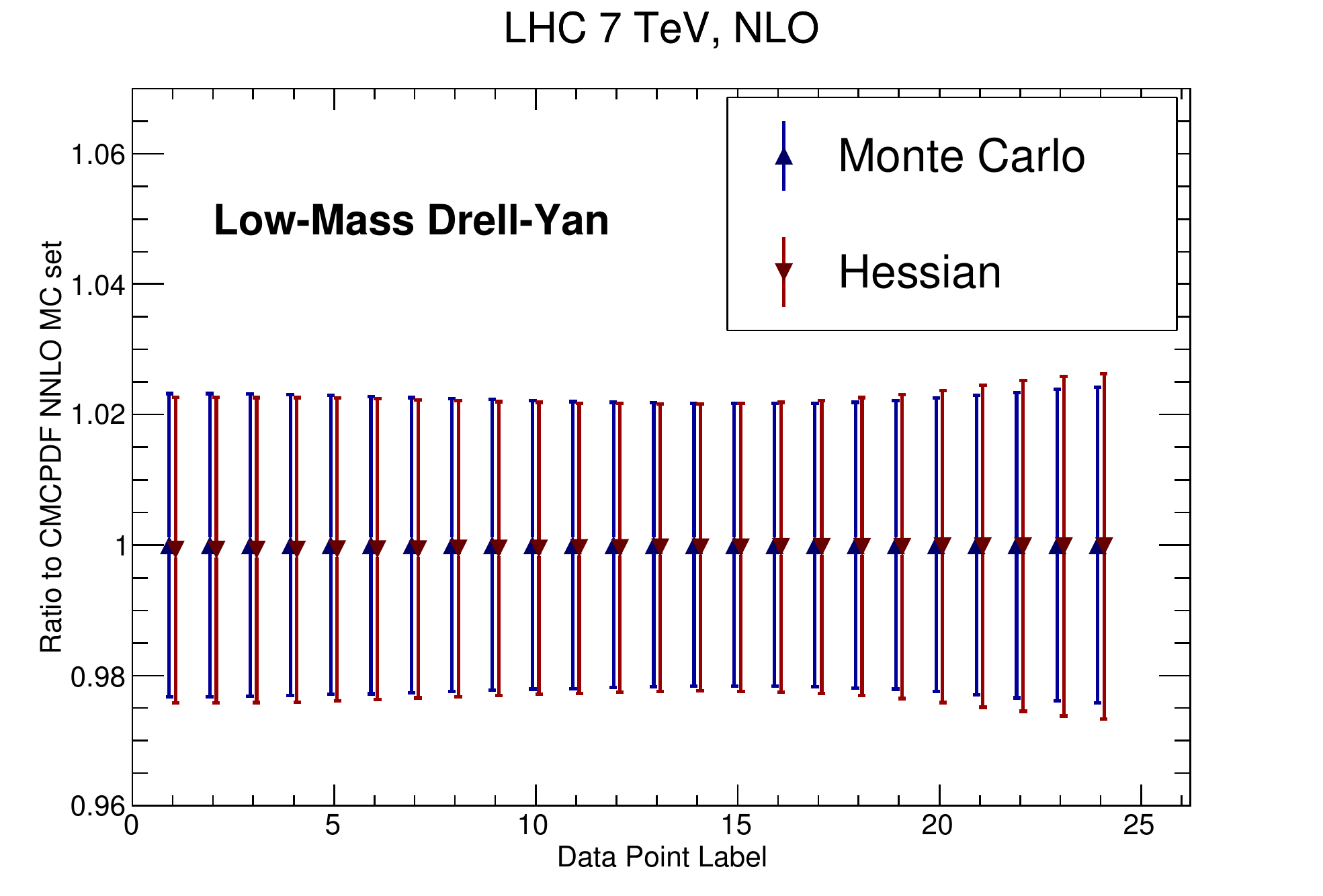}
  \includegraphics[width=0.42\textwidth]{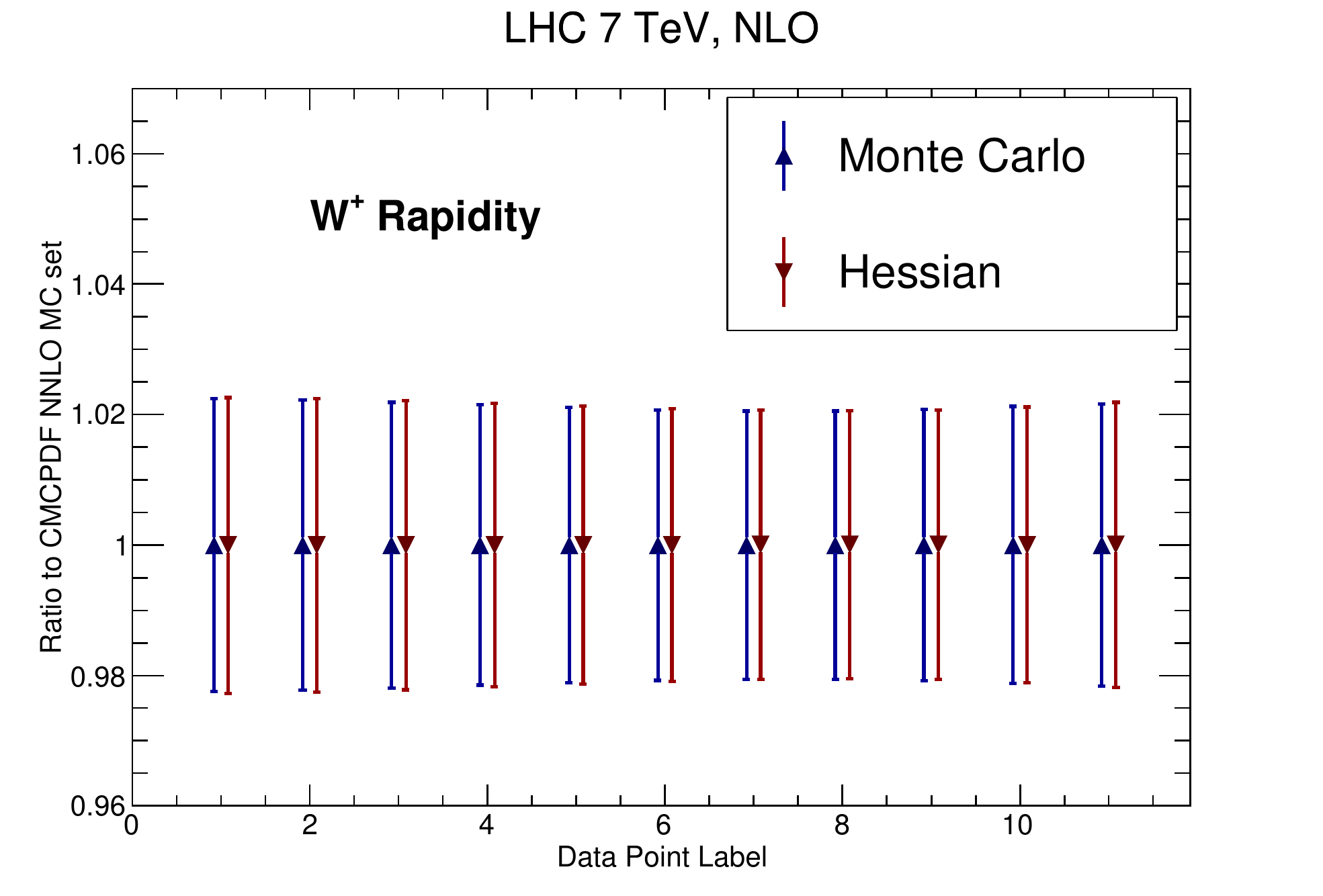}
  \includegraphics[width=0.42\textwidth]{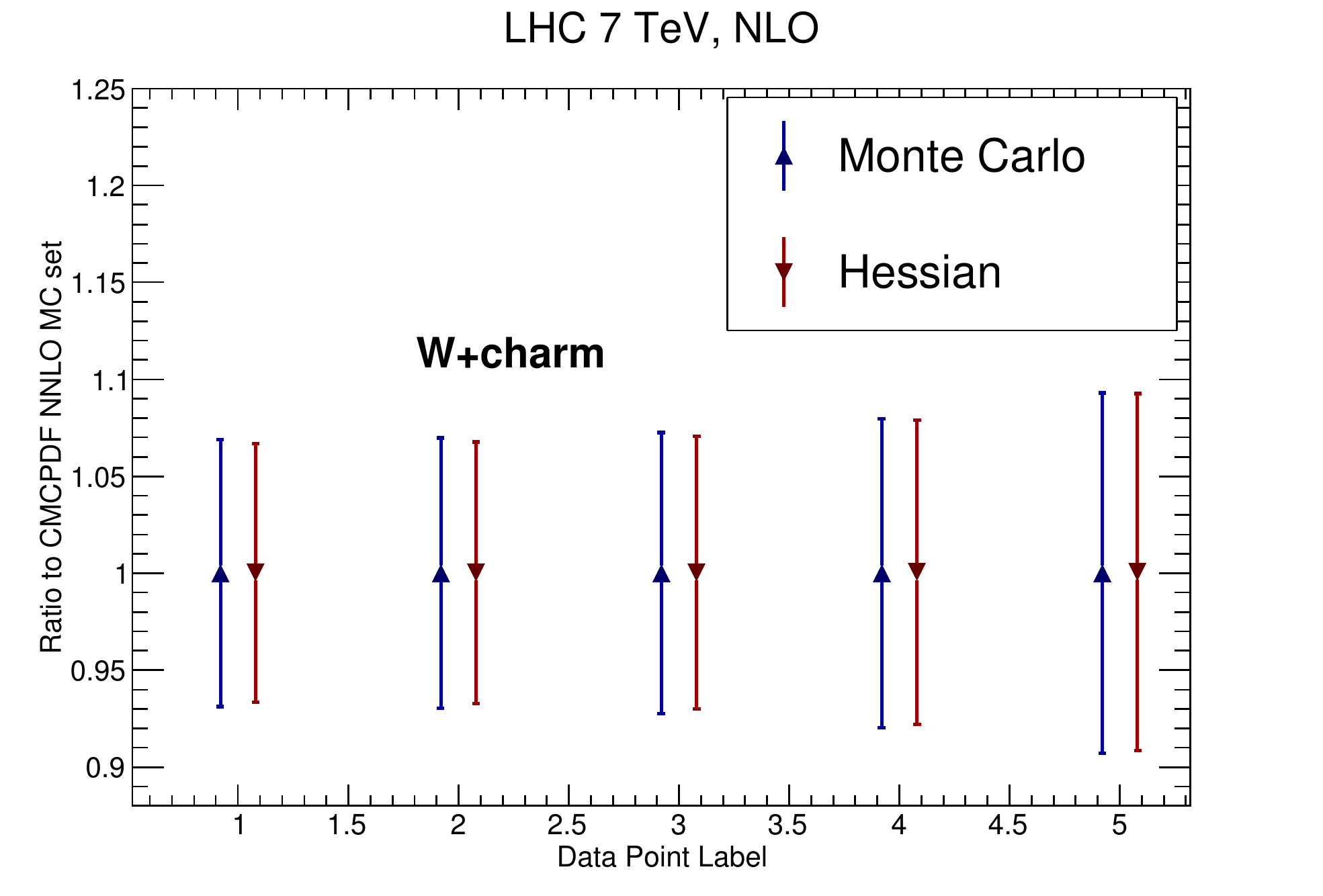}
  \includegraphics[width=0.42\textwidth]{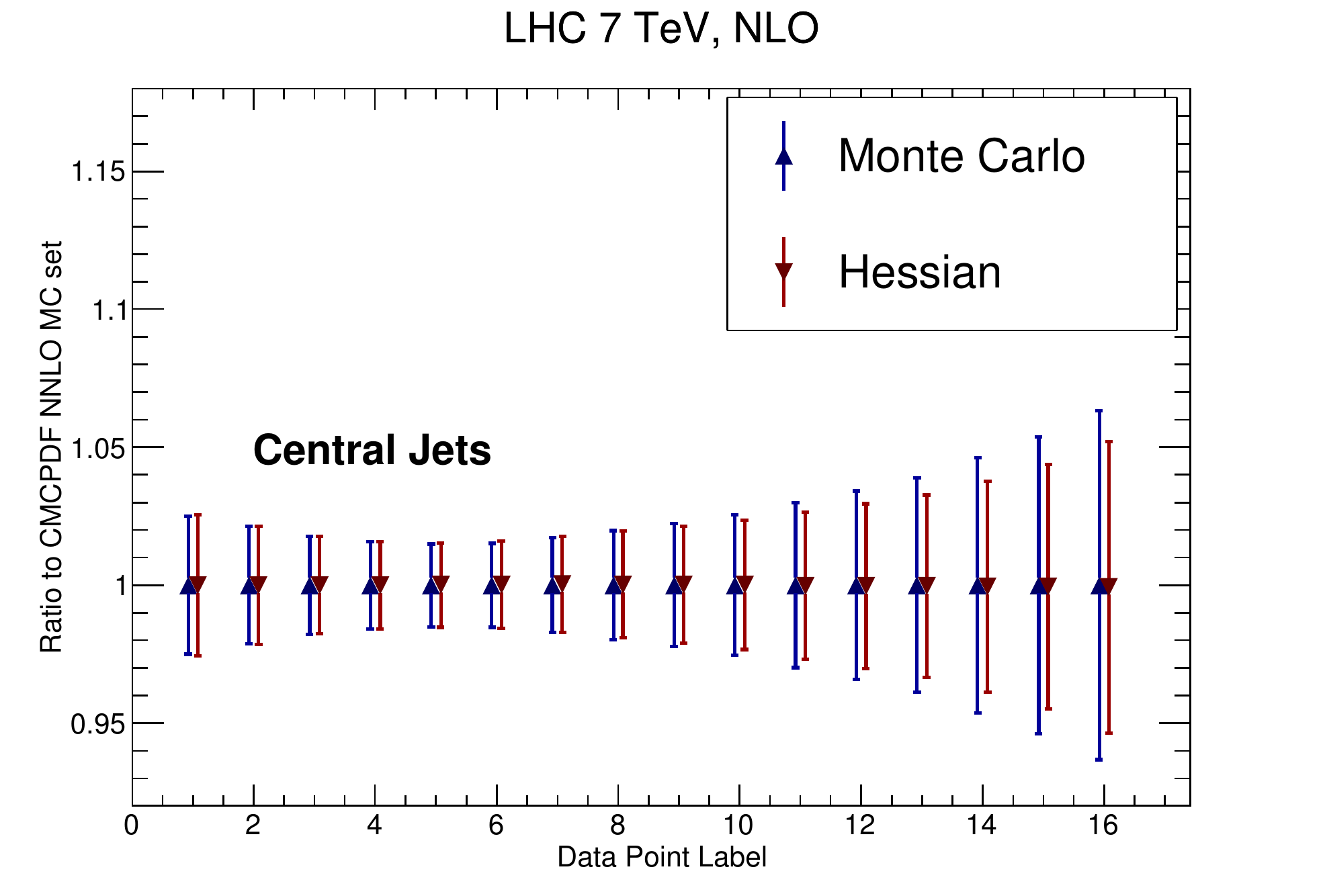}
  \includegraphics[width=0.42\textwidth]{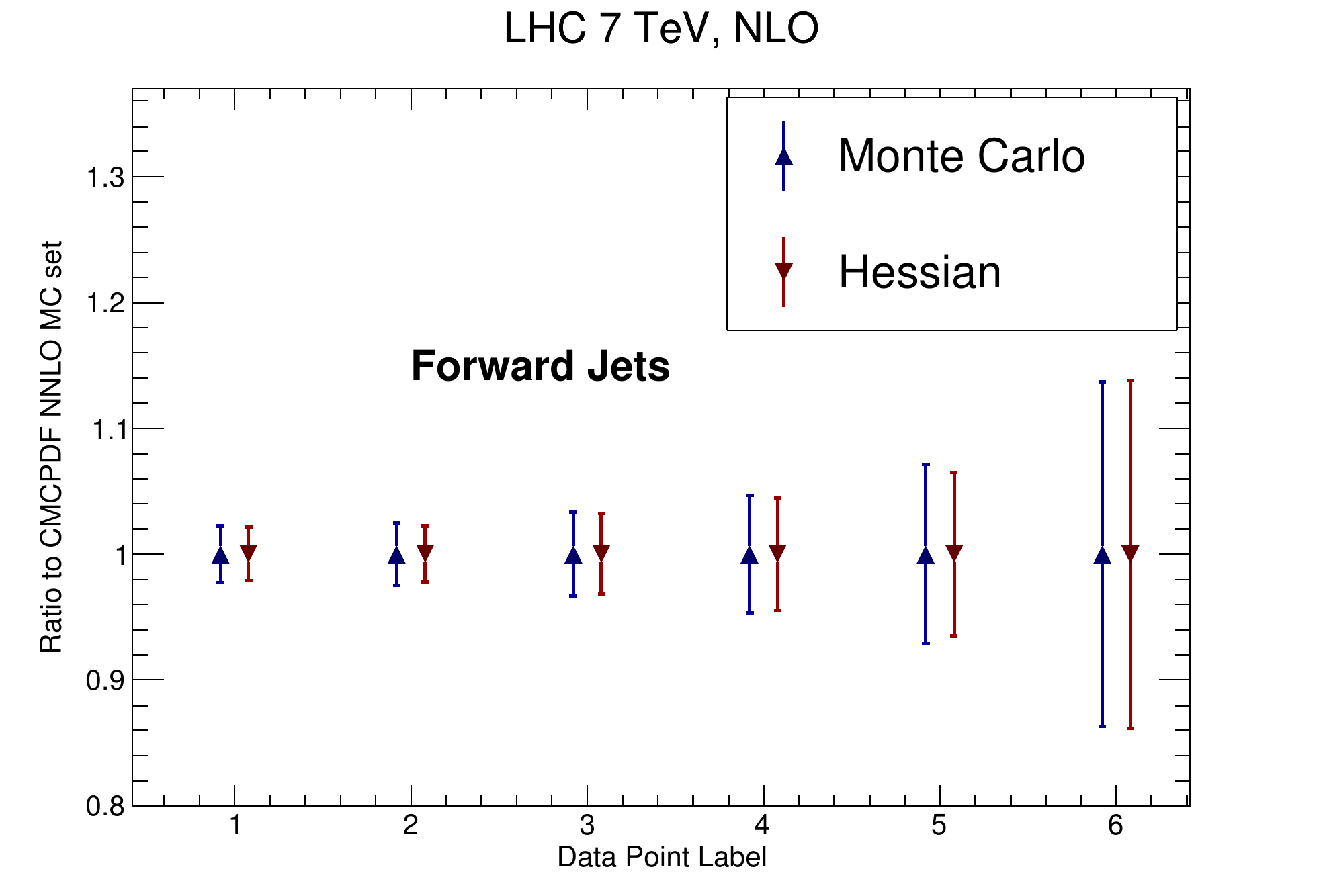}
\end{center}
\vspace{-0.3cm}
\caption{\small \label{fig:lhcmc}
  Same as Fig.~\ref{fig:lhcapplgrid1}, but now for the PDF sets
  of Fig.~\ref{fig:lumiscmc}.
}
\end{figure}

We have performed an extensive set of validation tests of these MC-H
PDFs, at the level of PDFs, luminosities
and physical observables, of which we now show some examples.
In Fig.~\ref{fig:lumiscmc} we compare
PDF luminosities at the LHC
with $\sqrt{s}=$13 TeV computed with the starting combined MC set and
the final MC-H Hessian set.
We find good agreement, with
differences below 10\% at the level of PDF uncertainties.
It is important to note that some disagreement is to be expected 
because the starting combined set is not
Gaussian,
in particular for regions of $x$ (such as
large and small $x$) and PDFs
 that are poorly constrained by experimental data: indeed, the
largest discrepancies are observed at low and high invariant masses $M_X$.
These differences thus signal an intrinsic limitation of the Hessian
representation, rather than  a failure of our methodology.

The good agreement at the level of PDF luminosities translates into
a good agreement at the level of physical observables. In 
Fig.~\ref{fig:lhcmc} we compare the processes that we discussed
in Sect.~\ref{sec:lhcpheno} (with the same settings), computed using
the starting combined MC PDF set and the final MC-H Hessian representation.
Again, results are normalized to the central value of the starting
combined set, and uncertainties bands correspond to the
one-sigma PDF uncertainty in each bin (recall that central values of
the starting and final PDF sets are the same by construction).
Again, discrepancies are below the 10\% level.

We concluded that the Hessian conversion algorithm presented in this
paper also provides a successful methodology for the construction of a
Hessian representation with a moderate number of eigenvectors of
combined Monte Carlo PDF sets. Differences between the starting MC set
and the final MC-H Hessian representation  can be used,
to a certain extent,
to quantify the degree of non-gaussianity which is present in the original
set.

\section{Delivery and outlook}
\label{sec:delivery}
We have provided a general purpose methodology for the Hessian
conversion of any Monte Carlo PDF set. When applied to a native Monte
Carlo set, this methodology provides an efficient Hessian
representation which is faithful to the extent that the starting set
is Gaussian. When applied to a Monte Carlo set obtained from a
starting Hessian, the methodology gives back the original set to very
good accuracy, but using the Monte Carlo replicas as a basis. Finally,
when applied to a combined Monte Carlo replica set it provides a
Hessian version (MC-H PDFs) of the recently proposed PDF compression
methodology (MC-PDFs~\cite{Carrazza:2015hva }), and an implementation
of the Meta-PDF idea~\cite{Gao:2013bia} which is  free of the
bias related to the choice of a specific functional form.

The main deliverable of this work is the {\tt mc2hessian}
code, which easily allows for the construction of a Hessian
representation of any given Monte
Carlo PDF set.
The {\tt mc2hessian} code is
written in {\tt Python} using the numerical implementations provided
by the {\tt NumPy/Scipy} packages~\cite{scipy}. 
This  code is publicly available from the GitHub repository
\begin{center}
  \url{https://github.com/scarrazza/mc2hessian}
\end{center}
and outputs results directly in the {\tt
  LHAPDF6} format~\cite{Buckley:2014ana}, so
that the new Hessian sets can be easily interfaced by
any other code.
However, it should be kept in mind that the Hessian representation
always requires careful validation, as some information loss is necessarily
involved in this transformation, and specifically any deviation from
Gaussianity is inevitably washed out.

The Hessian version of the NNPDF3.0 sets 
\begin{center}
\small
\tt NNPDF30\_nlo\_as\_0118\_hessian \\
NNPDF30\_nnlo\_as\_0118\_hessian
\end{center}
as well as the MC-H PDFs 
\begin{center}
\small
\tt MCH\_nlo\_as\_0118\_hessian \\
MCH\_nnlo\_as\_0118\_hessian 
\end{center}
will be made available in {\tt LHAPDF6}.
Future NNPDF releases will be provided both in the native Monte Carlo
and in the new Hessian representations.

An  interesting development of the methodology suggested here 
is that an unbiased Hessian representation could be used as a way to
single out the PDF flavours (and $x$-ranges) that provide the dominant
contribution to individual physics process, by picking the
dominant eigenvectors; along the lines of previous
suggestions~\cite{Pumplin:2009nm,Dulat:2013kqa}, but in a
parametrization-independent way. This could then be used to construct
tailored sets with a small number of eigenvectors
which, though not suitable for general-purposes
studies, could be useful for experimental profiling when
restricted to a small subset
of relevant processes.
These and related
issues will be the subject of future work.

\section*{Acknowledgments}

We are grateful to all the members of the NNPDF Collaboration for
fruitful discussions during the development of this project.
The work presented here was triggered by a number of provocative
suggestions at the PDF4LHC workshop and the Higgs cross-section
working group meeting in January 2015: in this
respect, we would like to thank Josh Bendavid, Andr\'e David,
Joey Huston, Jun Gao, Sasha Glazov and Pavel Nadolsky.
The work of J.~R. is supported by an STFC Rutherford Fellowship
ST/K005227/1 and by an European Research Council Starting Grant
``PDF4BSM''. S.~C. and S.~F. are supported in part by an Italian
PRIN2010 grant and by a European Investment Bank EIBURS
grant. S.~F. and Z.~K. are supported by the Executive Research Agency
(REA) of the European Commission under the Grant Agreement
PITN-GA-2012-316704 (HiggsTools).

\appendix

\section {Hessian representation through Singular Value Decomposition}
\label{sec:appendix}

The solution to the problem of finding a suitable Hessian representation of
MC PDF sets is not unique.
The main strength of the approach we have explored here is that a
representation in terms of the original MC replicas automatically
inherits a number of useful properties of the replicas, such as, for
instance,  the fact
that sum rules are automatically satisfied and  correlations are well
reproduced, as discussed in Sect.~\ref{sec:mch}.
This suggests an alternative approach, in which instead of
representing all starting replicas on a subset of them, we pick the
dominant combination of all replicas through
Singular Value Decomposition (SVD).
This alternative method is briefly discussed in this Appendix. So far
we have used it for validation of our main methodology, though it
might be especially suitable for future applications, such as the
construction of reduced eigenvector sets for specific physical processes.

Since the replicas of a MC set are continuous
functions with finite correlation length, 
the general problem of finding a Hessian representation
of a MC set can be interpreted as that of finding a representation of
a discrete covariance matrix of the form Eq.~(\ref{eq:covmat}):
the sampling in the space of
$x$ needs only to be fine grained enough that differences
between neighboring points are non-negligible.

Such  representation can be directly constructed in terms of Monte
Carlo replicas, in the following way.
We define the rectangular $(N_{x}N_{f})\times
N_{\rm rep}$ matrix
\be
X_{lk}= f^{(k)}_\alpha(x_i,Q_0) - f^{(0)}_\alpha(x_i,Q_0) \, ,
\ee
where we adopted the  same
convention for indices as in Eq.~(\ref{eq:matrixes}): 
$\alpha$ labels PDFs, $i$
points in the $x$ grid, $l \equiv N_x(\alpha-1) + i$ runs over all
$N_{x}N_{f}$ grid points, and 
 $k$ runs over all  MC replicas.
The covariance matrix Eq.~(\ref{eq:covmat}) is equal to
\be\label{eq:covtox}
{\rm cov}^{\rm pdf}_{ij,\alpha\beta}= \frac{1}{N_{\rm rep} - 1} XX^t .
\ee

A diagonal representation of the covariance matrix in terms of replicas is
found by  SVD of the matrix $X$. 
Namely, we can write $X$  as:%
\be
X=USV^t\ .
\ee
Assuming that  $N_{\rm pdf}N_x<N_{\rm rep}$,  $U$ is an orthogonal matrix of
dimensions $N_{\rm pdf}N_x\times N_{\rm rep}$ which contains 
the orthogonal eigenvectors of the
covariance matrix with nonzero eigenvalues; $S$ is a  diagonal matrix of real positive elements,
constructed out of the singular values of $X$, i.e. the  square
roots of the nonzero eigenvalues of $\rm{cov}^{\rm pdf}$ multiplied by
the normalization constant $(N_{\rm rep} - 1)^\frac{1}{2}$; and $V$ is
an orthogonal
$N_{\rm rep} \times N_{\rm rep}$ matrix of coefficients.

Because
\be
XX^t = US^2U^t = (US)(US)^t\ ,
\ee
the matrix 
\be
Z= US  
\ee
has the property that
\be\label{eq:ztox}
ZZ^t= XX^t .
\ee
But also,
\be\label{eq:zlin}
Z= XV
\ee
and thus $Z$ provides the sought-for representation of the covariance
 matrix as a linear combination of MC replicas.

We have thus arrived at an exact Hessian representation of the
covariance matrix in terms of replicas, but with the disadvantage that
the number of Hessian eigenvectors parameters is now equal to  $N^{(0)}_{\rm
eig}=N_{\rm pdf}N_x$, which is 
generally large.
However, in practice many of these eigenvectors
will lead to a very small contribution to the covariance matrix. We
can then select a smaller set of  $N_{\rm
eig}<N^{(0)}_{\rm eig}$ eigenvectors which still provides a good
approximation to the covariance matrix.

A possible strategy is, for example, to select the $N_{\rm eig}$
eigenvectors with largest singular values.
Denoting with  $u$,
$s$, and $v$  the $N_{\rm pdf}N_x\times N_{\rm eig}$, $N_{\rm eig}\times
N_{\rm eig}$ and $N_{\rm eig} \times N_{\rm rep}$ reduced matrices
computed using these eigenvalues,
for a given
value of $N_{\rm eig}$, using $v$ instead of $V$ in
Eq.~(\ref{eq:zlin}) 
minimizes the difference between the original and reduced  covariance
matrix
\be\label{eq:deltaind}
\Delta\equiv \left\Vert US^2U^t - us^2u^t  \right\Vert \, .
\ee

\begin{figure}[t]
\begin{center}
\includegraphics[width=0.49\textwidth]{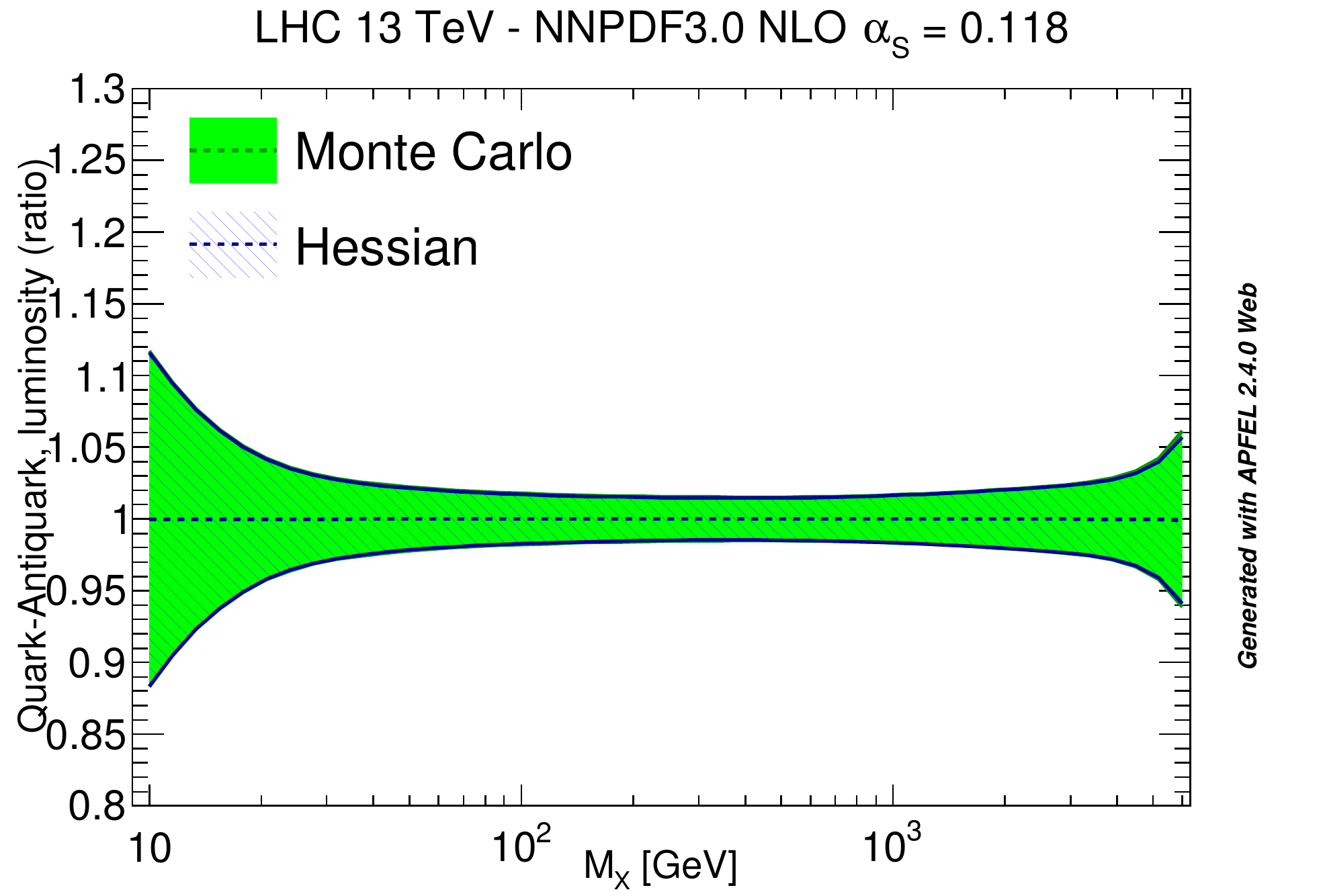}
\includegraphics[width=0.49\textwidth]{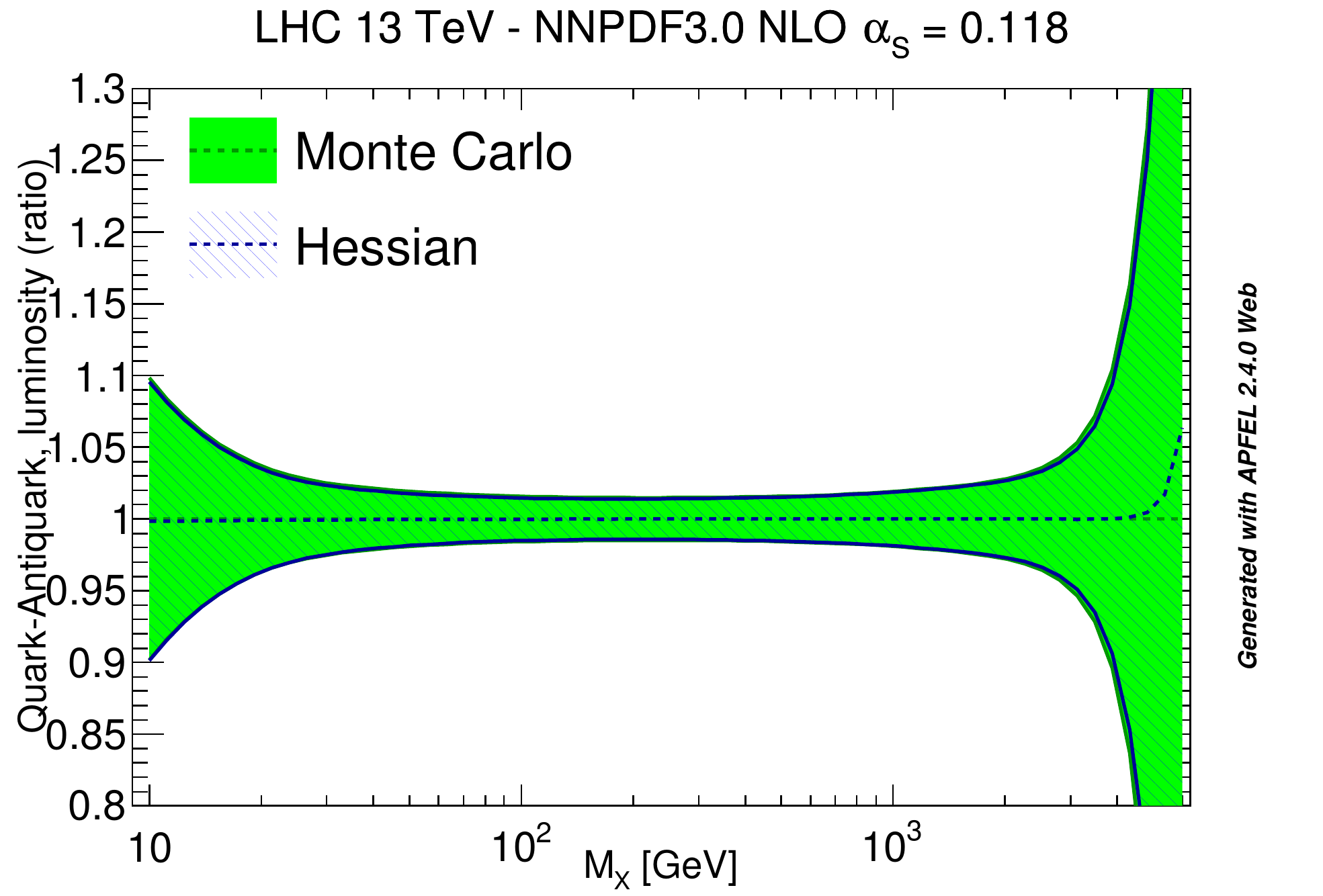}
\includegraphics[width=0.49\textwidth]{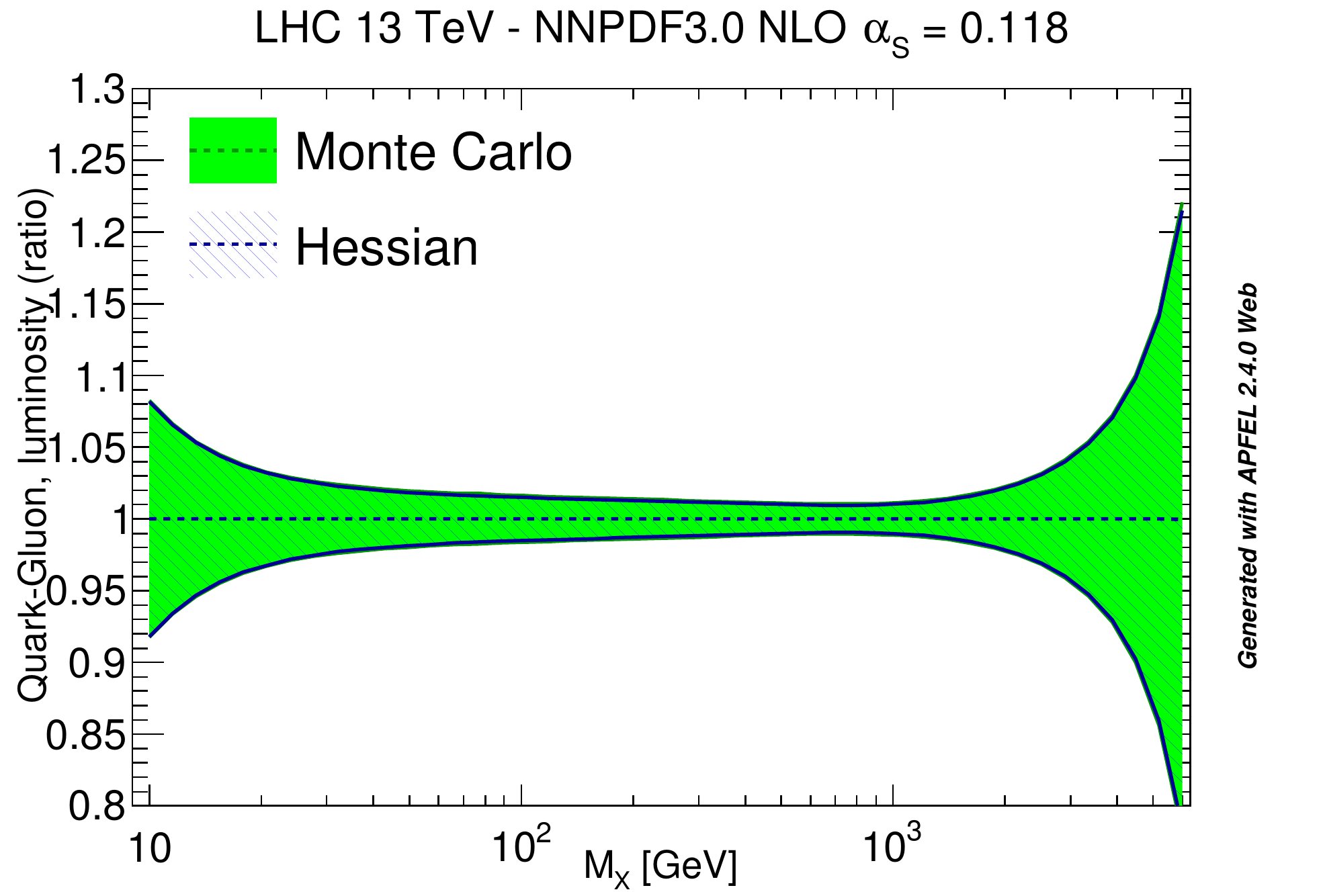}
\includegraphics[width=0.49\textwidth]{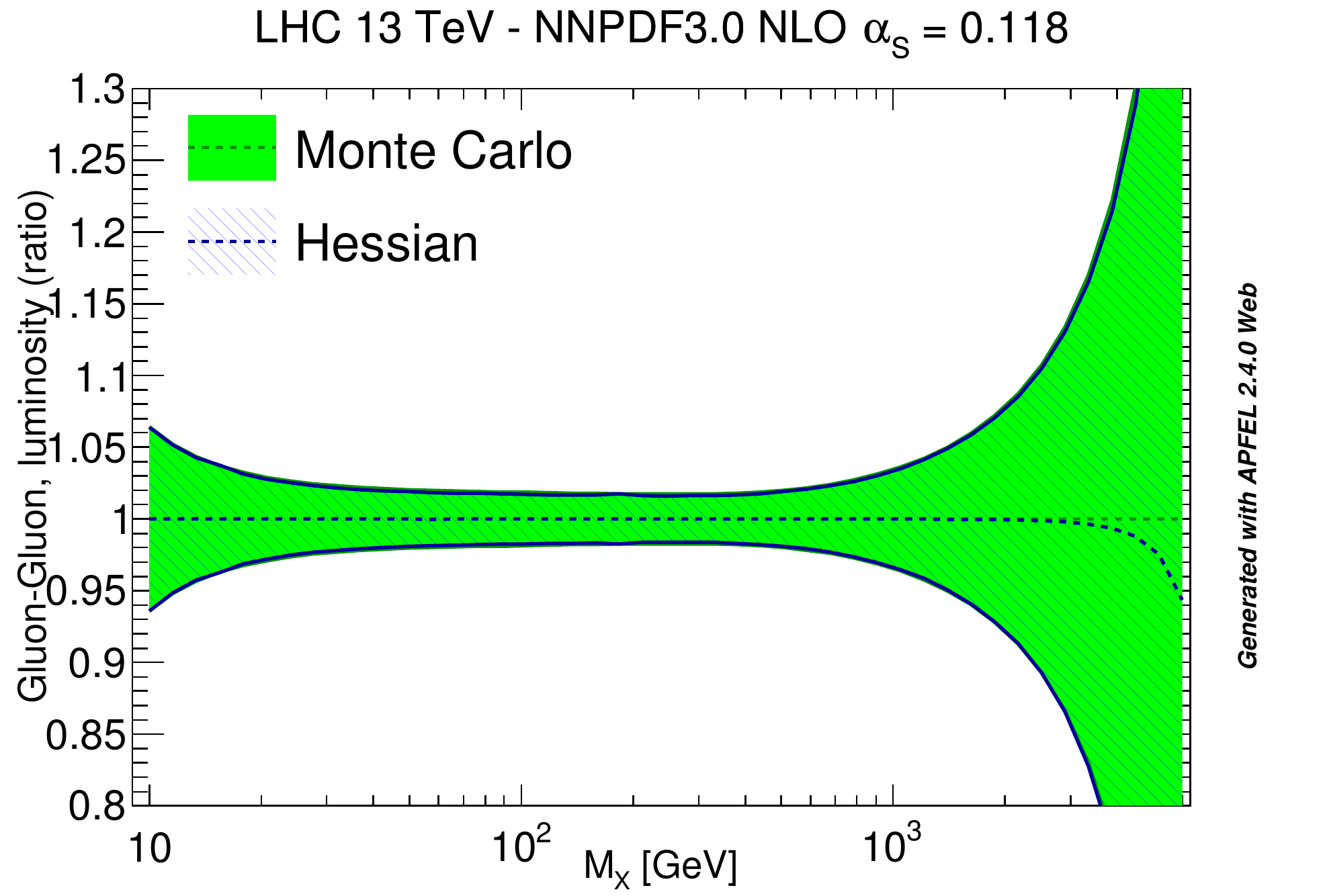}
\end{center}
\vspace{-0.3cm}
\caption{\small \label{fig:lumis-newmethod} Same as
  Fig.~\ref{fig:lumis} but now  with the Hessian representation
   constructed using the method discussed
   in this Appendix. }

\end{figure}

We have  verified that the method described in this Appendix
provides comparable results to those of the main strategy
discussed in the paper. To this purpose, we 
have selected the  set of $N_{\rm eig}=120$
eigenvectors that minimize the figure of merit
Eq.~(\ref{eq:deltaind}).  To illustrate the quality of the new method
in Fig.~\ref{fig:lumis-newmethod} we show
 a comparison of PDF luminosities, analogous to Fig.~\ref{fig:lumis},
and in Fig.~\ref{fig:lhcapplgrid1-newmethod} a comparison of LHC
 differential distributions, analogous to  Fig.~\ref{fig:lhcapplgrid1},
but now obtained
with the new method.
Is clear that the agreement is comparab le as that found with the
previous method.

\begin{figure}[t]
\begin{center}
 \includegraphics[width=0.42\textwidth]{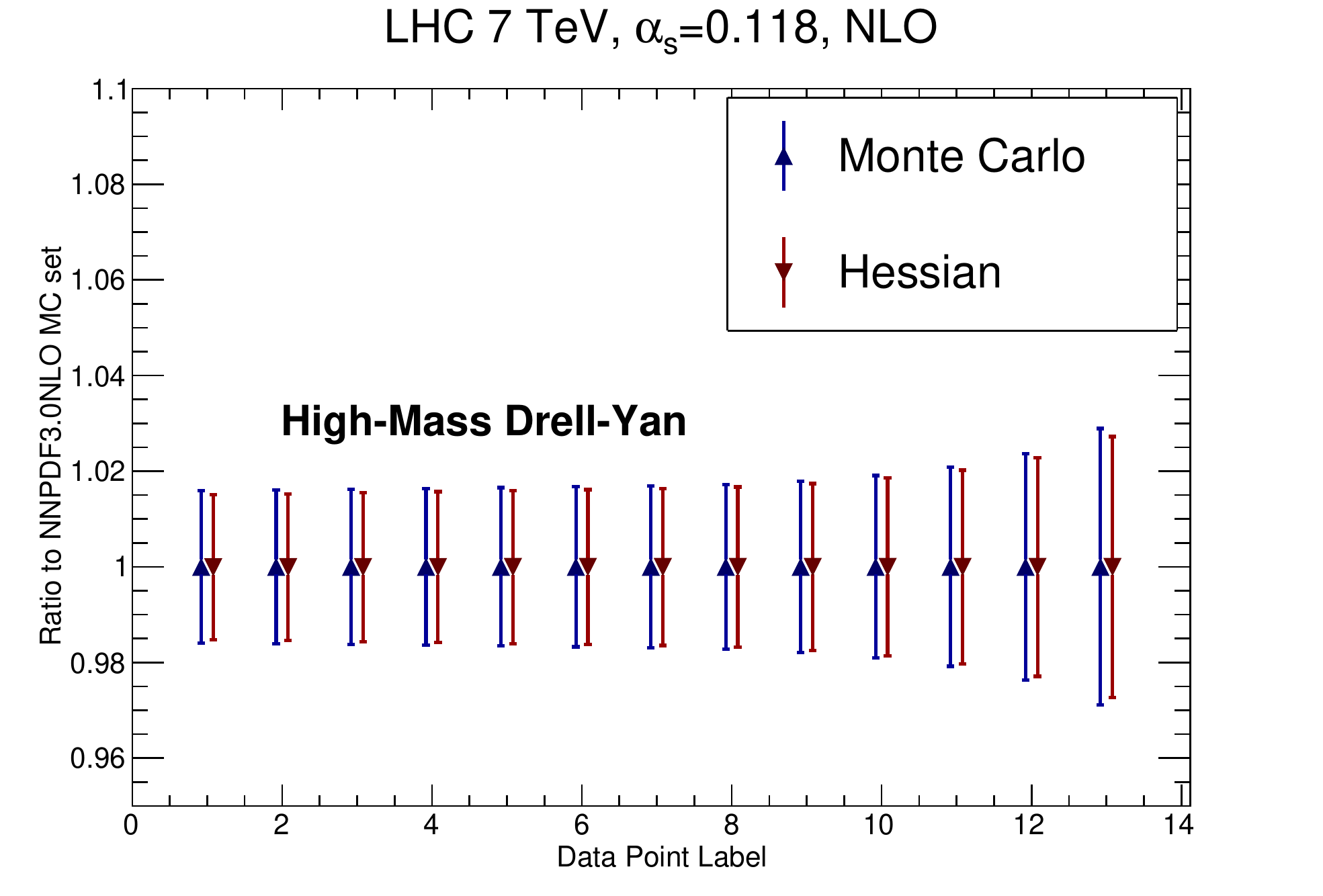}
 \includegraphics[width=0.42\textwidth]{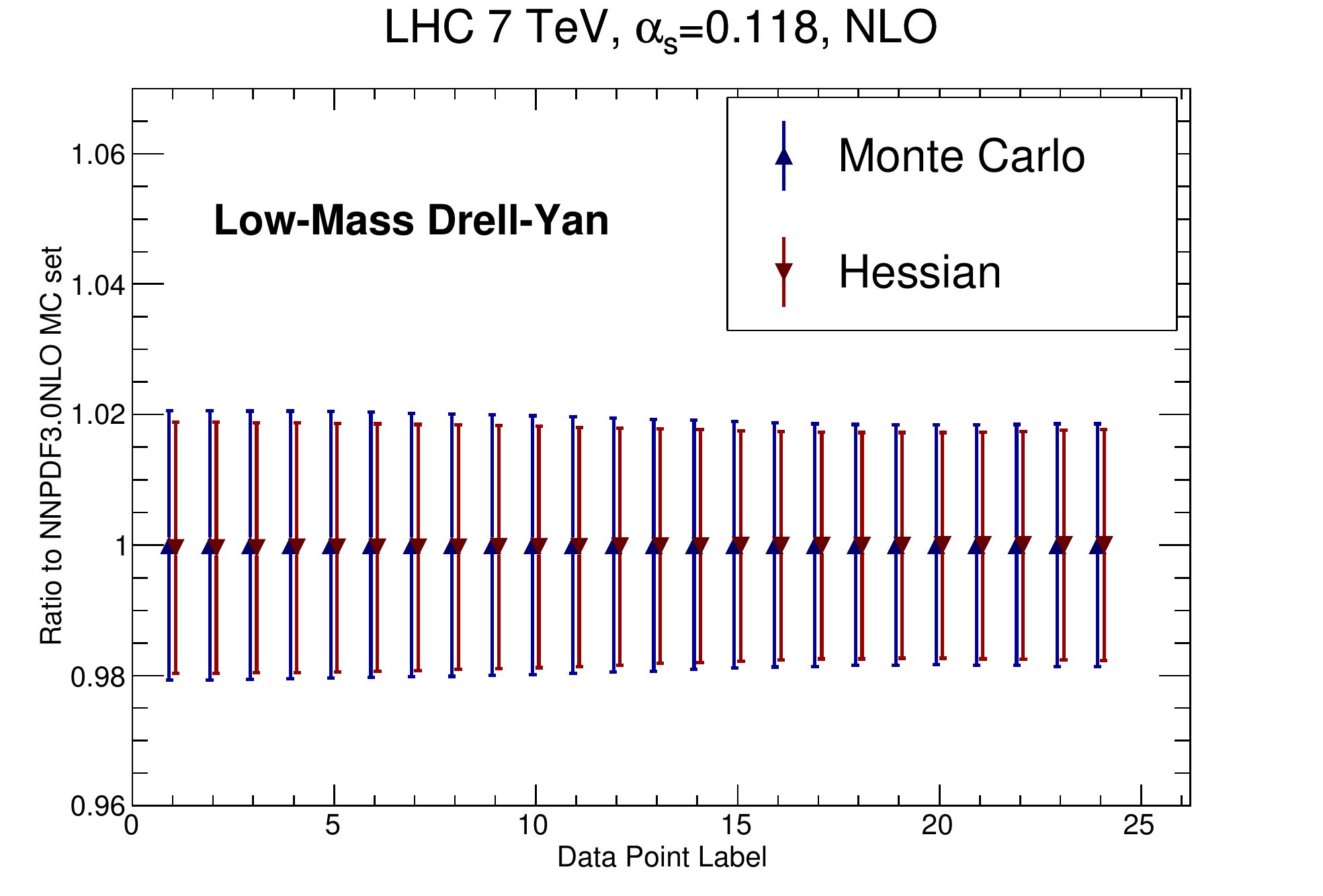}
 \includegraphics[width=0.42\textwidth]{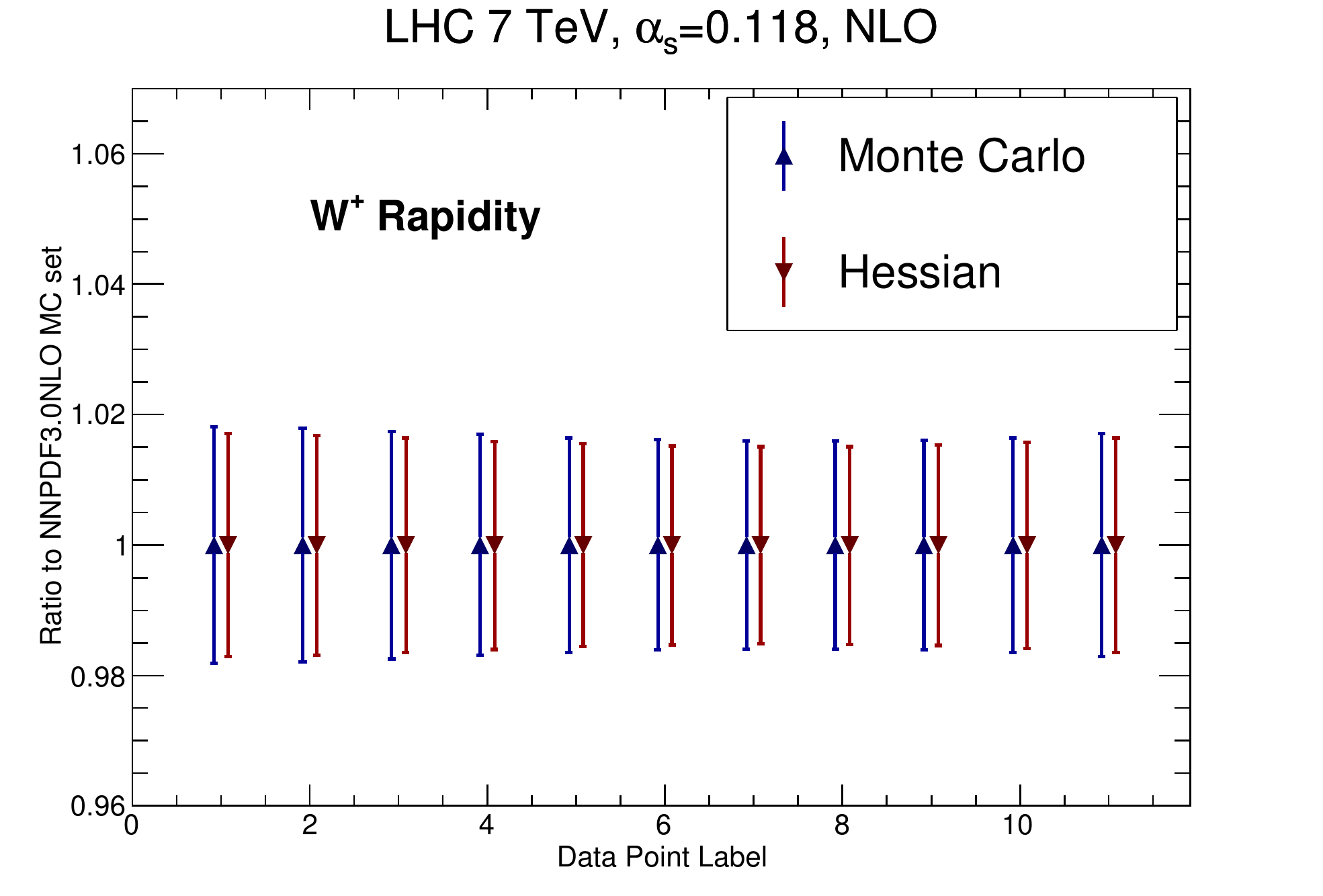}
 \includegraphics[width=0.42\textwidth]{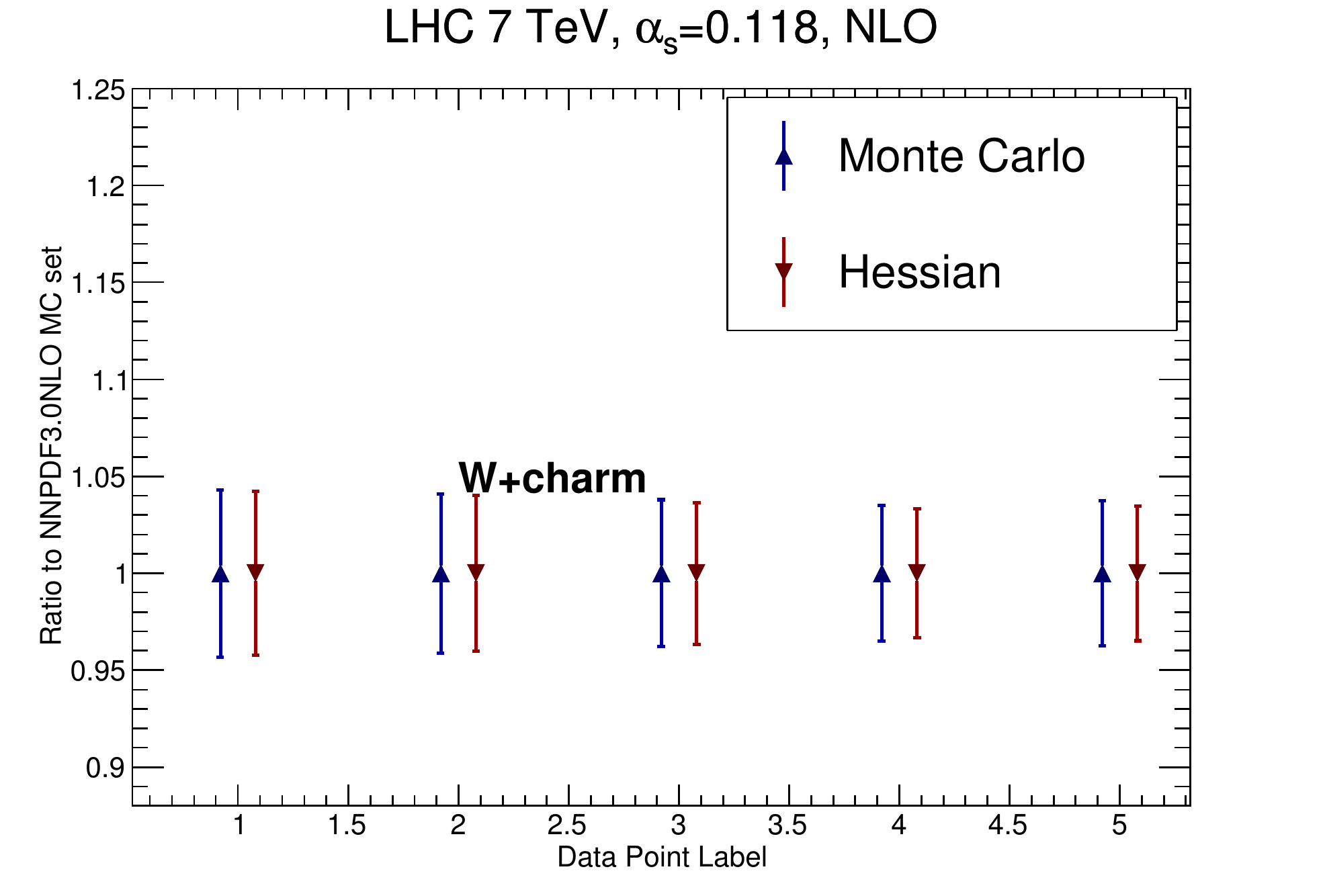}
 \includegraphics[width=0.42\textwidth]{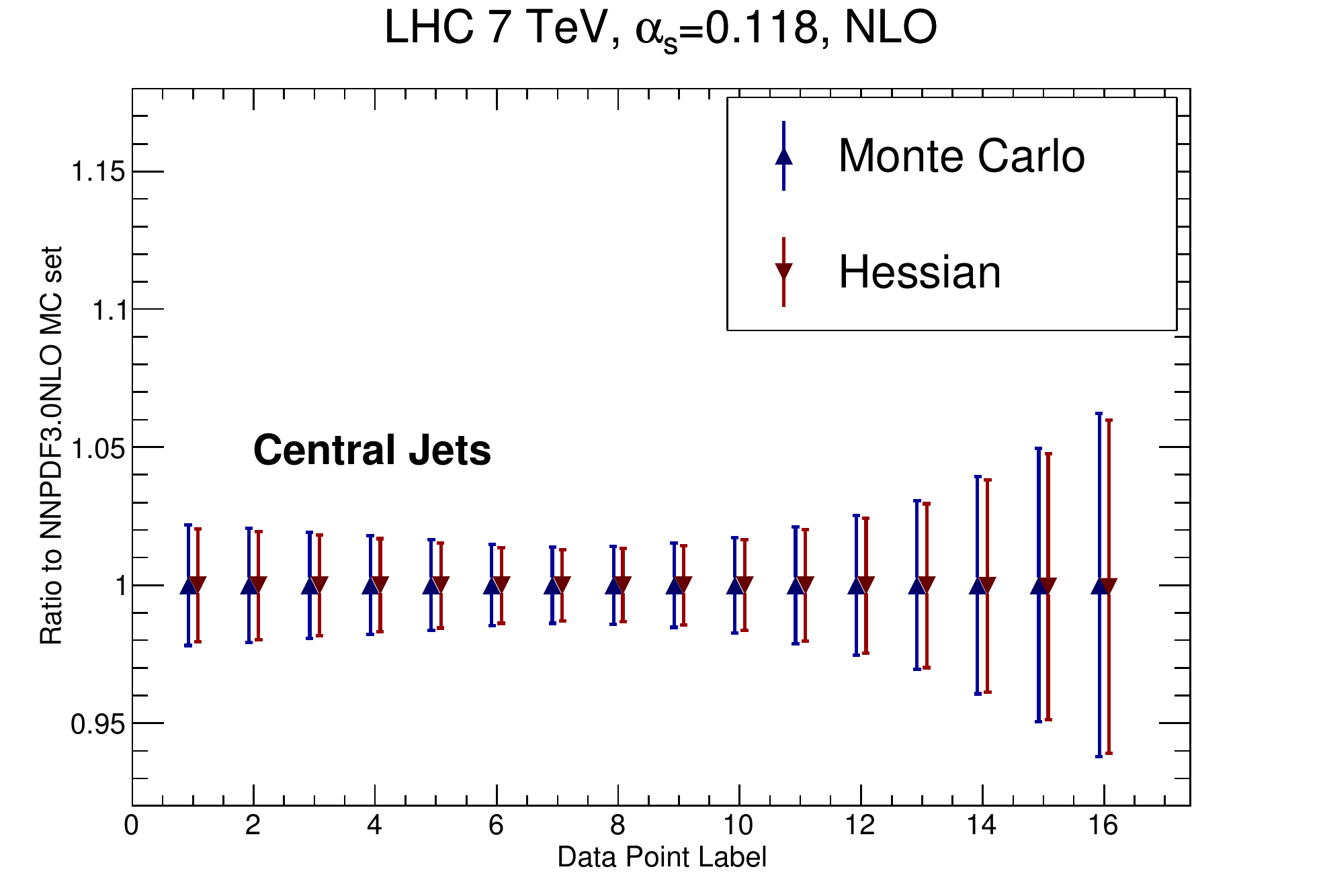}
 \includegraphics[width=0.42\textwidth]{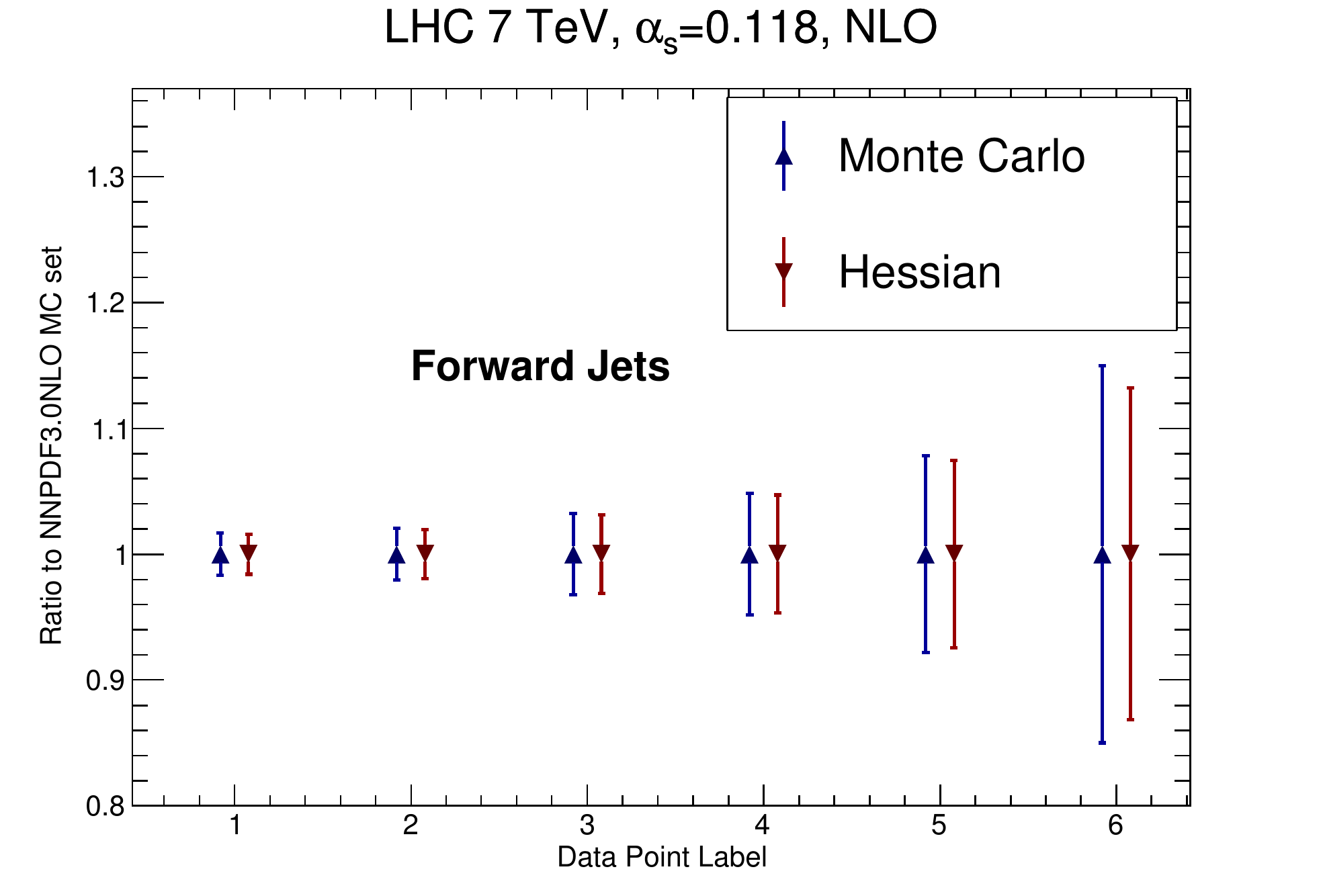}
\end{center}
\vspace{-0.3cm}
\caption{\small \label{fig:lhcapplgrid1-newmethod} Same as
   Fig.~\ref{fig:lhcapplgrid1} but now  with the Hessian representation
   constructed using the method discussed
   in this Appendix.}
\end{figure}

The main shortcoming of this method is that minimizing
Eq.~(\ref{eq:deltaind})  is not necessarily the best strategy for the
construction of an optimal general-purpose eigenvector set, in that it
does not always lead to also minimizing the 
figure of merit Eq.~(\ref{eq:estimator})
which was shown in Sect.~\ref{sec:numerical} to be physically
advantageous. Some tuning of the methodology would thus be required.

\providecommand{\href}[2]{#2}\begingroup\raggedright\endgroup

\end{document}